\documentclass[a4paper,12pt]{scrartcl}
\usepackage{cite}
\usepackage{xspace}
\usepackage{graphicx}
\usepackage{pslatex}
\usepackage{amsmath}
\usepackage{amsbsy}
\usepackage{txfonts}
\usepackage{color}

\def\nn{\nonumber}

\def\MSbar{{\ensuremath{\overline{\mbox{MS}}}}\xspace}

\def\as{\ensuremath{\alpha_s}\xspace}

\def\mt{\ensuremath{m_t}\xspace}
\def\mtt{\ensuremath{M_{t\bar t}}\xspace}
\def\CF{\ensuremath{C_F}\xspace}
\def\CA{\ensuremath{C_A}\xspace}
\def\muf{\ensuremath{\mu_f}\xspace}
\def\mur{\ensuremath{\mu_r}\xspace}
\def\AQCDLO{\ensuremath{{\cal A}_{\QCD}^{(0)}}\xspace}
\def\AQCDLOcc{\ensuremath{{{\cal A}_{\QCD}^{(0)}}^{\hspace{-0.3cm}\ast}\hspace{0.15cm}}\xspace}
\def\AQCDNLO{\ensuremath{{\cal A}_{\QCD}^{(1)}}\xspace}
\def\APhiiLO{\ensuremath{{\cal A}_{\phi_i}^{(0)}}\xspace}
\def\APhijLO{\ensuremath{{\cal A}_{\phi_j}^{(0)}}\xspace}
\def\APhixLO#1{\ensuremath{{\cal A}_{\phi_{#1}}^{(0)}}\xspace}
\def\APhiiNLO{\ensuremath{{\cal A}_{\phi_i}^{(1)}}\xspace}
\def\B{{\cal B}}
\def\FP#1{{{\cal F}^P_{#1}}}
\def\FS#1{{{\cal F}^S_{#1}}}
\def\fLOP#1{{{f}^{(0)}_{P#1}}}
\def\fLOS#1{{{f}^{(0)}_{S#1}}}
\def\ZPL{\ensuremath{Z_{\text{PL}}}\xspace}
\def\ZPLll{\ensuremath{Z_{\text{PL}}^{ll}}\xspace}
\def\ZPLlj{\ensuremath{Z_{\text{PL}}^{lj}}\xspace}

\def\vev{{\varv}}
\newcommand{\xsNLO}{\sigma^{\text{NLO}}}
\newcommand{\xsNLOm}{\sigma^{\text{NLO\{$m$\}}}}
\newcommand{\xshNLOm}{\hat{\sigma}^{\text{NLO\{$m$\}}}}
\newcommand{\xsNLOmp}{\sigma^{\text{NLO\{$m+1$\}}}}
\newcommand{\xsB}{\sigma^{\text{B}}}
\newcommand{\xsR}{\sigma^{\text{R}}}
\newcommand{\xsV}{\sigma^{\text{V}}}

\def\Re{{\rm Re}}
\def\Im{{\rm Im}}

\def\QCD{{\rm QCD}}
\def\GeV{{\rm GeV}}

\def\Refs#1{Refs.~\cite{#1}}
\def\Ref#1{Ref.~\cite{#1}}
\def\Eq#1{Eq.~(\ref{#1})}
\def\Fig#1{Fig.~\ref{#1}}
\def\Figs#1{Figs.~\ref{#1}}
\def\Sec#1{Sec.~\ref{#1}}
\def\Secs#1{Secs.~\ref{#1}}

\def\Tab#1{Table~\ref{#1}}
\def\CP{$CP$\xspace}

\newcommand{\ttbar}{t{\bar t}}

\newcommand{\vl}{\boldsymbol{\ell}}
\newcommand{\Nv}{\mathcal{N}_{\text{virt}}}
\newcommand{\Nr}{\mathcal{N}_{\text{real}}}
\newcommand{\Tr}{\text{Tr}}
\newcommand{\cancel}[1]{#1\!\!\!\!/}

\begin{document}

\begin{titlepage}
  \begin{flushright}
    HU-EP-15/52\\
    TTK-15-33 \\
  \end{flushright}
  \vspace{0.01cm}
  
  \begin{center}
    {\LARGE \bf Production of heavy Higgs bosons \\[0.3em] 
      and decay into top quarks at the LHC} \\
    \vspace{1.5cm}
    {\bf W. Bernreuther}\,$^{a,}$\footnote{\tt
      breuther@physik.rwth-aachen.de}, 
    {\bf P. Galler}\,$^{b,}$\footnote{\tt galler@physik.hu-berlin.de},
    {\bf C. Mellein} \,$^{a,}$\footnote{\tt
      mellein@physik.rwth-aachen.de}, 
    {\bf Z.-G. Si}\,$^{c,}$\footnote{\tt zgsi@sdu.edu.cn} 
    {\bf and  P. Uwer}\,$^{b,}$\footnote{\tt uwer@physik.hu-berlin.de}
    \par\vspace{1cm}
    $^a$Institut f\"ur Theoretische Teilchenphysik und Kosmologie, 
    RWTH Aachen University, \\ 52056 Aachen, Germany\\
    $^b$Institut f\"ur Physik, Humboldt-Universit\"at zu Berlin, 
    12489 Berlin, Germany\\
    $^c$School of Physics, Shandong University, Jinan, Shandong 250100, China
    \par\vspace{1cm}
    {\bf Abstract}\\
    \parbox[t]{\textwidth}
    {\small{
     We investigate the production of heavy, neutral Higgs boson resonances and their decays to top-quark top-antiquark $(\ttbar)$ pairs
    at the Large Hadron Collider (LHC) at next-to-leading order (NLO) in the strong coupling of quantum chromodynamics (QCD).
    The NLO corrections to heavy Higgs boson production and the Higgs-QCD interference are calculated
  in the large $m_t$ limit with an effective K-factor rescaling. The nonresonant $\ttbar$ background is taken into account at NLO QCD 
 including weak-interaction corrections. In order to consistently determine the total decay widths of the heavy Higgs bosons, we
  consider for definiteness the type-II two-Higgs-doublet extension of the standard model and choose three parameter scenarios
   that entail two heavy neutral Higgs bosons with masses above the $\ttbar$  threshold and unsuppressed Yukawa couplings to top quarks.
   For these three scenarios we compute, for the LHC operating at 13 TeV, 
   the $\ttbar$ cross section and the distributions of the 
   $\ttbar$ invariant mass, of the transverse top-quark momentum and rapidity, and of the cosine of the Collins-Soper angle
   with and without the two heavy Higgs resonances.
    For selected $\mtt$ bins we  estimate the significances for detecting a heavy Higgs signal  in the $\ttbar$ dileptonic and 
     lepton plus jets decay channels. 
   }}
    
  \end{center}
  \vspace*{0.7cm}
  
  PACS number(s): 12.38.Bx, 12.60.Fr, 14.65.Ha, 14.80.Ec 
\end{titlepage}

\setcounter{footnote}{0}
\renewcommand{\thefootnote}{\arabic{footnote}}
\setcounter{page}{1}

\section{Introduction} 
\label{sec:intro}

The investigations of the production and decays of the 125 GeV
spin-zero resonance \cite{Aad:2012tfa,Chatrchyan:2012xdj} at the Large
Hadron Collider (LHC) by the CMS \cite{Khachatryan:2014jba} and ATLAS
experiments \cite{Aad:2015gba} show that the properties of this
particle are consistent with those of the Higgs boson of the standard
model (SM). These results strongly support the Higgs mechanism of
electroweak symmetry breaking, but do not exclude the possibility that
additional Higgs bosons with masses below or around 1 TeV exist.
There are several well-known motivations for considering the existence
of an extended Higgs sector (for reviews see, for instance,
\Refs{Djouadi:2005gj,Branco:2011iw}), and major experimental efforts
at the LHC are being devoted to the search for additional
  spin-zero bosons. To date, these efforts include
investigations which were aimed at tracing additional neutral Higgs
bosons with significant couplings to the electroweak gauge bosons
\cite{ATLAS:2013nma,Khachatryan:2015lba,Aad:2014yja,Aad:2015wra,Khachatryan:2015cwa}
or enhanced couplings (in comparison to respective SM Higgs couplings)
to $b$ quarks \cite{Nagai:2013xwa,Khachatryan:2015yea} and charged
leptons \cite{Khachatryan:2014wca,Aad:2014vgg}.  These
searches---negative 
so far---have an impact on parts of the parameter spaces of
models with an extended Higgs sector but are, needless to say, by far
not exhaustive.
          
One possibility, which is notoriously difficult to explore
experimentally, is that one or several neutral Higgs bosons with
masses above the top-antitop quark $(\ttbar$) production threshold
exist, with significant Yukawa couplings to top quarks but suppressed
couplings to the weak gauge bosons and to $d$-type quarks and charged
leptons.  Very likely, the resonant production of such a Higgs boson
$\phi$ and its subsequent decay into $\ttbar$ at the LHC, $p p \to
\phi\to \ttbar X$, does not show up as a resonance bump in the
$\ttbar$ invariant mass distribution but as a distinctive peak-dip
structure.  This structure is caused by the interference of the signal and
the nonresonant $\ttbar$ background amplitudes if the narrow-width
approximation does not apply to $\phi$, which is very likely the case.
This was first shown in \Refs{Gaemers:1984sj, Dicus:1994bm} 
 for a scalar and pseudoscalar $\phi$, respectively, 
and in \Refs{Bernreuther:1993hq,Bernreuther:1997gs,Bernreuther:1998qv} also for a
charge parity (\CP) mixture.
In \Refs{Bernreuther:1993hq,Bernreuther:1997gs,Bernreuther:1998qv} effects of a heavy Higgs
boson on top-quark spin observables were also investigated. More
recently, the study of this $\phi$ decay channel was taken up again in
\Refs{Barger:2006hm,Frederix:2007gi,Barger:2011pu,Craig:2015jba,Jung:2015gta,
Bhattacherjee:2015sga}. All these investigations were performed at leading-order (LO) QCD.
              
First experimental attempts to search for heavy spin-zero states
$\phi$ in the $\ttbar$ decay channel at the LHC were made by the CMS
\cite{Chatrchyan:2013lca} and ATLAS \cite{Aad:2015fna} experiments.
These investigations are difficult, in particular because the
experimental resolution of the $\ttbar$ invariant mass is, in general,
larger than the expected width of $\phi$, which hampers a resolution
of the resonance region.  On the theoretical side, a more precise
description of the reaction $p p \to \phi\to \ttbar X$ is necessary.
In this paper we investigate the resonant production of heavy neutral
Higgs bosons $\phi$ and their decay into $\ttbar$ including an
approximate calculation of the next-to-leading order QCD corrections.
That is, the NLO Higgs production amplitudes are
computed in the limit of $\mt\to \infty$.  The interferences of the
signal and nonresonant $\ttbar$ background amplitudes are also
determined at NLO QCD. The nonresonant $\ttbar$ background is
computed at NLO QCD and the mixed QCD-weak interaction corrections are
included.  We analyze how heavy Higgs bosons affect the $\ttbar$ cross
section and several distributions. In particular, we determine the
shape of the $\ttbar$ invariant mass distribution in the vicinity of a
heavy Higgs resonance, which is of primary interest.  We have analyzed
the reaction at hand for arbitrary configurations of the $t$ and $\bar
t$ spins, but we defer the investigation of heavy Higgs boson effects
on $\ttbar$ spin correlations and on the $t$ and $\bar t$ polarization to
a future publication.

The NLO QCD analysis of $p p \to \phi\to \ttbar X$ presented here may be applied to a
number of SM extensions. Yet, in applications one has to fix the
number of heavy Higgs bosons which can be resonantly produced in the
$\ttbar$ channel.  Other important ingredients in this analysis are
the total widths of the Higgs bosons $\phi$ with mass larger than
twice the top-quark mass.  These widths must be computed in a concrete
model, in order to maintain the unitarity of the $S$ matrix.  For
definiteness we have chosen to analyze the process at hand within the
type-II two-Higgs-doublet extension (2HDM) of the standard model. This
model, which may be realized with or without Higgs sector \CP
violation, is among the simplest extensions of the SM. While
interesting per se, it may also be considered a prototype of an
extended Higgs/spin-zero sector.  (For a recent review see, e.g.,
\Ref{Branco:2011iw}.)  The spectrum of physical spin-zero particles
of 2HDMs consists of three neutral Higgs bosons $\phi_1, \phi_2,
\phi_3$ and a charged Higgs boson and its antiparticle, $H^\pm$. We
identify $\phi_1$ with the observed 125 GeV resonance and assume that the
masses $\phi_2$ and $\phi_3$ are larger than twice the top-quark mass.
The charged Higgs boson is assumed to be heavy enough as to play no
role in our analysis. The LHC results on the 125 GeV Higgs resonance
imply that the parameters of 2HDMs are constrained by the requirement
of $\phi_1$ being SM-like. In addition we are interested in parameter
scenarios where the top-quark Yukawa couplings of the heavy Higgs
bosons $\phi_2$, $\phi_3$ are somewhat enhanced or at least as large
as the SM top-quark Yukawa coupling---without simultaneous enhancement of
the Yukawa couplings of $d$-type quarks and charged leptons---and
where the couplings of $\phi_2$, $\phi_3$ to the weak gauge bosons are
severely suppressed. This is possible in the type-II 2HDM.  We define
three such parameter scenarios which are in accord  with
experimental constraints---two scenarios with \CP conservation and one
with \CP violation in the tree-level Higgs potential. For each of these parameter
scenarios, we compute the Higgs boson contributions to the $\ttbar$
cross section and to several top-quark distributions, in particular
the $\ttbar$ invariant mass distribution, for the LHC at 13 TeV.
In the context of our type-II 2HDM parameter scenarios the NLO
analysis is organized as follows: the nonresonant contribution of
$\phi_1$ is part of the mixed QCD-weak interaction corrections to the
nonresonant $\ttbar$ background. As to the amplitudes of $a b\to
\phi_j \to \ttbar X$ $(j=2,3)$, where $a,b$ denote partons,
their interference with the nonresonant $\ttbar$ amplitudes is taken
into account. In the case of neutral Higgs-sector \CP violation, the
$\phi_2$ and $\phi_3$ production amplitudes interfere also among each
other but, as shown below, this contribution is very small for our parameter scenarios 
 and can be neglected. The evaluation of our three
parameter scenarios exemplifies the significance of the QCD
corrections, but the effect of the two heavy Higgs bosons on the
top-quark distributions is, overall, rather small. Nevertheless,
statistically significant signals are possible.  We propose to
analyze existing and future LHC $\ttbar$ data by scanning the
distribution of the $\ttbar$ invariant mass, $\mtt$, with a sliding
$\mtt$ window of width $\sim 80$ GeV.  
                    
The paper is organized as follows. In Sec.~\ref{sec:2hdm} we briefly
recapitulate the main features of the type-II 2HDM with and without
Higgs-sector \CP violation that are relevant for our analysis.  In
Sec.~\ref{sec:3-2hdmsc} we define three parameter scenarios which we
use in the following: two sets associated with a \CP-conserving and
one set related to a \CP-violating Higgs potential.  We
discuss that our scenarios are compatible with existing
experimental constraints on the parameter space of the type-II 2HDM. 
Section \ref{sec:calculation} contains the  setup of our calculations.
For illustrative purposes we discuss first the resonant 
 production of heavy Higgs bosons and their decay to $\ttbar$ and the interference with
  the nonresonant background at LO QCD. Then we outline how we compute distributions 
   for this reaction including the Higgs-QCD interference at NLO QCD and we describe our 
    K-factor approximation adapted from \Ref{Kramer:1996iq}.
   The nonresonant $\ttbar$ background is incorporated at NLO QCD including the
   weak-interaction corrections. 
 Section \ref{sec:results} contains the results for $\ttbar$ production at the LHC (13 TeV)
  for the three parameter scenarios introduced in Sec.~\ref{sec:3-2hdmsc}. 
  First we compute  the inclusive $\ttbar$ cross section  and 
   the $\mtt$ distribution without and with the two heavy Higgs resonances. We 
   estimate, for appropriately chosen $\mtt$ intervals and for our  three parameter scenarios,
    the significance with which a heavy Higgs signal could be detected in the $\ttbar$ dileptonic and 
     lepton plus jets decay channels.
     Furthermore, we give the distributions of the rapidity and transverse momentum of the top quark and the
  distributions of the cosine of the Collins-Soper angle \cite{Collins:1977iv} for selected $\mtt$ windows. Moreover we show 
  that the negative experimental searches for heavy spin-zero resonances 
   in the $\ttbar$ channel \cite{Chatrchyan:2013lca,Aad:2015fna} at the LHC (8 TeV) do not 
    exclude our parameter scenarios.
  We conclude in Sec.~\ref{sec:summary}.
  Appendix~\ref{app:higgs_resonances}  contains a discussion of the complex mass scheme that
  we apply for the description of the heavy Higgs boson resonances. 
  The  cancellation of real and virtual nonfactorizable  QCD corrections to $gg \to \phi_j\to \ttbar$ 
  in the soft gluon approximation (SGA) is outlined in Appendix~\ref{app:SGA}.

\section{The type-II two-Higgs-doublet model}
\label{sec:2hdm}
As mentioned above, we choose the type-II two-Higgs-doublet model as an
example of a SM extension which allows one to describe, in a consistent
field-theoretic and phenomenologically viable way, both the 125 GeV
Higgs resonance and additional neutral heavy Higgs bosons. In this
section we shortly describe the salient features of several variants
of this model which are relevant for our analysis.
 
In 2HDMs the SM field content is extended by an additional Higgs
doublet. Both Higgs doublets $\Phi_1$ and $\Phi_2$ transform under
${\rm SU}(3)_c\times {\rm SU}(2)_L\times {\rm U}(1)_Y$ as $(1,2,1/2)$. \\
The type-II model is defined by its Yukawa coupling structure: the
doublet $\Phi_1$ is coupled to right-chiral down-type quarks and
charged leptons, while $\Phi_2$ is coupled to right-chiral up-type
quarks only.  The type-II model belongs to those 2HDMs where, by
construction, flavor-changing neutral currents are absent at tree
level.  The most general Hermitian, gauge invariant and renormalizable
2HDM potential reads (see, for instance, \cite{Branco:2011iw}):
\begin{eqnarray}
  V\left(\Phi_1,\Phi_2\right) & = & - \mu_1^2 |\Phi_1|^2 - \mu_2^2 |\Phi_2|^2 - \left( \mu_3^2 \Phi_1^{\dagger}\Phi_2 + {\rm h.c.} \right)   \nonumber \\
 &  & + \frac{\lambda_1}{2}|\Phi_1|^4 + \frac{\lambda_2}{2}|\Phi_2|^4 + \lambda_3 |\Phi_1|^2 |\Phi_2|^2 + \lambda_4 |\Phi_1^{\dagger}\Phi_2|^2  \nonumber \\
 &  & + \left[ \frac{\lambda_5}{2}\left(\Phi_1^{\dagger}\Phi_2\right)^2 + 
 \lambda_6 \left( \Phi_1^{\dagger}\Phi_1 \right)\left( \Phi_1^{\dagger}\Phi_2 \right) 
 + \lambda_7 \left( \Phi_2^{\dagger}\Phi_2 \right)\left( \Phi_1^{\dagger}\Phi_2 \right)  + {\rm h.c.} \right], \label{eq:2HDM_gen_pot}
\end{eqnarray}
with real $\mu_1^2$, $\mu_2^2$, $\lambda_{1,2,3,4}$ and complex
$\mu_3^2$, $\lambda_{5,6,7}$, adding up to 14 independent real parameters.
Three of the 14  parameters can be eliminated by a change of basis, $\Phi'_i=U_{ij}\Phi_j$, where $U$ is a unitary $2\times 2$ matrix. \\
After electroweak gauge symmetry breaking, using the unitary
gauge the two-Higgs doublets take the form
 \begin{eqnarray} 
\Phi_1 & =& \left(-H^+ \sin\beta,\, \frac{1}{\sqrt{2}}(\vev_1+\varphi_1-iA\sin\beta)\right)^T, \nonumber \\
\Phi_2 & = & e^{i\xi}\left(H^+\cos\beta,\, \, \frac{1}{\sqrt{2}}(\vev_2+\varphi_2+iA\cos\beta)\right)^T,
\label{eq:2HDM_phunit}
\end{eqnarray}
where $\tan\beta=\vev_2/\vev_1$ is the ratio of the vacuum expectation
values of the two-Higgs-doublet fields with $\vev=\sqrt{\vev_1^2+\vev_2^2}=246$
GeV. By a field redefinition the phase $\xi$ can be put to zero
without loss of generality. The field $H^+$ describes the physical
charged Higgs boson of the model, while $\varphi_{1,2}$ and $A$ are
the physical neutral \CP-even and \CP-odd field degrees of freedom. The
Higgs potential \eqref{eq:2HDM_gen_pot} leads to mixing among these
neutral fields. We denote the neutral Higgs fields and the associated
Higgs particles of definite mass by $\phi_j, j=1,2,3$. They are
related to $\varphi_{1,2}$ and $A$ by an orthogonal transformation,
$(\phi_1,\phi_2,\phi_3)^T= R(\varphi_1,\varphi_2, A)^T.$ We
parametrize the orthogonal matrix $R$ as follows:
\begin{equation}\label{Rparametr}
R = \left(\begin{array}{ccc}c_1 c_2& c_2s_1 & s_2
      \\-c_ 1s_2 s_3 -c_3 s_1 & c_1 c_3 -s_1 s_2 s_3 & c_2 s_3
   \\ -c_1c_3s_2 + s_1 s_3 & -c_1 s_3 - c_3 s_1 s_2 & c_2 c_3 \end{array}\right) \; ,
\end{equation}
where $c_i$ and $s_i$ $(i=1,2,3)$ are the cosines and sines of the
mixing angles $\alpha_i$. Due to the symmetries of $R$  \cite{ElKaffas:2006nt} 
 one can restrict the range of the angles to $-\pi/2\leq \alpha_i<\pi/2$. \\
The Higgs potential in its general form \eqref{eq:2HDM_gen_pot}
violates the \CP symmetry and, as a consequence, the \CP-even and \CP-odd
Higgs fields mix and the mass eigenstates $\phi_{j}~(j=1,2,3)$ do not
have a definite \CP. In this general \CP-violating
version of the type-II 2HDM, the Higgs potential has, as mentioned
above, 11 independent parameters.\footnote{For a general discussion,
  see \Ref{Grzadkowski:2014ada}.}  For phenomenological studies it
is useful to choose as independent parameters a different set, in
terms of which the parameters of the potential can be expressed. This
set includes the following nine parameters which we call phenomenological parameters:
\begin{equation} \label{eq:setP}
  m_1, \,  m_2,  \,  m_3, \,  m_+ ,\,  \alpha_1 , \,  \alpha_2 , \, \alpha_3 , \,  \tan\beta , \,  \vev \, ,
\end{equation}
where $m_j~(j=1,2,3)$ are the masses of the three neutral Higgs bosons and
$m_+$ is the mass of $H^+$. The value of $\vev$ is of course fixed by
experiment.
  
In order to motivate the type-II 2HDM construction by a symmetry
argument, one often imposes a $Z_2$ symmetry onto the model which may
be softly broken by the Higgs potential. This approximate symmetry
implies that the hard $Z_2$ symmetry-breaking terms in
\eqref{eq:2HDM_gen_pot} are absent, $\lambda_6 = \lambda_7=0$.  This
variant of the model still allows for neutral Higgs sector \CP
violation \cite{Bernreuther:1992dz}, i.e., for the mixing of
$\varphi_{1,2}$ and $A$, if ${\rm Im}(\lambda_5/\mu_3^4)\neq 0$.
One can show \cite{Inoue:2014nva} that in this case the number of
independent phenomenological parameters listed in \eqref{eq:setP} is
reduced by one.  In the model with approximate $Z_2$ symmetry where
${\rm Im}\lambda_5\neq 0$, one can, for instance, express the mass
$m_3$ of the neutral Higgs boson $\phi_3$ in terms of $m_1, m_2,$
$\tan\beta$, and the Higgs mixing angles $\alpha_i~(i=1,2,3)$.  In our
applications below, where we consider also the case of Higgs sector \CP
violation, we however refer to the more general case where
\eqref{eq:setP} is an independent parameter set, in order to avoid a
theoretical bias.
    
The type-II 2HDM where the Higgs potential conserves \CP is a special
case and is also of great phenomenological interest.  In this case
only $\varphi_{1}$ and $\varphi_2$ mix. This is described by the
mixing angle $\alpha_1$, while $\alpha_2=\alpha_3=0$. The matrix $R$
is now block diagonal with $R_{13}=R_{23}=R_{31}=R_{32} = 0$ and
$R_{33} = 1.$ The two \CP-even and the \CP-odd neutral mass eigenstates
are often conventionally denoted by $\phi_1=h$, $\phi_2=H$, and
$\phi_3=A$. The type-II 2HDM with \CP-conserving Higgs potential may be
motivated by assuming an approximate $Z_2$ symmetry.  Then the
potential has eight independent parameters, but they are reduced by one by
the condition that the ground state of the model does not break \CP
spontaneously. Thus, one may choose as independent parameters the
following set:
\begin{equation} \label{eq:setPrime}
  m_1, \, m_2, \, m_3, \, m_+, \, \alpha_1,\,  \tan\beta,\,  \vev \, .
\end{equation} 

In the following we identify the 125 GeV Higgs resonance with the
Higgs boson $\phi_1$ and assume that the masses of the two other
neutral states $\phi_{2,3}$ lie above the $\ttbar$ threshold. In
addition, we assume that the mass of the charged Higgs boson $H^+$ is
of the order of the masses of $\phi_{2,3}$, so that the two-body
decays $\phi_{2,3} \to W^\pm H^\mp$ are kinematically excluded.  Fits
to LHC data \cite{Khachatryan:2014jba,Aad:2015gba} imply that the
couplings of the 125 GeV Higgs boson to the third-generation fermions
and gauge bosons are not too different from those of the SM Higgs
boson. In the context of the \CP-conserving 2HDM this means that the
model is close to the so-called alignment limit\footnote{Often the
  parametrization of the mixing matrix $R$ used in the literature is
  different from \eqref{Rparametr}. For instance, in the
  parametrizations of \cite{Inoue:2014nva,Chen:2015gaa},
  $\alpha=\alpha_1-\pi/2$; i.e., the alignment limit is then given by
  $\beta-\alpha=\pi/2$.}  $\beta=\alpha_1$. Yet, at present the
uncertainties on the measured couplings of the 125 GeV resonance are
still rather large, $\sim 20\% - 30 \%$. In particular it is not yet
experimentally shown that the 125 GeV resonance is a pure \CP-even state---a
pseudoscalar admixture is still a possibility. Nevertheless, in the
scenarios which we define in the next section and apply subsequently
we shall choose parameters in or close to the alignment limit.
   
For our analysis below we need the interactions of the neutral Higgs
bosons with fermions and weak gauge bosons, and their interactions
among each other.  We parametrize the couplings of the $\phi_j$ to the
quarks and charged leptons $f=q,\ell$ and to weak gauge boson pairs in
the type-II model with Higgs sector \CP violation as follows:
\begin{equation} \label{eq:LfWZ}
  {\cal L}_1= -\frac{m_f}{\vev}\left( a_{jf} {\bar f}f
    - b_{jf}{\bar f} i\gamma_5 f\right) \phi_j  
  + f_{jVV} \phi_j\left(\frac{2 m_W^2}{\vev}W^-_\mu W^{+\mu} 
    + \frac{m_Z^2}{\vev} Z_\mu Z^\mu \right) \, ,
\end{equation}
where a sum over $f$ and $j=1,2,3$ is implicit.  The reduced scalar
and pseudoscalar Yukawa couplings $a_{jf}$ and $b_{jf}$ and the
reduced couplings $f_{jVV}$ depend on the values of $\tan\beta$ and on
the elements $R_{ij}$ of the Higgs mixing matrix \eqref{Rparametr} and
are listed in \Tab{tab:LfWZ}.  The couplings $f_{jVV}$ satisfy
the sum rule
\begin{equation} \label{eq:sumV}
  \sum\limits_{j=1}^3 f_{jVV}^2  =  1 \, .
\end{equation}
The interaction of neutral Higgs bosons $\phi_j, \phi_l$ $(j\neq l)$
with the $Z$ boson is given by
\begin{equation} \label{eq:LZij}
  {\cal L}_2 =  - \frac{g_W}{2\cos\theta_W} f_{Zjl} Z^\mu \phi_j
  \stackrel{\leftrightarrow}{\partial_\mu}\phi_l \, ,
\end{equation}
where $g_W$ is the ${\rm SU}(2)_L$ gauge coupling and $\theta_W$
denotes the weak mixing angle.  For computing the rates of the decays
$\phi_j \to Z \phi_1$ $(j=2,3)$, we need the couplings $f_{Zj1}$. They are given by
\begin{equation} \label{eq:couZj1}
  f_{Zj1} =  (-\sin\beta R_{11}+\cos\beta R_{12})R_{j3} 
  - (-\sin\beta R_{j1}+\cos\beta R_{j2})R_{13} \, .
\end{equation}
Moreover, for computing the decays $\phi_j \to 2 \phi_1$ $(j=2,3)$ the
trilinear neutral Higgs couplings are required, ${\cal L}_3 = g_{ijl}
\phi_i \phi_j \phi_l$. In the \CP-violating type-II 2HDM the couplings
$g_{ijl}$ have a quite complicated form. They were derived in terms of
the potential parameters in \cite{Chen:2015gaa} and in terms of the
phenomenological parameters in \cite{Mell15}.

\vspace{2mm}
\begin{table}[htbp]
\begin{center}
  \caption{Reduced couplings to quarks, leptons and gauge bosons of
    the neutral Higgs bosons $\phi_j$ of the type-II 2HDM in the
    parametrization \eqref{eq:LfWZ}. The labels $t,b$, and $\tau$ are
    place holders for $u$-type, $d$-type quarks, and charged leptons.}
  \vspace{1mm}
\begin{tabular}{cccc|c}
  $a_{jt}$  & $a_{jb} = a_{j\tau}$  &    $b_{jt}$  & $b_{jb} = b_{j\tau}$  &    $f_{jVV}$                     \\ \hline \hline
  $R_{j2}/\sin\beta$ & $R_{j1}/\cos\beta$ & $R_{j3}\cot\beta $  &  $R_{j3}\tan\beta $ & $\cos\beta R_{j1}+ \sin\beta  R_{j2}$ 
\end{tabular}
\label{tab:LfWZ}
\end{center}
\end{table}

\section{Three type-II 2HDM scenarios}
\label{sec:3-2hdmsc}

In this section we define three phenomenologically viable parameter
sets for the type-II 2HDM, two for the \CP-conserving and one for the
\CP-violating variant, which we use in the next sections. 
We compute for these parameter sets the total widths of the heavy
neutral Higgs bosons $\phi_2, \phi_3$, including QCD corrections.  As
already mentioned above, in all three cases we identify the 125 GeV
Higgs boson with $\phi_1$ and choose the parameters such that its
couplings to fermions and gauge bosons are SM-like. For the parameter
sets below, the total width $\Gamma_1$ of $\phi_1$ is of the order of
4 MeV. This width plays no role in the computations of
\Secs{sec:calculation}--\ref{sec:results}.
For the masses of the two other neutral Higgs bosons we assume
$m_2, m_3 >2 m_t$.  The resonant production of $\phi_2, \phi_3$ has a
chance of being experimentally detectable in the $\ttbar$ decay channel
only if the Yukawa couplings are not suppressed in comparison with the SM
top-Yukawa coupling.  Therefore we choose $\tan\beta=0.7$ for all
three scenarios below. Experimental constraints from $B$ physics
require that for $\tan\beta<1$ the mass $m_+$ of the charged Higgs
boson is large. In order to satisfy these constraints
\cite{Mahmoudi:2009zx,Hermann:2012fc,Eberhardt:2013uba} we assume that
$m_+ > 720$ GeV.  This implies for the parameter scenarios below that
the decays $\phi_{2,3}\to W^\pm H^\mp$ cannot occur.
 
We determine the total widths of $\phi_2$ and $\phi_3$ by computing
the sum of the largest two-body decay rates.  The QCD corrections to
the decays $\phi_j\to q{\bar q}$ and $\phi_j\to g g$ may be computed
with the computer program {\tt HDECAY} \cite{Djouadi:1997yw} or {\tt
2HDMC} \cite{Eriksson:2009ws}. Both codes apply only to \CP-conserving
2HDM but can be easily adapted to \CP-violating scenarios. Since
both programs use the renormalization scale
 $\mu=m_{\phi}$, which is not suited for our applications (see below),
  we use the expressions for the partial decay rates to quarks and to gluons
   including QCD corrections from  \Refs{Braaten:1980yq,Drees:1990dq}
    and \Ref{Spira:1995rr}, respectively.
 Expressions for the other decay rates  (including the \CP-violating case)
can also be found in the literature, see e.g. \Ref{Chen:2015gaa}. Where possible
we checked our results with {\tt 2HDMC}.
In these computations we use the top-quark mass
$m_t$ in the on-shell scheme and put $m_t=173.34$ GeV. Moreover,
in order to be consistent with our computations of \Secs{sec:calculation}--\ref{sec:results}
 we consider the following values and variations of the renormalization
scale $\mu$:
\begin{equation} \label{eq:rensc-var}
  \frac{\mu_0}{2} \leq   \mu \leq 2 \mu_0 \;, \qquad \mu_0 = \frac{m_2 + m_3}{4} \, .
\end{equation} 

Electroweak corrections to the decays $\phi_j\to f{\bar f}$
\cite{Dabelstein:1991ky} and $\phi_j\to VV$ \cite{Fleischer:1980ub}
are sizeable, especially for heavy Higgs bosons.  Because we consider
parameter scenarios where the $\phi_j VV$ couplings are suppressed for
$j=2,3$, these corrections to $\phi_j\to VV$ will not affect the total
widths $\Gamma_{2,3}$ very much. The electroweak corrections to
$\phi_j\to t{\bar t}$ can be as large as the QCD corrections. Because
we analyze in \Sec{sec:calculation} resonant hadronic heavy Higgs boson production
and decay to $\ttbar$ only to lowest order in the non-QCD couplings,
we omit the contributions of the electroweak
corrections to $\Gamma_{2,3}$.
       
\subsection{Scenario 1}
\label{susec:sc1}
We consider the type-II model with a \CP-conserving Higgs potential. We
choose
\begin{equation} \label{eq:angle-sc1}
  \tan\beta=0.7   \;,\quad    \alpha_1 = \beta = 0.611 \;,\quad 
  \alpha_2 = \alpha_3 = 0 \, ,
\end{equation}
and choose the following Higgs boson masses, with $\phi_2$ and
$\phi_3$ being nearly mass degenerate:
\begin{equation}\label{eq:mass-sc1}
  m_1 = 125 {\rm\,GeV}\;,\quad m_2 = 550 {\rm\,GeV}\;,\quad 
  m_3 = 510 {\rm\,GeV} \;,\quad  m_{+} > 720 {\rm\,GeV} \; .
\end{equation}
That is, $\phi_1$ and $\phi_2$ are chosen to be pure scalars while
$\phi_3$ is a pseudoscalar Higgs boson, cf. \Tab{tab:Yuksc1}.
The reduced Yukawa couplings and couplings to the weak gauge bosons,
which result from \eqref{eq:angle-sc1} and the formulas of
\Tab{tab:LfWZ}, are given in \Tab{tab:Yuksc1}. Recall that
a pseudoscalar Higgs boson has no tree-level couplings to $W^+W^-$ and
$ZZ$ ($f_{3VV}=0$). The choice $\alpha_1 = \beta$ puts the model into
the alignment limit. Therefore we have also $f_{2VV}=0$.

\vspace{2mm}
\begin{table}[htbp]
\begin{center}
\caption{Values of the reduced couplings to the third-generation
  fermions and gauge bosons of the neutral Higgs bosons $\phi_j$ for
  the parameter set \eqref{eq:angle-sc1} which is used in scenarios 1
  and 2.}  \vspace{1mm}
\begin{tabular}{c|cccc|c}
 & $a_{jt}$  & $a_{jb} = a_{j\tau}$  &    $b_{jt}$  & $b_{jb} = b_{j\tau}$  &    $f_{jVV}$                     \\ \hline \hline
$\phi_1$ & 1 & 1 & 0 & 0 & 1 \\
$\phi_2$ & 1.429 & $ -0.700 $ & 0 & 0& 0 \\
$\phi_3$ & 0 & 0 & 1.429 & 0.700 & 0 \\ \hline
\end{tabular}
\label{tab:Yuksc1}
\end{center}
\end{table}

\vspace{2mm}
\begin{table}[htbp]
\begin{center}
\caption{Dominant partial decay widths and the total widths of the two heavy, neutral Higgs bosons $\phi_2$ 
  and $\phi_3$ in our \CP-conserving 2HDM scenario 1,
  defined in \eqref{eq:angle-sc1}--\eqref{eq:mass-sc1}. The super- and subscripts denote the shift
  due to the scale variations \eqref{eq:rensc-var}.}
{\renewcommand{\arraystretch}{1.4}
\renewcommand{\tabcolsep}{0.2cm}
\begin{tabular}{l|ll}
 & $\Gamma_2$ [GeV] & $\Gamma_3$ [GeV]\\
\hline
$\phi_i\rightarrow t\bar{t}$ & $34.48^{+0.33}_{-0.28}$ & $49.15^{+0.38}_{-0.32}$\\
$\phi_i\rightarrow gg$ & $0.08^{+0.01}_{-0.01}$ & $0.13^{+0.02}_{-0.02}$\\
\hline
Total & $34.56^{+0.34}_{-0.28}$ & $49.28^{+0.40}_{-0.34}$\\
\end{tabular}}
\label{tab:res_widths_sc1}
\end{center}
\end{table}

We compute the largest two-body decay rates of the scalar $\phi_2$ and
the pseudoscalar $\phi_3$ and determine their total widths by adding
up these rates. The leading-order decay-rate formulas for $\phi_j\to
f{\bar f}, VV, Z\phi_l, 2 \phi_l, gg$, where $V=W,Z$, are well-known
and are not reproduced here (see, e.g., \cite{Eriksson:2009ws}).
The relevant tree-level couplings are given in Sec.~\ref{sec:2hdm}.
The trilinear neutral Higgs couplings are taken from \cite{Chen:2015gaa,Mell15}
 and we use the QCD corrections to $\phi_{2,3}\to t{\bar t}, gg$ given in 
\Ref{Braaten:1980yq,Drees:1990dq,Spira:1995rr}. The results
are listed in \Tab{tab:res_widths_sc1}.  The uncertainties
result from varying the renormalization scale according to
\eqref{eq:rensc-var}.  The partial decay widths of $\phi_j\to f{\bar
  f}$ $(f\neq t),$ $\phi_j\to \gamma\gamma$, and $\phi_j\to Z\gamma$
are a few $10^{-3}$ GeV or smaller and are neglected in the total widths
$\Gamma_2, \Gamma_3$. In particular, the decay rate to $b\bar{b}$ is small
because the Yukawa couplings to down-type quarks are not enhanced
(\Tab{tab:Yuksc1}) due to our choice of $\tan\beta<1$.
Moreover, to lowest order in the non-QCD couplings the partial decay rates for 
$\phi_i\rightarrow VV$, $\phi_i\rightarrow \phi_1Z$, and $\phi_i\rightarrow \phi_1\phi_1$ are zero.

\subsection{Scenario 2}
\label{susec:sc2}
We choose again a \CP-conserving neutral Higgs sector scenario with the
same set of angles, Eq.~\eqref{eq:angle-sc1}, as in scenario 1.  Here
we assume that the pseudoscalar $\phi_3$ is significantly heavier than
the scalar $\phi_2$. We choose
\begin{equation}\label{eq:mass-sc2}
  m_1 = 125 {\rm\,GeV}\;,\quad m_2 = 550 {\rm\,GeV}\;,\quad 
  m_3 = 700 {\rm\,GeV} \;,\quad m_{+} > 720 {\rm\,GeV} \; .
\end{equation}
The values of the reduced couplings of $\phi_2, \phi_3$ to the
third-generation fermions and gauge bosons given in
\Tab{tab:Yuksc1} apply also here.
 The results for the most important two-body decay widths and the total
widths of $\phi_2, \phi_3$ are given in
\Tab{tab:res_widths_sc2}. To lowest order in the non-QCD couplings the partial decay rates for 
$\phi_i\rightarrow VV$, $\phi_i\rightarrow \phi_1Z$, $\phi_i\rightarrow \phi_1\phi_1$,
and $\phi_3\rightarrow \phi_1\phi_2$ are zero.

Notice that the decay widths for $\phi_{2}\rightarrow VV$, $\phi_{2}
\rightarrow \phi_1\phi_1$ and $\phi_{3}\rightarrow Z\phi_1$ vanish in
scenarios 1 and 2, respectively, because we have chosen
the exact alignment limit $\alpha_1=\beta$.  The more one moves away
from this limit, the larger the contributions of these modes will be
to the total decay widths.

\vspace{2mm}
\begin{table}[htbp]
\begin{center}
\caption{Dominant partial decay widths and total widths  of the two heavy, neutral Higgs bosons $\phi_2$ 
  and $\phi_3$ in our \CP-conserving 2HDM scenario 2, defined in
  \eqref{eq:angle-sc1} and \eqref{eq:mass-sc2}. The super- and subscripts denote the shift
  due to the scale variations \eqref{eq:rensc-var}.} \vspace{1mm}
{\renewcommand{\arraystretch}{1.2}
\renewcommand{\tabcolsep}{0.2cm}
\begin{tabular}{l|ll}
 & $\Gamma_2$ [GeV] & $\Gamma_3$ [GeV]\\
\hline
$\phi_i\rightarrow t\bar{t}$ & $34.41^{+0.32}_{-0.26}$ & $71.97^{-0.15}_{+0.13}$\\
$\phi_i\rightarrow \phi_2Z$ & 0 & 3.14\\
$\phi_i\rightarrow gg$ & $0.08^{+0.01}_{-0.01}$ & $0.17^{+0.01}_{-0.01}$\\
\hline
Total & $34.49^{+0.33}_{-0.27}$ & $75.28^{-0.14}_{+0.11}$\\
\end{tabular}}
\label{tab:res_widths_sc2}
\end{center}
\end{table}

\subsection{Scenario 3}
\label{susec:sc3}
We consider a type-II model with a \CP-violating Higgs potential
(without approximate $Z_2$ symmetry) such that \eqref{eq:setP} forms a
set of independent parameters.  For $\tan\beta$ and the Higgs mixing
angles we choose
\begin{equation}\label{eq:angle-sc3}
  \tan\beta=0.7 \; , \quad \alpha_1 = \beta \;,\quad 
  \alpha_2 = \frac{\pi}{15} \;,\quad \alpha_3 = \frac{\pi}{4} \, .
\end{equation}
We set the Higgs boson masses to the values
\begin{equation}\label{eq:mass-sc3}
  m_1 = 125 {\rm\,GeV}\;,\quad m_2 = 500 {\rm\,GeV}\;,\quad 
  m_3 = 800 {\rm\,GeV}\; , \quad m_{+} > 720 {\rm\,GeV} \, .
\end{equation}
Using \Tab{tab:LfWZ} and the parameters \eqref{eq:angle-sc3} we 
get the values 
of the reduced Yukawa couplings and couplings to the weak gauge bosons 
listed in \Tab{tab:Yuksc3}.
As this table shows, in this scenario the three neutral Higgs bosons 
$\phi_{1,2,3}$ are \CP mixtures.
   
\vspace{2mm}
\begin{table}[htbp]
\begin{center}
\caption{Values of the reduced couplings to the third-generation
  fermions and gauge bosons of the neutral Higgs bosons $\phi_j$ for
  the parameter set \eqref{eq:angle-sc3} of scenario 3.}  \vspace{1mm}
\begin{tabular}{c|cccc|c}
  & $a_{jt}$  & $a_{jb} = a_{j\tau}$  &    $b_{jt}$  & $b_{jb} = b_{j\tau}$  &    $f_{jVV}$                     \\ \hline \hline
  $\phi_1$ & 0.978 & 0.978 & 0.297 & 0.146 & 0.978 \\
  $\phi_2$ &0.863 & $-0.642$ & 0.988 & 0.484 & $-0.147$  \\
  $\phi_3$ & $-1.157$ & 0.348 & 0.988 & 0.484 & $-0.147$ \\ \hline
\end{tabular}
\label{tab:Yuksc3}
\end{center}
\end{table}

\vspace{2mm}
\begin{table}[htbp]
  \begin{center}
    \caption{Dominant partial decay widths and total widths of the two heavy, neutral Higgs bosons
      $\phi_2$ and $\phi_3$ in our \CP-violating 2HDM scenario 3,
      defined in \eqref{eq:angle-sc3}--\eqref{eq:mass-sc3}. The super- and subscripts denote the shift
  due to the scale variations \eqref{eq:rensc-var}.} \vspace{1mm}

{\renewcommand{\arraystretch}{1.2}
\renewcommand{\tabcolsep}{0.2cm}
\begin{tabular}{l|ll}
 & $\Gamma_2$ [GeV] & $\Gamma_3$ [GeV]\\
\hline
$\phi_i\rightarrow t\bar{t}$ & $32.31^{+0.31}_{-0.26}$ & $85.05^{-0.30}_{+0.25}$\\
$\phi_i\rightarrow VV$ & 1.12 & 5.11\\
$\phi_i\rightarrow \phi_1Z$ & 0.65 & 3.24\\
$\phi_i\rightarrow \phi_2Z$ & 0 & 31.28\\
$\phi_i\rightarrow \phi_1\phi_1$ & 2.38 & 3.00\\
$\phi_i\rightarrow \phi_1\phi_2$ & 0 & 0.31\\
$\phi_i\rightarrow gg$ & $0.08^{+0.01}_{-0.01}$ & $0.17^{+0.01}_{-0.01}$\\
\hline
Total & $36.55^{+0.32}_{-0.27}$ & $128.16^{-0.29}_{+0.24}$\\
\end{tabular}}
\label{tab:res_widths_sc3}
\end{center}
\end{table}

The results for the most important two-body decay widths and the total
widths of $\phi_2, \phi_3$ are given in
\Tab{tab:res_widths_sc3}.
In this scenario without approximate $Z_2$ symmetry
we have to specify also the values of the complex parameters $\lambda_6$,
$\lambda_7$ in the 2HDM potential. These parameters only enter the calculation of the
decay widths which involve trilinear Higgs couplings, i.e. only the
decays $\phi_i\rightarrow\phi_1\phi_1$ and $\phi_i\rightarrow\phi_1\phi_2$
listed in \Tab{tab:res_widths_sc3}. 
We choose $\Re\lambda_6=0$, $\Im\lambda_6=-3.677$ and $\lambda_7=0$.

\subsection{Experimental constraints}
\label{suse:expconst}

The Higgs couplings and masses chosen in the three scenarios above are
compatible with existing experimental constraints. As to the couplings
of $\phi_1$, in scenarios 1 and 2 the couplings of $\phi_1$ are
identical to those of the SM Higgs boson. In the \CP-violating
scenario $\phi_1$ has a pseudoscalar component. Nevertheless, also in
this scenario the couplings of $\phi_1$ given in
\Tab{tab:Yuksc3} lie within the respective coupling ranges of
the 125 GeV boson determined by the CMS \cite{Khachatryan:2014jba} and
ATLAS \cite{Aad:2015gba} experiments, which allow $\sim 20\% -30 \%$
deviations from the respective SM Higgs couplings.  Because in our
scenarios the tree-level couplings of $\phi_2, \phi_3$ to $b$ quarks,
$\tau$ leptons, weak gauge bosons, and to $\phi_1$ are very small or
even zero, the negative searches, so far, at the LHC for nonstandard
heavy resonances decaying to $W^+W^-$ \cite{Khachatryan:2015cwa}, $ZZ$
\cite{Khachatryan:2015cwa,ATLAS:2013nma}, $Z\phi_1$
\cite{Aad:2015wra,Khachatryan:2015lba}, $\phi_1\phi_1$
\cite{Aad:2014yja,Khachatryan:2015yea}, $\tau^-\tau^+$
\cite{Khachatryan:2014wca,Aad:2014vgg}, $b \bar b$
\cite{Nagai:2013xwa} do not exclude the above scenarios. We comment in 
 Sec.~\ref{sec:results}  on the CMS \cite{Chatrchyan:2013lca} and ATLAS
\cite{Aad:2015fna} searches in the $\ttbar$ decay channel.

The nonobservation of deviations from the SM predictions in B-physics
data places limits on $\tan{\beta}$ and on the mass $m_{+}$ of the
charged Higgs boson.  These parameters control the contribution of
$H^\pm$ to loop-induced observables like the rare decays $B\rightarrow
X_s+\gamma$ and neutral $B_d^0-\bar{B}_d^0$ mixing. In general the
loop effects of charged Higgs bosons decrease with increasing mass
$m_{+}$ and vice versa. Our choice $\tan\beta=0.7$ and $m_+ > 720$ GeV
in our three 2HDM scenarios falls into the allowed parameter range
which is determined by comparing experimental data with 2HDM
predictions \cite{Mahmoudi:2009zx,Hermann:2012fc,Eberhardt:2013uba}.

The allowed parameters of 2HDM are also restricted by the experimental
values of the oblique electroweak parameters.  The general expressions
for the oblique parameters $S$, $T$, and $U$ in 2HDM and discussion of
resulting constraints can be found in
\cite{Branco:2011iw,Grimus:2007if,Grimus:2008nb,Haber:2010bw}. As we
have chosen $\phi_1$ to be SM-like, $\phi_2,\phi_3$ and $H^\pm$ to be
heavy, and the mixing angles such that the couplings of $\phi_2,
\phi_3$ to the weak gauge bosons are suppressed, our scenarios are
compatible with the constraints from electroweak precision
measurements.
  
The experimental upper limits on the electric dipole moments of the
neutron \cite{Baker:2006ts} and the electron \cite{Baron:2013eja} put
limits on the angles associated with Higgs sector \CP-violation in
2DHM. Our choice of Higgs mixing angles \eqref{eq:angle-sc3} in the
\CP-violating scenario 3 which implies the Higgs couplings given
\Tab{tab:Yuksc3} and our choice of Higgs boson masses
\eqref{eq:mass-sc3} lies within the allowed parameter range of
\CP-violating 2HDM without approximate $Z_2$ symmetry as derived, for
instance, in \cite{Chen:2015gaa}.

\section{Theoretical predictions beyond the QCD Born approximation}
\label{sec:calculation}

As mentioned in the introduction the aim of this article is to study,
within the type-II two-Higgs-doublet model, the resonant production of
heavy neutral Higgs bosons and their decay into $\ttbar$,
\begin{equation}
pp\rightarrow\phi_{2,3}\rightarrow\ttbar X \, ,
\label{eq:resonant_ttbar}
\end{equation}
including the nonresonant SM $\ttbar$ background
\begin{equation}
pp\rightarrow\ttbar X
\label{eq:background_ttbar}
\end{equation}
and the interference of the resonant and nonresonant amplitudes. Specifically,  we consider
the three scenarios introduced in \Sec{sec:3-2hdmsc} and derive 
predictions for the cross section and distributions including all relevant NLO QCD corrections.

\subsection{Setup and leading-order results}
\label{subsec:LO}
\begin{figure}[htbp]
  \begin{center}
    \includegraphics[width=0.22\columnwidth]{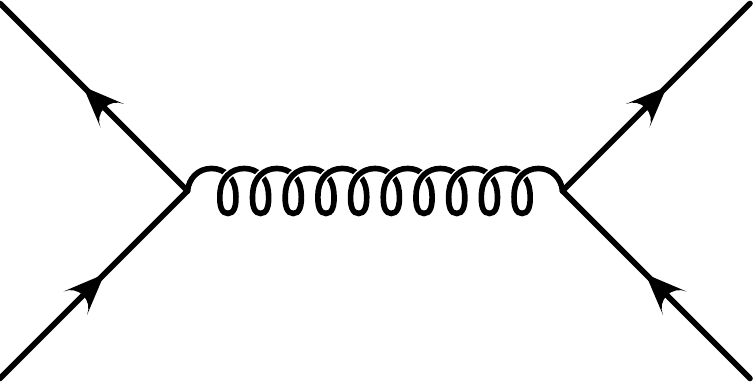}\quad
    \includegraphics[width=0.72\columnwidth]{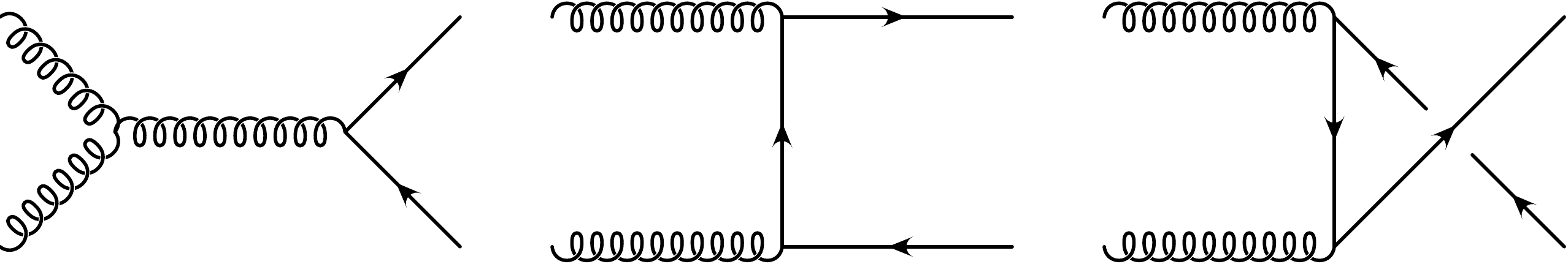}
    \caption{Dominant SM processes contributing to top-quark pair production.}
    \label{fig:QCD-Born}
  \end{center}
\end{figure}

\begin{figure}[htbp]
  \begin{center}
    \leavevmode
    \includegraphics[width=0.5\columnwidth]{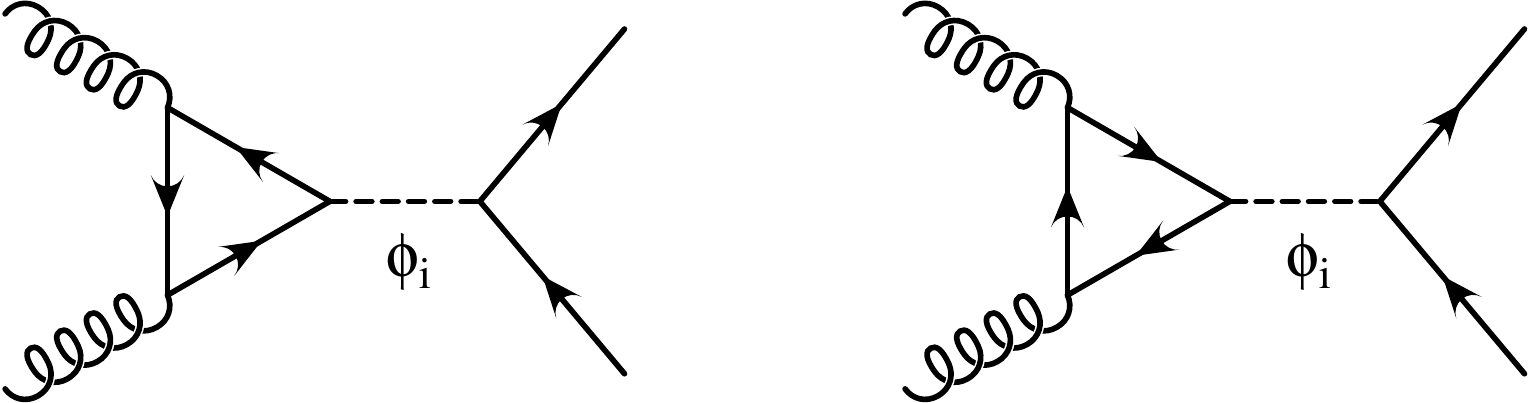}\
    \caption{$s$-channel Higgs boson contribution to top-quark pair production.}
    \label{fig:Phi-Born}
  \end{center}
\end{figure}

 In order to set up our conventions and introduce some abbreviations we start
with a short review of the leading-order results. For a detailed
discussion we refer to \Refs{Dicus:1994bm,Bernreuther:1997gs} and
\Ref{Craig:2015jba} where detector effects have been included in the
simulation. As usual, the hadronic cross section is calculated from
the partonic cross section through a convolution with the parton
distribution functions (PDF),
\begin{equation}
  d\sigma_{pp\to t\bar tX} = \sum_{i,j}\int dx_i dx_j\,\,
  F_{i/p}(x_i,\muf)  F_{j/p}(x_j,\muf)\, d\hat{\sigma}_{ij\to t\bar tX}(x_i
  P_1,x_j P_2,\muf,\mu_r),
  \label{eq:hascatXsec}
\end{equation}
where the $F_{i/p}(x,\mu_f)$ denote the PDF,
\muf and $\mu_r$ are the factorization  and renormalization scales, and
$\hat{\sigma}_{ij\to t\bar tX}$ denote the partonic cross sections. The sum
runs over all parton configurations contributing to top-quark pair
production. Within QCD the dominant contributions are due to gluon
fusion ($gg\to t\bar t$) and quark-antiquark annihilation ($q\bar q
\to t\bar t$). The leading-order Feynman diagrams are shown in
\Fig{fig:QCD-Born}.
The relative contributions of these two processes to  \eqref{eq:hascatXsec} depend on the energy scale,
 with the contribution from
gluon fusion dominating at low energies \cite{Kuhn:2013zoa}.  In \Fig{fig:Phi-Born} the
additional contribution due to    the $s$-channel exchange of a  Higgs boson $\phi_i$ is shown. 
In principle, all quarks contribute in the fermion triangles. However in
the scenarios considered here we have set $\tan\beta=0.7$. Thus the
bottom Yukawa coupling is not enhanced (as would be the case in
scenarios with $\tan\beta\gg 1$). That is why in these scenarios
the bottom loop contribution is is more
  suppressed than in the SM with respect to the contribution
from the top-quark loop. Contributions from quarks lighter than the
$b$ quark are even smaller. Hence we can safely neglect all
quark flavor contributions to the triangle which are lighter than
the top quark.
As we consider here an extended Higgs sector, we have to sum
over the (neutral) Higgs bosons $\phi_1$, $\phi_2$, and $\phi_3$.
We identify $\phi_1$ with the SM Higgs boson which contributes
to the nonresonant background (\ref{eq:background_ttbar}).
Since we are mainly interested in the resonant production of the
heavy Higgs bosons $\phi_2$ and $\phi_3$ we take at leading order
only the contribution from the diagrams shown in \Fig{fig:Phi-Born}
into account and neglect nonresonant Yukawa corrections due to
$\phi_{2,3}$ exchange.

\begin{figure}[htbp]
  \begin{center}
    \leavevmode
    \includegraphics{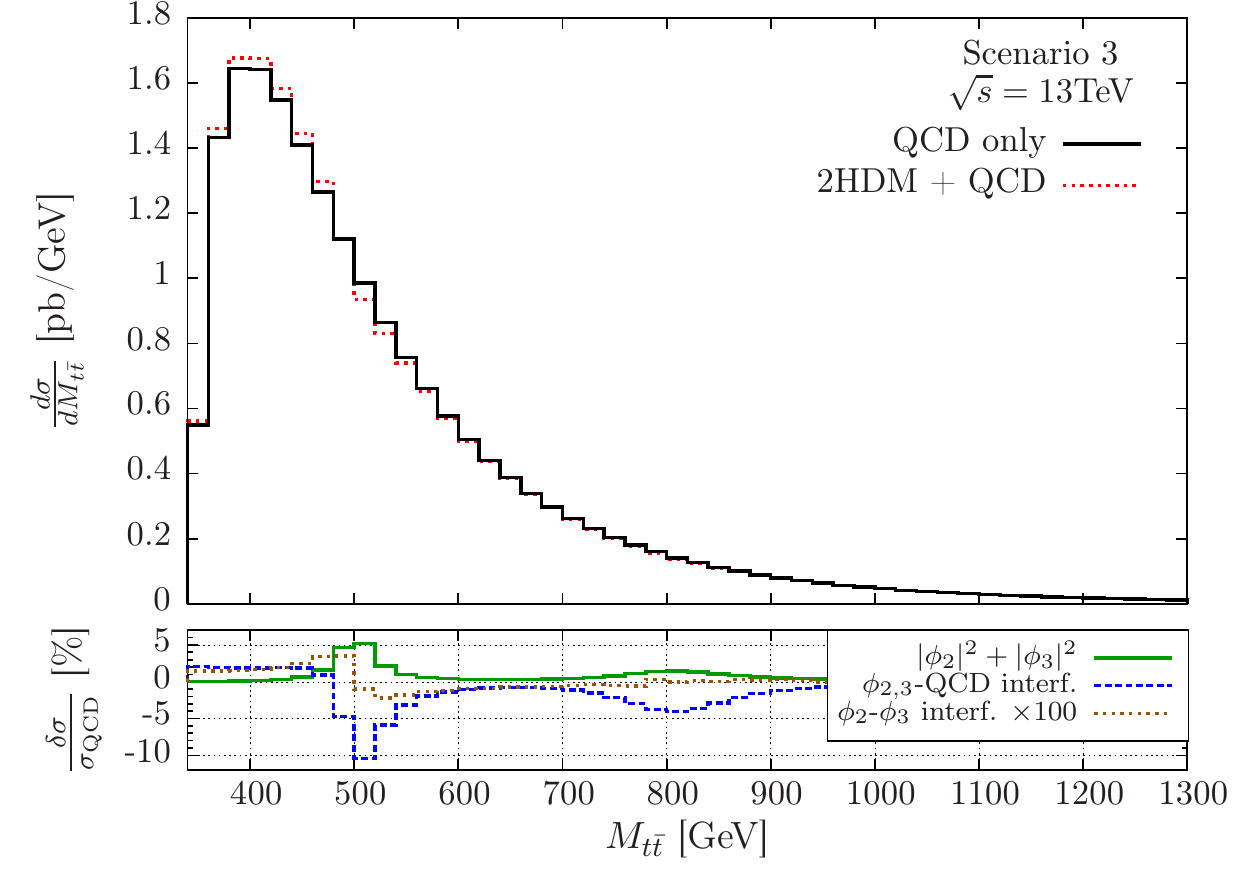}    
    \caption{Impact of different contributions to the hadronic differential
     cross section ${d\sigma}/{d\mtt}$ at leading order.
     Upper pane: The solid black curve shows the QCD background
     and the dotted red  curve displays the sum of QCD background, 2HDM contribution
     and the interference term. Lower pane: Relative contributions with respect to the
     QCD background. The solid green curve shows the sum of the contributions from
     $\phi_2$ and $\phi_3$, the dashed blue curve depicts the interference
      of the $\phi_2, \phi_3$ production amplitudes 
     with the QCD background,  and the dotted brown  curve the $\phi_2$-$\phi_3$ interference
     (multiplied with a factor 100 for better visibility).}
    \label{fig:loresult}
  \end{center}
\end{figure}

\begin{figure}[htbp]
  \begin{center}
    \leavevmode
    \includegraphics{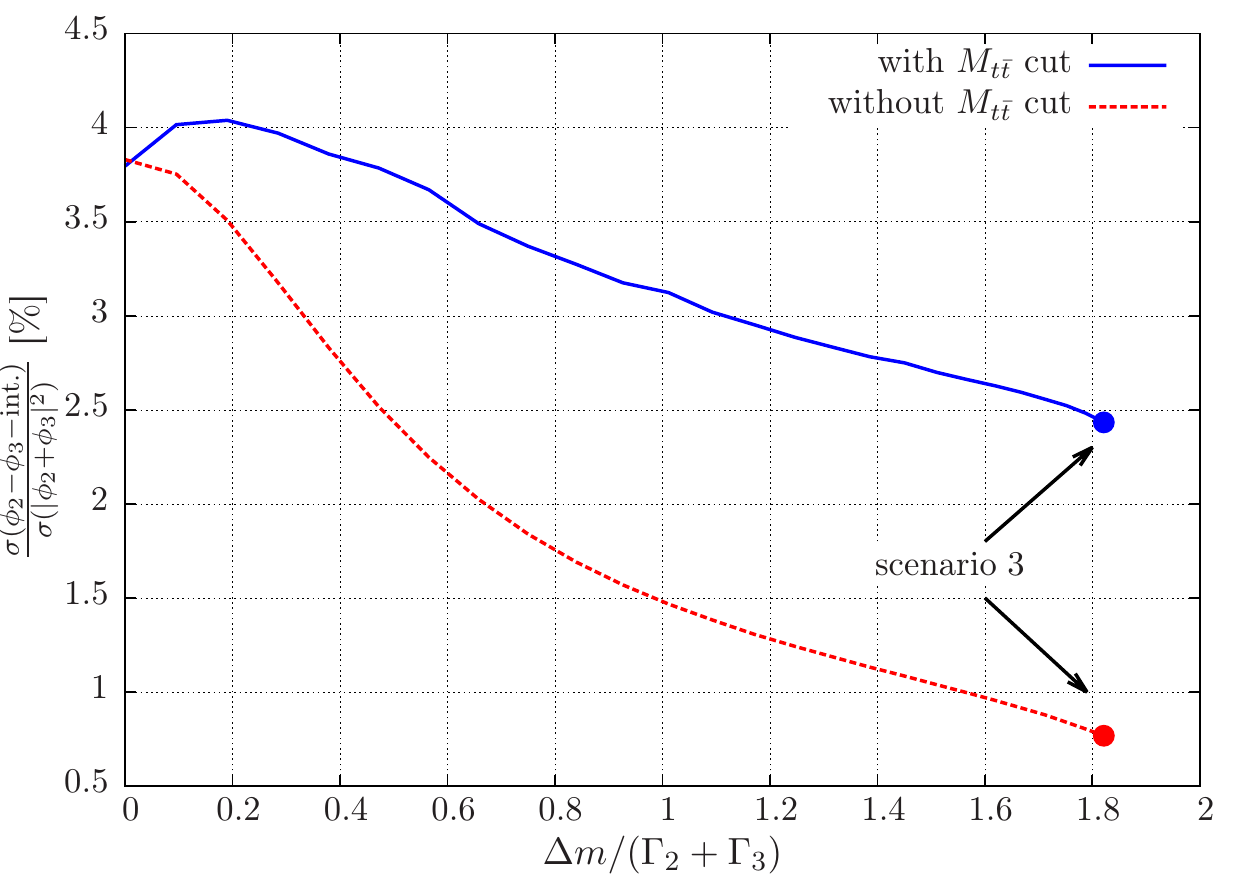}
    \caption{Relative contribution of the $\phi_2$-$\phi_3$-interference
    term (\ref{eq:PhiPhiInterference}) with respect to the total hadronic 
     $pp\to \phi_2,\phi_3\to \ttbar$  cross section 
     without QCD background as a function of the mass
    separation $\Delta m=m_3-m_2$ normalized to $\Gamma_2+\Gamma_3$.
    The parameters used for this plot
    correspond to those of scenario 3 except for the masses and widths of
    $\phi_2$ and $\phi_3$ which have been varied accordingly. The end of
    each curve corresponds exactly to scenario 3 with $m_2=500$ GeV,
    $m_3=800$ GeV and corresponding widths as given in
    \Tab{tab:res_widths_sc3}.}
    \label{fig:PhiPhiInterference}
  \end{center}
\end{figure}

For $gg\to t\bar t$ the scattering amplitude may be
written as
\begin{equation}
  {\cal A} = \AQCDLO + \sum_i \APhiiLO
\end{equation}
and the partonic cross section can be calculated using
\begin{equation}
  \label{eq:xsection}
  d\sigma_{gg\to t\bar t} = {1\over 32\pi} {\beta_t\over s} 
  \overline{|{\cal A}|^2}d\cos(\theta),
\end{equation}
where $s$ denotes the partonic center-of-mass energy squared,
$\cos(\theta)$  is the cosine of the scattering angle in the parton
center-of-mass system (cms), 
\begin{equation}
 \beta_t=\sqrt{1-{4 \mt^2\over s}} 
\end{equation}
is the velocity of the top quark in the partonic cms, and   $\mt$ denotes
the top-quark mass renormalized in the on-shell scheme. We overline
the squared matrix element in \Eq{eq:xsection} to indicate that it is
averaged/summed over the incoming/outgoing spin and color degrees of
freedom.  Squaring the amplitude for $gg\to t\bar t$ leads to
\begin{equation}
  \label{eq:Asquared}
  |\AQCDLO|^2 + 2 \sum_i \Re(\AQCDLO {\APhiiLO}^\ast)
   + 2\sum_{i>j}\Re( {\APhiiLO} {\APhijLO}^\ast) + \sum_i |{\APhiiLO}|^2.
\end{equation}
For the scenarios 1 -- 3 discussed in \Sec{sec:3-2hdmsc} sizeable
effects due to the extended Higgs sector can only be expected well
above the threshold for top-quark pair production in the vicinity of
the respective resonances $\phi_2, \phi_3$. (At these energies, the 
 contribution of the SM-like  Higgs boson $\phi_1$ is tiny and will, in fact, 
  be incorporated in the following sections into the nonresonant background 
   contributions.)
The expression for the pure QCD matrix element squared $|\AQCDLO|^2$ in
\Eq{eq:Asquared} can be found in \Refs{Nason:1987xz,Beenakker:1988bq}.
For the two heavy Higgs bosons the interference with the QCD amplitude
reads
\begin{equation}
  \label{eq:QCDPhiInterference}
  2 \sum_{i=2,3}\overline{ \Re(\AQCDLO {\APhiiLO}^\ast)}
  = - {16 \pi \as\over \CA\CF} {\mt^2\over \vev} {s\over
  1-\beta_t^2\*z^2}
\sum_{i=2,3} \left[ a_{it}\beta_t^2 \Re\left(P_i(s)\FS{i}\right) 
  - 2b_{it}\Re\left(P_i(s)\FP{i}\right) \right],
\end{equation}
with the propagator of the Higgs boson $\phi_i$ defined by
\begin{equation}
  P_i(s) = {1\over s - m_i^2 + i m_i \Gamma_i},
\end{equation}
and the $s$ dependent vertex factors given by
\begin{eqnarray}
  \FS{i} &=& -{\as a_{it}\over 8\pi\vev} \tau
  \left[(1-\tau)f(\tau)-1\right],\\
  \FP{i} &=& {\as b_{it}\over 16\pi\vev} \tau f(\tau),
\end{eqnarray}
where
\begin{equation}
  \tau = 1-\beta_t^2 = {4\mt^2\over s}.
\end{equation}
The loop function $f$ is given by
\begin{equation}
  f(\tau) = \left\{
    \begin{array}[htbp]{ll}
      {1\over 4} \left[\ln\left({1+\sqrt{1-\tau}\over 
            1-\sqrt{1-\tau}}\right)-i\pi\right]^2 & \tau < 1\\
      -\arcsin^2\left(\sqrt{1\over \tau}\right) & \tau\ge 1
    \end{array}
  \right. .
\end{equation}
The $a_{it}$ ($b_{it}$) denote the reduced (pseudo-)scalar Yukawa couplings defined
in \Sec{sec:3-2hdmsc}. The $\Gamma_i$ are the constant widths of the $\phi_i$ as
given in \Sec{sec:3-2hdmsc}. $\as\equiv\as(\mur)$ is the coupling constant of the strong
interaction evaluated at the  renormalization scale $\mur$. \CA and
\CF are the Casimir invariants of the SU(N) gauge group ($N=3$ for QCD),
\begin{equation}
  \CA = N,\quad \CF = {1\over 2N}(N^2-1) \, ,
\end{equation}
and $z$ is the cosine of the scattering angle of the top quark defined in the
parton cms,
\begin{equation}
  z = \cos(\theta_t).
\end{equation}
The two remaining contributions in \Eq{eq:Asquared} read
\begin{equation}
  \label{eq:PhiSquared}
  \sum_{i=2,3} \overline{|{\APhiiLO}|^2}
  = {2\over \CF}{\mt^2\over \vev^2}s^3 
  \sum_{i=2,3} |P_i(s)|^2\left(|\FS{i}|^2+4|\FP{i}|^2\right) 
  \left(a_{it}^2\beta_t^2+
  b_{it}^2\right),
\end{equation}
\begin{equation}
  \label{eq:PhiPhiInterference}
  2\overline{\Re( {\APhixLO{2}} {\APhixLO{3}}^\ast)}
  = {4\over \CF}{\mt^2\over \vev^2}s^3 \Re\left[P_2(s)P_3(s)^\ast
    \left(\FS{2}\FS{3}^\ast+4 \FP{2}\FP{3}^\ast\right) \right]
  \left(a_{2t}a_{3t}\beta_t^2+b_{2t}b_{3t}\right).
\end{equation}
 The interference term 
(\ref{eq:PhiPhiInterference}) is nonzero only
 in \CP-violating scenarios, where $\phi_2$,
$\phi_3$ have both scalar  and pseudoscalar components.
Although a detailed phenomenological discussion including higher
order corrections will be presented in \Sec{sec:results} it is useful
to study the  size  of the different contributions \eqref{eq:Asquared}
 to the LO differential cross section  \eqref{eq:hascatXsec}.
  The cross section
for top-quark pair production differential in the invariant mass \mtt of
the top-quark pair is shown in \Fig{fig:loresult} for scenario 3. 
As anticipated, sizeable effects due to the extended Higgs sector are
only visible in the vicinity of the resonances. A distinctive peak-dip
structure, which is due to the interference of the Higgs signal with
the QCD background amplitude, is visible for the lighter of the two heavy Higgs resonances 
 (which is $\phi_2$ in scenario 3). The heavier Higgs boson produces a
similar peak-dip structure which is, however, less pronounced
and hidden by the effect of the lighter resonance.  As a consequence of the peak-dip
structure the contribution of the extended Higgs sector to the total $\ttbar$
cross section is very small and below the currently attainable
experimental accuracy.  The effect can be enhanced by introducing a
cut on \mtt\ and evaluating the cross sections in the regions $2\mt <
\mtt < 480 $ GeV and $\mtt > 480$ GeV. However, even in this case
the effects are only at the level of a few percent. Qualitatively similar
results are obtained for scenarios 1 and 2.\\
Contrary to scenarios 1 and 2 the $\phi_2$-$\phi_3$ interference term 
 is nonzero in scenario 3.  As  shown in \Fig{fig:loresult} this
 term is small compared to the other contributions; it is less than 0.4 
 per mill with respect to the QCD background in each \mtt bin.  There are two
reasons for this interference to be so small. First, in scenario 3 the
 Higgs boson mass separation $\Delta m=m_3-m_2>\Gamma_2+\Gamma_3$ leads to a
suppression caused by the product of the two Higgs boson propagators in
\Eq{eq:PhiPhiInterference}, while in the case $\Delta m=0$ there is no
such suppression. This behavior is displayed in
\Fig{fig:PhiPhiInterference}, where the red curve shows the relative
contribution of the $\phi_2$-$\phi_3$ interference term 
 to the LO hadronic
 $pp\to \phi_2,\phi_3\to \ttbar$  cross section without the QCD background as a function of the mass
separation $\Delta m=m_3-m_2$.
 Second, even in the case $\Delta m=0$ the $\phi_2$-$\phi_3$  interference term is still
small (about 4\% compared to  the Higgs-only cross section). This is the
result of our choice of parameters for scenario 3, which leads to
Yukawa couplings with opposite signs. This is in turn  responsible
for cancellations in \Eq{eq:PhiPhiInterference} and renders the
interference term small.  The blue curve in
\Fig{fig:PhiPhiInterference} shows in addition
 the behavior of this interference term as a function of the Higgs mass separation
  using an additional cut, namely $m_2-150\,\text{GeV}\le\mtt\le m_2$. This cut 
   enhances the $\phi_2$-$\phi_3$
interference because it prevents the partial cancellation due to the
peak-dip structure which is also present in this 
interference term. We apply similar cuts in the NLO analysis presented below. Even
with this cut the interference contribution \Eq{eq:PhiPhiInterference}
is small compared to the total Higgs boson contributions to the $\ttbar$
 cross section and can be safely
neglected in phenomenological applications.

In order to find or put limits on heavy Higgs boson production and decay to $\ttbar$ 
 the experimental
analysis must focus on the resonance region. Since the mass(es) of the heavy resonance(s)
 is (are) not known beforehand, an analysis with a sliding
 $\mtt$ window combined with a sideband analysis should be applied. From
the theoretical point of view NLO corrections are required to
provide reliable predictions in the vicinity of the resonances. 
The evaluation of the NLO corrections
will be discussed in the next subsection.

 Besides we remark that, depending on the mass, couplings and width of the heavy Higgs resonance,
  a pure dip rather than a peak dip may show up in the $\mtt$ spectrum,
  or even a complete washout of the resonance signal can occur. For a recent analysis, see \Ref{Jung:2015gta}.
  For the parameter scenarios described in Sec.~\ref{sec:3-2hdmsc}  the lighter of the two heavy Higgs resonances produces
   always a peak-dip structure in the  $\mtt$ distribution; see Sec.~\ref{sec:results}.
  
\subsection{Next-to-leading order QCD corrections}
\label{subsec:NLOQCD}
For the calculation of the NLO QCD corrections to the process in
\Eq{eq:resonant_ttbar} we apply two approximations. First,
as far as the Higgs coupling to the gluons is concerned, we work
in the heavy top-quark limit. Second, we take advantage
of the fact that the theoretical predictions are only required in the
vicinity of the resonances because the experimental analysis will focus
on this region. This allows us to work in the pole
approximation, keeping only contributions that are resonant in the
intermediate Higgs boson(s).  This is  consistent with our approach
 to work at leading order in the Yukawa couplings. The heavy Higgs resonances 
  are described by a Breit-Wigner propagator with constant width. In our 
  approach this amounts to using the so-called complex mass scheme
\cite{Denner:1999gp,Denner:2005fg,Nowakowski:1993iu}, as outlined in Appendix~\ref{app:higgs_resonances}. 

In order to fix our conventions we briefly review the approximations 
in the following. 
\paragraph*{Effective theory approach:}
In the heavy top-quark limit the triangles shown in \Fig{fig:Phi-Born} are
shrunk to an effective vertex.\footnote{The virtual NLO QCD corrections for scalar and pseudoscalar Higgs-boson 
production in gluon fusion were calculated for arbitrary quark masses in \Ref{Harlander:2005rq}.}
Relying on Lorentz covariance and
gauge invariance, the effective vertex can be described within an
effective field theory approach with the relevant contribution to the
Lagrangian given by
\begin{equation}
  {\cal L}_{\text{eff}} = \sum_{j=2}^3\bigg[f_{Sj}G_{\mu\nu}^aG_a^{\mu\nu}
  +f_{Pj}\epsilon_{\mu\nu\alpha\beta}G_a^{\mu\nu}G_a^{\alpha\beta}\bigg]\phi_j \, ,
  \label{eq:EffectiveLagrangian}
\end{equation}
where $f_{Sj}$ and $f_{Pj}$ are the Wilson coefficients for the coupling
 of the \CP-even and \CP-odd component of the Higgs boson $\phi_j$ to gluons.
The effective Lagrangian leads to the following
Feynman rules ($g_s = \sqrt{4\pi\as}$)
\begin{eqnarray}
\begin{minipage}[htbp]{3.5cm}
\hspace{1cm}\includegraphics[scale=0.6]{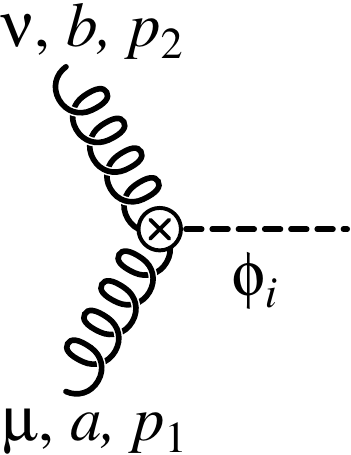}
\end{minipage}
& = & -4i\delta^{ab}\left[f_{Sj}\left(g_{\mu\nu}(p_1\cdot p_2)
-p_{2\mu}p_{1\nu}\right)-2f_{Pj}\epsilon_{\mu\nu\rho\sigma}p_1^{\rho}p_2^{\sigma}
\right]\label{Feyn_ggH},
\end{eqnarray}
\begin{eqnarray}
\begin{minipage}[htbp]{3.5cm}
\vspace{0.18cm}
\includegraphics[scale=0.6]{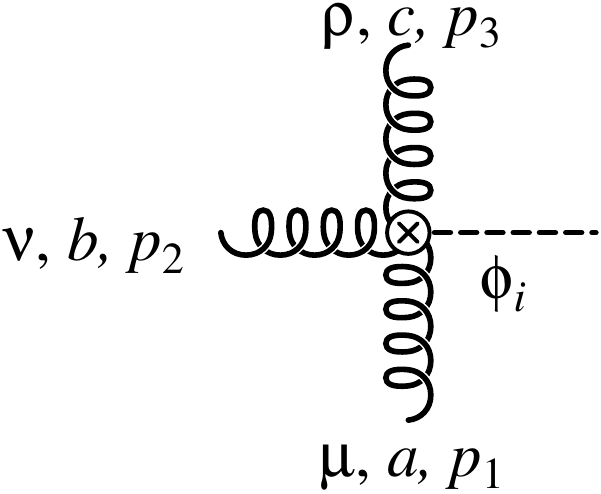}
\end{minipage}
& = & 4g_sf^{abc}\Bigg\{f_{Sj}\left[g_{\mu\nu}(p_1-p_2)_{\rho}
+ g_{\nu\rho}(p_2-p_3)_{\mu} +g_{\rho\mu}(p_3-p_1)_{\nu}\right]\nonumber\\
&-&2f_{Pj}(p_1+p_2+p_3)^{\alpha}\epsilon_{\alpha\mu\nu\rho}\Bigg\}.
\label{Feyn_gggH}
\end{eqnarray}
The momenta are taken to be incoming and we use the convention
$\epsilon_{0123}=+1$.
 The Lagrangian ${\cal L}_{\text{eff}}$ contains also an effective
$gggg\phi_j$ coupling, which is however not needed at the order of perturbation theory 
in which we are working. The effective couplings  $f_{Sj}$ and $f_{Pj}$ can be further
expanded in powers of \as:
\begin{equation}
  f_{Sj}=\frac{\as}{\pi}\left(f_{Sj}^{(0)}+\frac{\alpha_s}{\pi}f_{Sj}^{(1)}
    + \ldots \right), \quad 
  f_{Pj}=\frac{\as}{\pi}\left(f_{Pj}^{(0)}
    +\frac{\alpha_s}{\pi}f_{Pj}^{(1)}
    +\ldots\right).
\end{equation}
Comparing the results obtained in the full theory as presented in the previous
subsection with the results in the effective theory approach, it is
straightforward to obtain the well-known result:
\begin{eqnarray}
  {\as \over \pi}f^{(0)}_{Sj} &=& \lim_{\mt \to \infty} \FS{j} 
  = {\as \over\pi}{a_{jt}\over 12 \vev},\\ 
  {\as \over \pi}f^{(0)}_{Pj} &=& \lim_{\mt \to \infty} \FP{j}
  = -{\as \over\pi}{b_{jt}\over 16 \vev}.
\end{eqnarray}
Using the expressions from \Ref{Djouadi:1991tka,Harlander:2001is,
Chetyrkin:1998mw,Harlander:2002vv} adapted to our
conventions we get
for $f^{(1)}_{Sj}$ and $f^{(1)}_{Pj}$:
\begin{eqnarray}
  f_{Sj}^{(1)} & = & \frac{(4\pi)^{\epsilon}}{\Gamma(1-\epsilon)}
  \frac{a_{jt}}{12\vev}\Big(\frac{11}{4}-\frac{\beta_0}{\epsilon}\Big)
  =\frac{(4\pi)^{\epsilon}}{\Gamma(1-\epsilon)}f_{Sj}^{(0)}\Big(\frac{11}{4}
  -\frac{\beta_0}{\epsilon}\Big),\\
  f_{Pj}^{(1)} & = & \frac{(4\pi)^{\epsilon}}{\Gamma(1-\epsilon)}
  \frac{b_{jt}}{16\vev}\frac{\beta_0}{\epsilon}=
  -\frac{(4\pi)^{\epsilon}}{\Gamma(1-\epsilon)}f_{Pj}^{(0)}
  \frac{\beta_0}{\epsilon},
\end{eqnarray}
where $\epsilon=(4-d)/2$ is the dimensional regulator working in $d$
dimensions, and 
\begin{equation}
  \beta_0 = \frac{1}{2}\left(\frac{11}{6}C_A-\frac{2}{3}T_RN_F \right)
\end{equation}
denotes the leading coefficient of the QCD $\beta$-function with $N_F$
being the number of massless quark flavors and $T_R=1/2$. One may
question the validity of the heavy top-quark limit given that we apply
the effective theory approach above the threshold of top-quark pair
production. Since the effective theory approach has been shown to
provide also reliable estimates for energies well above its naive
domain of validity---if the full  leading-order result is kept 
and an appropriate reweighting is performed \cite{Kramer:1996iq}---we 
believe that reliable estimates of the NLO corrections can be obtained
within this approximation.
We outline our reweighting method adapted from \Ref{Kramer:1996iq} at the end of this section.

\paragraph{Pole contribution and soft gluon approximation:}
As sizeable effects of the extended Higgs sector can be expected
only in the heavy Higgs boson resonance region it is sufficient to
restrict the theoretical predictions to  \mtt intervals  around the Higgs boson masses. 
This can be achieved by the pole approximation
 where only the contributions enhanced by a resonant propagator are
kept. 
\begin{figure}[htbp]
  \begin{center}
    \leavevmode    
    \includegraphics[width=0.2\columnwidth]{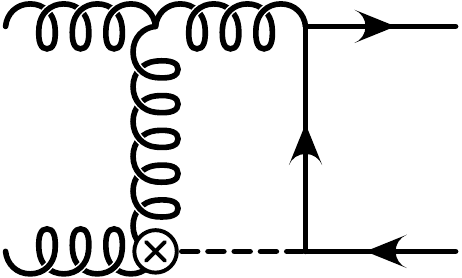}
    \caption{Diagram to be calculated in the soft gluon approximation.}
    \label{fig:nonresonant}
  \end{center}
\end{figure}
\begin{figure}[htbp]
  \begin{center}
    \leavevmode    
    \includegraphics[width=0.3\columnwidth]{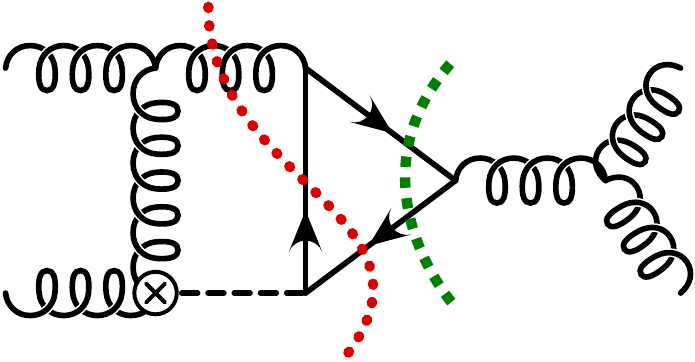}
    \caption{Virtual (green dashed) and real (red dotted) 
      contributions which cancel in the soft gluon approximation.}
    \label{fig:SoftGluonCancelation}
  \end{center}
\end{figure}
For Higgs propagators appearing within loop diagrams the soft gluon
approximation can be used to extract the leading pole contribution.
Taking as an example the diagram shown in \Fig{fig:nonresonant} it is
clear that a resonant contribution at $\mtt \approx m_i$ is only possible if the
additional gluon in the loop is soft. This is the essence of the soft
gluon approximation in which the integrand of the loop integral is
approximated by the contribution from soft gluons. Note that the loop
integration is not restricted within this approximation. For more
details we refer to \Refs{Fadin:1993dz,Melnikov:1995fx,Beenakker:1997ir,Dittmaier:2014qza}.  
 Evidently, the same approximation has to be applied also to the real corrections.
In fact, the virtual corrections alone are infrared divergent because
the soft gluon is coupled to external lines. Only the sum of the
virtual and real corrections  yields a finite result. In the soft
gluon limit the amplitudes simplify significantly and it turns out
that the appropriate real corrections completely cancel the
corresponding virtual contributions once sufficiently inclusive
observables are studied.  An example of this cancellation is
illustrated in \Fig{fig:SoftGluonCancelation}.  This was checked
for the various contributions on a case by case study. More details about
the example shown in \Fig{fig:SoftGluonCancelation} are given in
Appendix~\ref{app:SGA}. The corrections of the type displayed in 
Figs.~\ref{fig:nonresonant}--\ref{fig:SoftGluonCancelation}
 are also called nonfactorizable corrections because the gluon radiation
  connects Higgs boson production and decay to $\ttbar$.
 We also stress that gauge invariance is only
guaranteed if the soft gluon approximation is consistently applied.

\begin{figure}[htbp]
  \begin{center}
    \leavevmode
    \includegraphics[width=0.5\columnwidth]{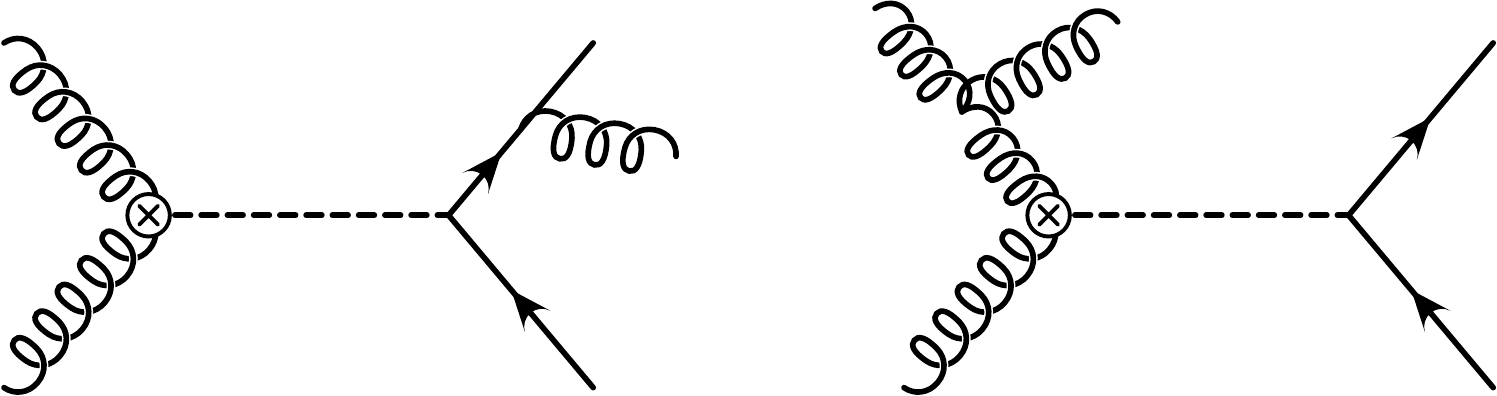}
    \caption{Sample diagrams for initial and final state real corrections
    to the process\hspace{\textwidth}$gg\rightarrow\phi_{2,3}\rightarrow\ttbar$.}
    \label{fig:phi-phi-real}
  \end{center}
\end{figure}

After these introductory remarks we now briefly summarize the various
 terms that contribute, within the aforementioned approximations, to 
  resonant $\phi_j$ production and interference with the nonresonant 
   $\ttbar$ amplitudes at NLO QCD, i.e., at order $\alpha_s^3$.
\begin{figure}[htbp]
  \begin{center}
    \leavevmode
    \includegraphics[width=0.9\columnwidth]{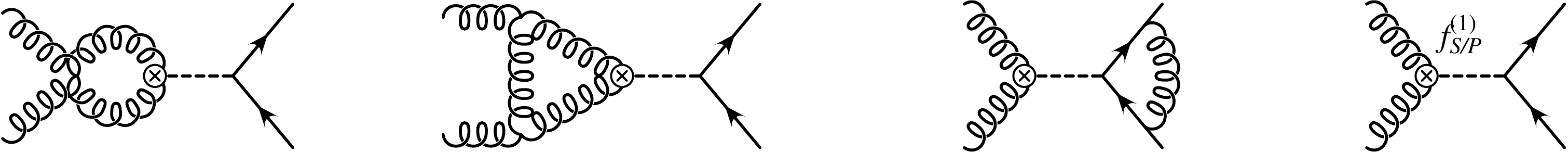}
    \caption{QCD corrections to top-quark pair production via resonant Higgs exchange.}
    \label{fig:AphiNLO}
  \end{center}
\end{figure}
In \Fig{fig:AphiNLO} the factorizing QCD corrections for top-quark pair production
through Higgs exchange are shown and we do not display diagrams that
vanish in dimensional regularization. 
 Denoting the amplitude for this
contribution with \APhiiNLO and the QCD corrections to 
nonresonant top-quark pair production (without any Higgs contribution) with \AQCDNLO,
the virtual corrections to the squared matrix element read
\begin{equation}
  \sum_{i,j}2\Re\left(\APhiiNLO {\APhijLO}^\ast \right)
  + \sum_{i}2\Re\left(\APhiiNLO \AQCDLOcc \right)\quad
  + \sum_{i}2\Re\left(\AQCDNLO {\APhiiLO}^\ast \right).
\end{equation}
We refrain from discussing the  NLO QCD  corrections to $\ttbar$ production here, because they
 have been known for a long time \cite{Nason:1987xz,Nason:1989zy,Beenakker:1988bq,
  Beenakker:1990maa,Bernreuther:2001rq,Bernreuther:2004jv}. (In \cite{Bernreuther:2001rq,Bernreuther:2004jv}
  the full spin dependence of the top quarks
  is kept.)  After renormalization the virtual corrections are
  ultraviolet (UV) finite. However, they still contain infrared (IR)
  singularities which are canceled only after the inclusion of the
  real corrections. Sample diagrams for $gg\to \phi_i\to t\bar t g$
  are shown in \Fig{fig:phi-phi-real}.
The square of this amplitude cancels the IR singularities in
$2\Re\left(\APhiiNLO {\APhijLO}^\ast \right)$.  Because of
the color structure  there is no interference between initial and final
state radiation in this case. The amplitude for $gg\to \phi_i\to t\bar
t g$ interferes also with the QCD amplitude $gg\to t\bar t g$. This
contribution cancels the IR singularities in
$\sum_{i}2\Re\left(\APhiiNLO \AQCDLOcc \right)$ and
$\sum_{i}2\Re\left(\AQCDNLO {\APhiiLO}^\ast \right)$. We stress that
the contribution to be combined with $\sum_{i}2\Re\left(\APhiiNLO
  \AQCDLOcc \right)$ is evaluated using the soft gluon
approximation.  

\begin{figure}[htbp]
  \begin{center}
    \leavevmode
    \includegraphics[width=0.25\columnwidth]{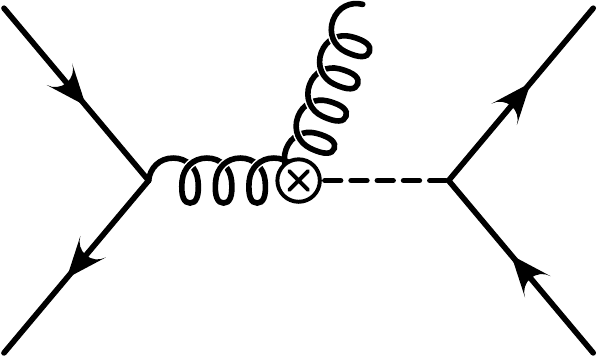}
    \caption{Resonant heavy Higgs boson production in $q\bar q \to t \bar t g$.}
    \label{fig:qq-phi-ttg}
  \end{center}
\end{figure}

 At order $\as^3$ heavy Higgs bosons can also be produced by
  $q\bar q$  annihilation with an amplitude that is not suppressed by light
   quark Yukawa couplings. The respective Feynman diagram for 
    $q{\bar q} \to  t \bar t g$ is shown in  \Fig{fig:qq-phi-ttg}.
 Besides the square of this amplitude also the interference with the 
 QCD amplitude for $q\bar q \to t \bar t g$ contributes.
Both contributions are free of IR singularities.

In addition, resonant heavy Higgs boson production and decay to $\ttbar$
 occurs also by (anti)quark gluon fusion, $qg \to t {\bar t} q$ and $g\bar q \to t {\bar t} \bar  q$.
 The corresponding amplitudes are obtained from the amplitude depicted in 
 \Fig{fig:qq-phi-ttg} by crossing.
 The square of the amplitude with the intermediate Higgs bosons and the interference
with the respective QCD amplitude lead to double and single resonant 
 terms. Contrary  to the $q\bar q$ case both contributions
suffer from initial state singularities.

\paragraph{Some technical aspects:} 
 The calculation of the NLO corrections described above is
conceptually straightforward. Since the full results are rather
lengthy we briefly outline the calculation and show explicit results
for some illustrative examples only. As far as the virtual corrections are
concerned we employ the Passarino-Veltman reduction \cite{Passarino:1978jh}
to reduce the one-loop tensor integrals to scalar integrals. Including the
contribution from UV renormalization we obtain for example for the interference
of the diagrams shown in \Fig{fig:AphiNLO}  and the leading-order
Higgs boson exchange diagrams
\begin{eqnarray}
  \sum_{i,j}2\overline{\Re\left(\APhiiNLO {\APhijLO}^\ast \right)}
  &=& \frac{\alpha_s}{\pi}\Bigg\{
  -\frac{(4\pi)^{\epsilon}}{\Gamma(1-\epsilon)}
  \left\{\Re\left[\left(\frac{\mu^2}{-s-i0}\right)^{\epsilon}\right]
    \frac{C_A}{\epsilon^2}+\frac{2\beta_0}{\epsilon}\right\}
  \B
  \nonumber\\
  &+&\left(\frac{4\pi\mu^2}{m_t^2}\right)^{\epsilon}
  \Gamma(1+\epsilon)C_F\frac{1}{\epsilon}\Re\left(a_{-1}\right)
  \B\nonumber\\
  &+&\frac{11}{2}\left.\B\right|_{f_P=0} 
  +2C_A\left.\B\right|_{f_S=0}\nonumber\\
  &+&C_F\left[\Re(a_0)\left.\B\right|_{b_t=0}
    +\Re(\bar{a}_0)\left.\B\right|_{a_t=0}
  \right] 
  +\mathcal{O}(\epsilon)\Bigg\} \, ,
\label{eq:M_phi_virt}
\end{eqnarray}
where $\B$ denotes the contribution from
\begin{displaymath}
\sum_{i=2,3} \overline{|{\APhiiLO}|^2}
+2\overline{\Re( {\APhixLO{2}} {\APhixLO{3}}^\ast)}
\end{displaymath}
evaluated in the heavy top-quark limit:
\begin{eqnarray}
  \B &=& {2\over \CF}\left({\as\over \pi}\right)^2{\mt^2\over
    \vev^2}s^3 
  \Bigg\{
  \sum_{i=2,3} |P_i(s)|^2\left(|\fLOS{i}|^2+4|\fLOP{i}|^2\right) 
  \left(a_{it}^2\beta_t^2+
  b_{it}^2\right)\nn\\
&+& 2\Re\left[P_2(s)P_3(s)^\ast
    \left(\fLOS{2}\fLOS{3}^\ast+4 \fLOP{2}\fLOP{3}^\ast\right) \right]
  \left(a_{2t}a_{3t}\beta_t^2+b_{2t}b_{3t}\right)\Bigg\}.
\end{eqnarray}
The coefficients $a_{-1}$, $a_0$ and $\bar a_0$ are related to the
one-loop $\phi_j \to \ttbar$ form factor and can be taken for example from \Ref{Bernreuther:2005gw}.
The poles in the dimensional regulator $\epsilon$ are due to IR singularities.
Note that we keep the polarization of the incoming gluons in four
dimensions. As a consequence $\B$ does not depend on $\epsilon$. The IR singularities 
 are handled with the Catani-Seymour dipole
subtraction  \cite{Catani:1996vz,Catani:2002hc}. In this formalism local
counterterms are added to the real corrections such that on the one hand the
soft and collinear singularities are canceled pointwise in phase space and, on
the other hand, the subtraction is simple enough to be integrated analytically
over the $d$-dimensional phase space of the unobserved parton. The
cross section reads  schematically \cite{Catani:1996vz}:
\begin{eqnarray}
  \xsNLO & = & \xsNLOmp(p)+\xsNLOm(p)+\int_0^1dx\,d\xshNLOm(x;xp)\nn\\
  & = & \int_{m+1}\left[\left(\xsR(p)\right)_{\epsilon=0}
-\left(\sum_{\text{dipoles}}d\xsB(p)\otimes \left(dV_{\text{dipole}}+
dV_{\text{dipole}}'\right)\right)_{\epsilon=0}\right]\nn\\
& + & \int_m\left[d\xsV(p)+d\xsB(p)\otimes\mathbf{I}\right]_{\epsilon=0}
+\int_0^1dx\int_m\left[d\xsB(xp)\otimes\left(\mathbf{K}+\mathbf{P}\right)(x)\right].
\label{eq:CSMaster}
\end{eqnarray}
Here $\xsB$, $\xsV$, and $\xsR$ denote the Born cross section and the 
 contributions from the virtual
and real corrections, respectively. Since the real corrections
combined with the subtraction terms are finite, they can be evaluated
numerically in four dimensions. The singularities are thus made manifest
in the integrated subtraction terms which are added back 
[last line in \Eq{eq:CSMaster}].
The subtraction term to be
combined
with the real corrections is constructed as a sum of individual dipoles:
\begin{displaymath}
  \sum_{\text{dipoles}}d\xsB(p)\otimes \left(dV_{\text{dipole}}+
dV_{\text{dipole}}'\right).
\end{displaymath}
The analytically integrated dipoles are collected in the operators 
$\mathbf{I}$, $\mathbf{K}$, and $\mathbf{P}$ which may introduce
correlations in spin and color. Combining for the case at hand the
virtual corrections with the contribution from  the
$\mathbf{I}$
operator an IR finite result is obtained:
\begin{eqnarray}
  d\sigma_V + d\sigma_{\mathbf{I}}&=&
  {1\over 32\pi}\frac{\alpha _s}{\pi }{\beta_t\over s} 
  \Bigg\{\Bigg[2 \beta _0+\frac{67}{18}C_A
  -\frac{10}{9} N_F T_R\nonumber\\
  &+&C_F \Bigg(3-\frac{\beta_t ^2+6 \beta_t +1}{12 \beta_t }\pi^2
  -\frac{1+\beta_t ^2}{4 \beta_t }\ln^2 y
  -\frac{1+\beta_t ^2}{2 \beta_t }\ln\left(\frac{1+\beta_t ^2}{2}\right)
  \ln\left(y\right)\nonumber\\
  &+&\left(\frac{1+\beta_t^2}{2 \beta_t } \ln\left(y\right)
    +\frac{3}{2}\right) \ln\left(\frac{1-\beta_t ^2}{2(1+\beta_t^2)}\right)
  +\mathcal{V}^{(\text{NS})}_q\Bigg)+2 \beta_0 \ln\left(\frac{\mu
      ^2}{s}\right)\Bigg]\B \nonumber\\
  &+&2 C_A \left.\B\right|_{f_S=0}
  +\frac{11}{2}\left.\B\right|_{f_S=0}\nonumber\\
  &+&C_F\left[\Re(a_0)\left.\B\right|_{b_t=0}
    +\Re(\bar{a}_0)\left.\B\right|_{a_t=0}\right]
  \Bigg\}d\cos(\theta)
\end{eqnarray}
with 
\begin{eqnarray}
\mathcal{V}^{(\text{NS})}_q&=&
\frac{3}{2}\ln\left(\frac{1}{2} \left(1+\beta_t ^2\right)\right)\nonumber\\
&+&\frac{1+\beta_t ^2}{2\beta_t} \left(
  2\ln\left(y\right)\ln\left(\frac{2 \left(1+\beta_t ^2\right)}{(1+\beta_t)^2}\right)
 +2\text{Li}_2\left(y^2\right)
 -2 \text{Li}_2\left(\frac{2 \beta_t }{1+\beta_t}\right)
 -\frac{\pi ^2}{6}\right)\nonumber\\
&+&\ln\left(1-\frac{1}{2}\sqrt{1-\beta_t^2}\right)
-2 \ln\left(1-\sqrt{1-\beta_t ^2}\right)
-\frac{1-\beta_t ^2}{1+\beta_t ^2} 
\ln\left(\frac{\sqrt{1-\beta_t ^2}}{2-\sqrt{1-\beta_t^2}}\right)\nonumber\\
&-&\frac{\sqrt{1-\beta_t ^2}}{2-\sqrt{1-\beta_t ^2}}
+2\frac{ \left(1-\sqrt{1-\beta_t ^2}- \beta_t ^2\right)}{1+\beta_t ^2}
+\frac{\pi ^2}{2}
\end{eqnarray}
and
\begin{eqnarray}
y & = & \frac{1-\beta_t}{1+\beta_t}.
\end{eqnarray}
The sum of the contributions from the integrated dipoles responsible for initial
state singularities  and the  QCD factorization  reads 
\begin{eqnarray}
d\sigma_{\mathbf{KP}}&=&\int_0^1dx\int_m\left[d\xsB(xp)\otimes\left(\mathbf{K}+\mathbf{P}\right)(x)\right]
\nonumber\\
&=&\frac{\alpha_s}{\pi}\int_0^1 dx\int d\Pi_2\left\{
\overline{K}^{gg}(x)-K_{\text{F.S.}}^{gg}(x)+\widetilde{K}^{gg}(x)
-P^{gg}(x)\ln\left(\frac{\mu_F^2}{xs}\right)\right\}
\B(xs).
\end{eqnarray}
In the evaluation of $\B(xs)$ the partonic center-of-mass energy
squared is set to $xs$.  Since we factorize the initial state
singularities in the \MSbar scheme the contribution
$K_{\text{F.S.}}^{gg}(x)$ encoding the scheme dependence for schemes
different from the \MSbar scheme is zero. For the definition of the
remaining contributions we refer to \Ref{Catani:1996vz}. 

For the real corrections
very short expressions can be derived for the squared Higgs boson
contribution.  In the reaction
\begin{equation}
g(p_1) + g(p_2) \rightarrow t(k_1) + \bar{t}(k_2) + g(q)
\end{equation}
 the $\ttbar$ pair is produced in a color singlet state. Therefore, the squared
  matrix element is given by an incoherent sum of two terms associated with gluon radiation 
   from the initial and final state. We find
\begin{eqnarray}
 \overline{|{\cal A}(gg\stackrel{\phi_{2},\phi_3}{\longrightarrow}t\bar t
 g)|^2_{\rm ISR}}  
 &=& 
 \frac{16\alpha_s^3C_A}{\pi C_F}\frac{m_t^2}{\vev^2}
 \frac{(s+t+u)^4+s^4+t^4+u^4}{stu}\nonumber\\
 &\times&\Bigg\{\sum_{j=2,3}\left((f_{Sj}^{(0)})^2+4(f_{Pj}^{(0)})^2\right)
 \Big(-a_{jt}^2 t_{12}+b_{jt}^2 s_{12}\Big)
 \left|P_j(s_{12})\right|^2\nonumber\\
 &+&2\left(f_{S2}^{(0)}f_{S3}^{(0)}
   +4f_{P2}^{(0)}f_{P3}^{(0)}\right)
 \Big(-a_{2t}a_{3t}t_{12}+b_{2t}b_{3t}s_{12}\Big)
 \Re\big[P_{23}(s_{12})\big]\Bigg\}
\end{eqnarray}
with 
\begin{equation}
s  =  (p_1 + p_2)^2,\quad
t  =  (p_1 - q)^2,\quad
u  =  (p_2 - q)^2,\quad
s_{12}  =  (k_1 + k_2)^2,\quad
t_{12} = (k_1-k_2)^2,
\end{equation}
and
\begin{equation}
  P_{23}(s)= P_2(s)P_3(s)^\ast.
\end{equation}
The contribution from final state gluon radiation reads
\begin{eqnarray}
\overline{|{\cal A}(gg\stackrel{\phi_{2},\phi_3}{\longrightarrow}t\bar t
g)|^2_{\rm FSR}} 
& = & 
\frac{16 \alpha_s^3 s^2}{\pi s_{13}^{'2} s_{23}^{'2}}\frac{m_t^2}{\vev^2}
\Bigg\{\sum_{j=2,3}\Big[(f_{Sj}^{(0)})^2+4(f_{Pj}^{(0)})^2\Big]
\Big[a_{jt}^2 K_a+b_{jt}^2 K_b\Big]\left|P_j(s)\right|^2\nonumber\\
&+&2\Big[f_{S2}^{(0)}f_{S3}^{(0)}+4 f_{P2}^{(0)}f_{P3}^{(0)}\Big]
\Big[a_{2t} a_{3t} K_a+b_{2t} b_{3t} K_b\Big]\Re\big[P_{23}(s)\big]\Bigg\}
\end{eqnarray}
with
\begin{eqnarray}
K_a & = & 8 m_t^4 (s-s_{12})^2
-2 m_t^2 \left(s (s_{13}'+s_{23}')^2+4 s_{12} s_{13}'s_{23}'\right)
+s_{13}' s_{23}'\left(s^2+s_{12}^2\right),\\
K_b & = & s_{13}'s_{23}' \left(s^2+s_{12}^2\right)
-2 m_t^2 s(s-s_{12})^2,
\end{eqnarray}
and
\begin{equation}
s_{13}'  =   2k_1\cdot q,\quad s_{23}'  =   2k_2\cdot q.  
\end{equation}

For the quark initiated processes
\begin{equation}
  q(p_1) + \bar q(p_2) \to t(k_1) + \bar t(k_2) + g(q)
\end{equation}
and
\begin{equation}
  q(p_1) + g(p_2) \to t(k_1) + \bar t(k_2) + q(q)  
\end{equation}
the squared amplitudes  without the QCD interference are given  by
\begin{eqnarray}
\overline{|{\cal A}(qg \stackrel{\phi_{2},\phi_3}{\longrightarrow}t\bar t q)|^2}& = &
\frac{16 \alpha_s^3}{\pi}\frac{m_t^2}{\vev^2}\frac{s^2+u^2}{-t}
\Bigg\{\sum_{j=2,3}
\Big[(f_{Sj}^{(0)})^2+4(f_{Pj}^{(0)})^2\Big]
\big(-a_{jt}^2t_{12} +b_{jt}^2s_{12} \big)\left|P_j(s_{12})\right|^2\nonumber\\
& + & 2 \Big[f_{S2}^{(0)}f_{S3}^{(0)}+4f_{P2}^{(0)}f_{P3}^{(0)}\Big]
\big(-a_{2t}a_{3t}t_{12} +b_{2t}b_{3t}s_{12} \big)\Re[P_{23}(s_{12})] \Bigg\}
\end{eqnarray}
and
\begin{eqnarray}
\overline{|{\cal A}(q \bar q \stackrel{\phi_{2},\phi_3}{\longrightarrow} 
t\bar t g)|^2} & = &
\frac{32 \alpha_s^3 C_F}{\pi} \frac{m_t^2}{\vev^2}\frac{t^2+u^2}{s}
 \Bigg\{\sum_{j=2,3}
\Big[(f_{Sj}^{(0)})^2+4(f_{Pj}^{(0)})^2\Big]
\big(-a_{jt}^2 t_{12} + b_{jt}^2s_{12}\big)\left|P_j(s_{12})\right|^2\nonumber\\
&+&2 \Big[f_{S2}^{(0)}f_{S3}^{(0)}+4f_{P2}^{(0)}f_{P3}^{(0)}\Big]
\big(-a_{2t}a_{3t}t_{12}+b_{2t}b_{3t}s_{12}\big)\Re[P_{23}(s_{12})]\Bigg\}.
\end{eqnarray}
The interferences of the signal amplitudes with the nonresonant QCD amplitudes
produce rather lengthy expressions and we do not reproduce them here.
As far as the NLO QCD and weak corrections to
 $\ttbar$ without contributions from $\phi_2$ and
$\phi_3$ are concerned, we use the results of our previous work \cite{Bernreuther:2001rq,Bernreuther:2004jv,Bernreuther:2010ny}.
As mentioned above, the (small) contribution from the SM-like Higgs boson
$\phi_1(125{\rm GeV})$ is part of the order $\alpha_s^2\alpha$ mixed QCD-weak corrections.

As already emphasized at the beginning of this section we apply the
large top-quark mass limit for the computation of the NLO QCD 
 corrections to heavy Higgs production. Because we work above the top-quark pair production 
threshold we use this approximation somewhat beyond its range of
validity. However, in \Ref{Kramer:1996iq} it was  shown  that the range
of application can be extended to $m_{\phi}>2m_t$ within rather small
uncertainties ($\approx 10\%$), provided the higher order corrections
are rescaled by an appropriate K factor.  This procedure has been examined 
 for the production of a heavy Higgs boson by gluon fusion in the SM,  $pp \to \phi + X$.
 In \Ref{Kramer:1996iq} it was also
shown that the main contribution to the QCD NLO corrections is due to
soft and collinear gluon radiation which does not resolve the
gluon-Higgs coupling. Because the QCD corrections to 
\begin{equation}
pp\rightarrow\phi_2,\phi_3\rightarrow t\bar{t},
\label{ttbarprodwithHiggs}
\end{equation}
especially the factorizing ones, are very similar to those of  Higgs
production by gluon fusion in the SM, it is well motivated to assume the same behavior with
respect to the effective gluon-Higgs couplings that we use. The K-factor
method applied to our  calculation should  therefore lead to reliable
results.  Compared to the case of inclusive SM Higgs production there is,
however, a major difference: the interference with the QCD background.
In the following we briefly describe the K-factor method adapted to
the situation studied in this article.  The total $\ttbar$ cross section is
 written as
\begin{equation}
\sigma_{\text{NLO}}=\sigma_{\text{NLO}}^{\text{QCDW}}
+\sigma_{\text{NLO}}^{\text{approx.}},
\end{equation}
where $\sigma_{\text{NLO}}^{\text{QCDW}}$ denotes the 
 nonresonant $\ttbar$ cross section including 
 the NLO QCD and weak corrections  and $\sigma_{\text{NLO}}^{\text{approx.}}$
 is the  contribution of the heavy Higgs bosons 
 including the interference with the QCD background at NLO.
 We calculate it as follows\footnote{The formulas are presented here only for the inclusive
  cross section. However, they can also be applied to each individual
  bin of a distribution.}:
\begin{eqnarray}
\sigma_{\text{NLO}}^{\text{approx.}}=\sum_{j=2,3}\left(
\sigma_{\text{full},j}^{(0)}+\sigma_{\text{full},j,\text{QCD}}^{(0)}
+K_j\sigma_{\text{eff},j}^{(1)}
+\sigma_{\text{eff},j,\text{QCD}}^{(1)}\right)
\label{kfactormaster}
\end{eqnarray}
with
\begin{equation}
K_j=\frac{\sigma_{\text{full},j}^{(0)}}{\sigma_{\text{eff},j}^{(0)}}.
\label{kfactor}
\end{equation}
Here $\sigma_{\text{full},j}^{(0)}$ denote the leading-order cross
sections for
\begin{equation}
pp\rightarrow\phi_j\rightarrow t\bar{t} \, ,
\label{phijexchange}
\end{equation}
 and $\sigma_{\text{full},j,\text{QCD}}^{(0)}$ is the interference term of
  the $\phi_j$ production amplitude with the background at LO.
   The full top-quark mass dependence is kept
    in these two contributions.
  The terms $\sigma_{\text{eff},j}^{(1)}$ ($\sigma_{\text{eff},j}^{(0)}$)
and $\sigma_{\text{eff},j,\text{QCD}}^{(1)}$
represent the NLO (LO) contributions to (\ref{phijexchange}) and the interference
with the QCD background at NLO, respectively, in the effective theory. 
The K factor is not applied to  the NLO interference term  in \eqref{kfactormaster}. 
 Using a K factor  for this term
        analogous to \Eq{kfactor},
\begin{equation}
\tilde{K}_j=\frac{\sigma_{\text{full},j,\text{QCD}}^{(0)}}
{\sigma_{\text{eff},j,\text{QCD}}^{(0)}} \, ,
\label{kfactorinterference}
\end{equation}
would lead to a singular behavior whenever
$\sigma_{\text{eff},j,\text{QCD}}^{(0)}$ vanishes. This can happen for
differential cross sections, e.g. for the $M_{t\bar{t}}$ distribution.

\section{Results}
\label{sec:results}
For the numerical evaluation of the (differential) cross section discussed
 in the previous section we use, apart from the 2HDM parameter scenarios introduced in Sec.~\ref{sec:3-2hdmsc},
 the following input values.
The masses of the top quark (in the on-shell scheme) and of the SM-like Higgs boson $\phi_1$ are set to
\begin{equation}
  \mt = 173.34\, \GeV, \quad m_1 = 125\, \GeV. 
\end{equation}
We use the following values of  the electromagnetic fine structure constant and  the gauge boson masses:
\begin{equation}
  \alpha ={1\over 129}, \qquad m_W = 80.385\, \GeV,\quad m_Z = 91.1876\, \GeV \, .
\end{equation}
We employ the  PDF set
 CT10nlo \cite{Lai:2010vv} which provides also the value of the strong coupling at a scale $\mu$.
 To estimate the impact of uncalculated higher
orders we set the renormalization scale \mur equal to the
factorization scale \muf ($\muf=\mur\equiv\mu$) 
and vary $\mu$ by a factor 2 up and down. As the central scale we use 
\begin{displaymath}
  \mu_0 = {m_2+m_3\over 4}.
\end{displaymath}
This choice is motivated by the choice $\mu_0=m_H/2$ in the SM case; see, for instance, \cite{Dittmaier:2011ti}.

\begin{figure}
\centering
\includegraphics{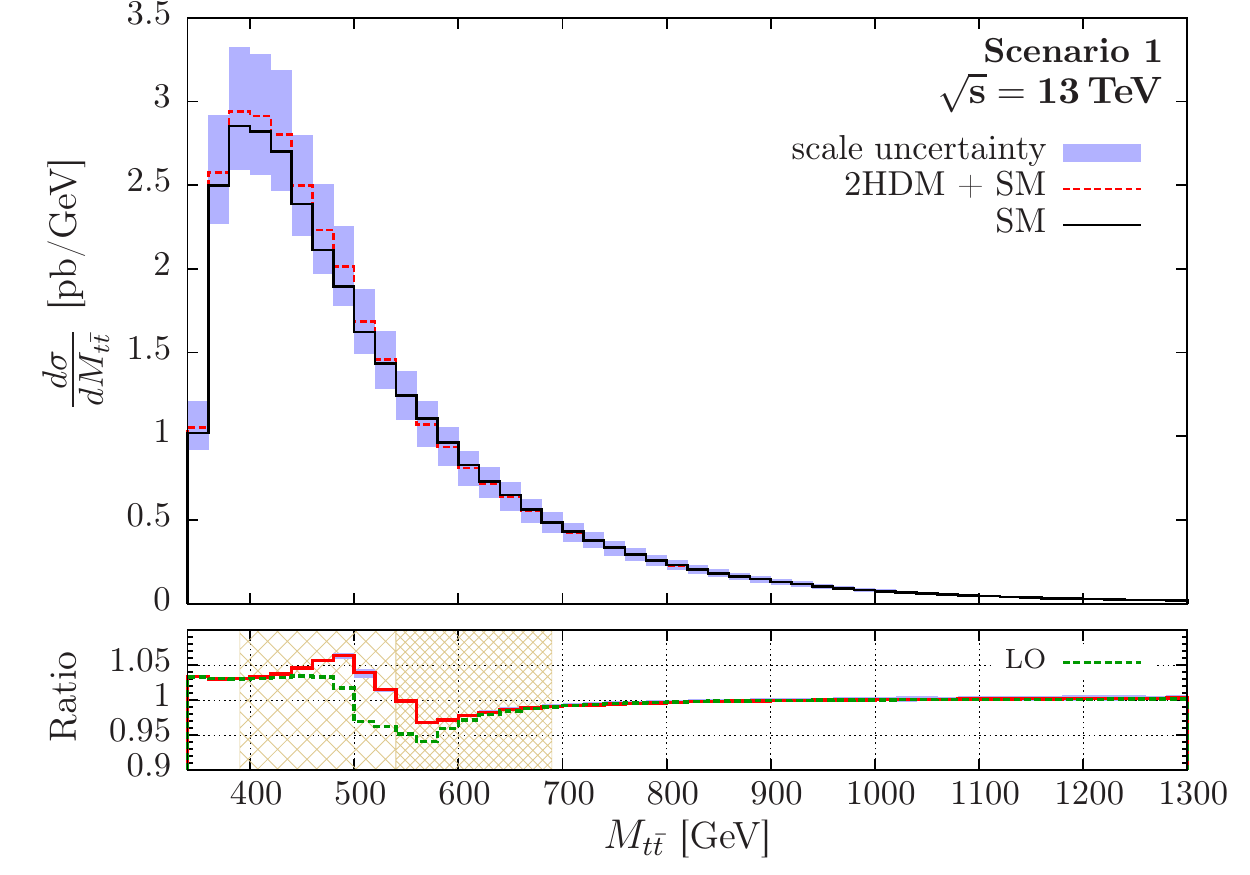}
\caption{Invariant mass distribution $M_{t\bar{t}}$ of $\ttbar$ for scenario 1 at NLO.
The upper pane shows the SM contribution (solid black) which includes NLO QCD
and weak corrections. The dashed red  line shows the sum of SM and
2HDM contributions at NLO QCD. The shaded blue  area represents the scale uncertainty
when varying $\mu_R=\mu_F=\mu_0$ by a factor of 2 or 1/2, respectively,
in the combined SM + 2HDM result. The lower pane shows the sum of SM and 2HDM
contributions normalized to the SM contribution at NLO (solid red) and
LO (dashed green). The scale uncertainty (shaded blue) of the ratio is very
small and is invisible for most of the bins in this plot.
The hatched regions in the lower pane indicate the range of the \mtt cuts.
}
\label{mttsc1}
\end{figure}

\begin{figure}
\centering
\includegraphics{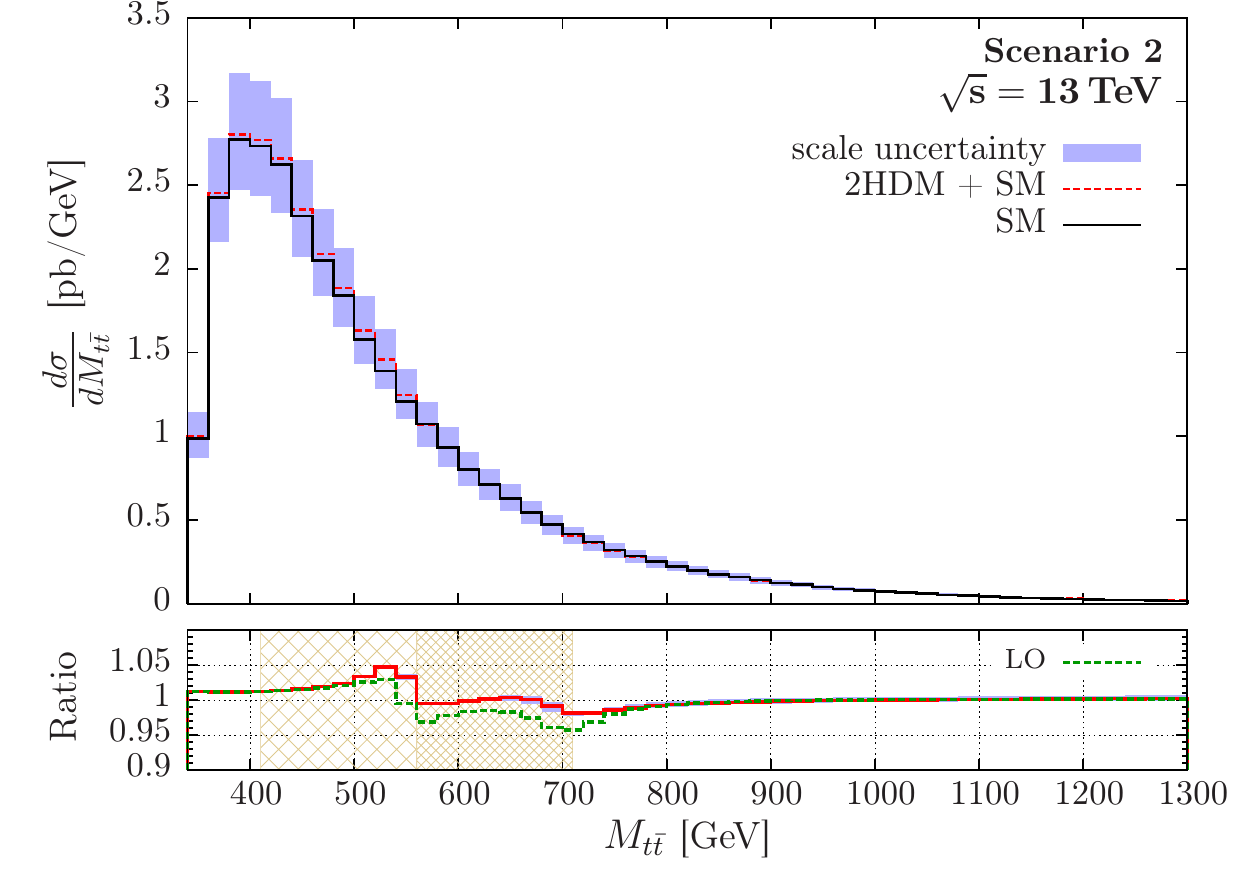}
\caption{Same as \Fig{mttsc1}, but for scenario 2.}
\label{mttsc2}
\end{figure}

\begin{figure}
\centering
\includegraphics{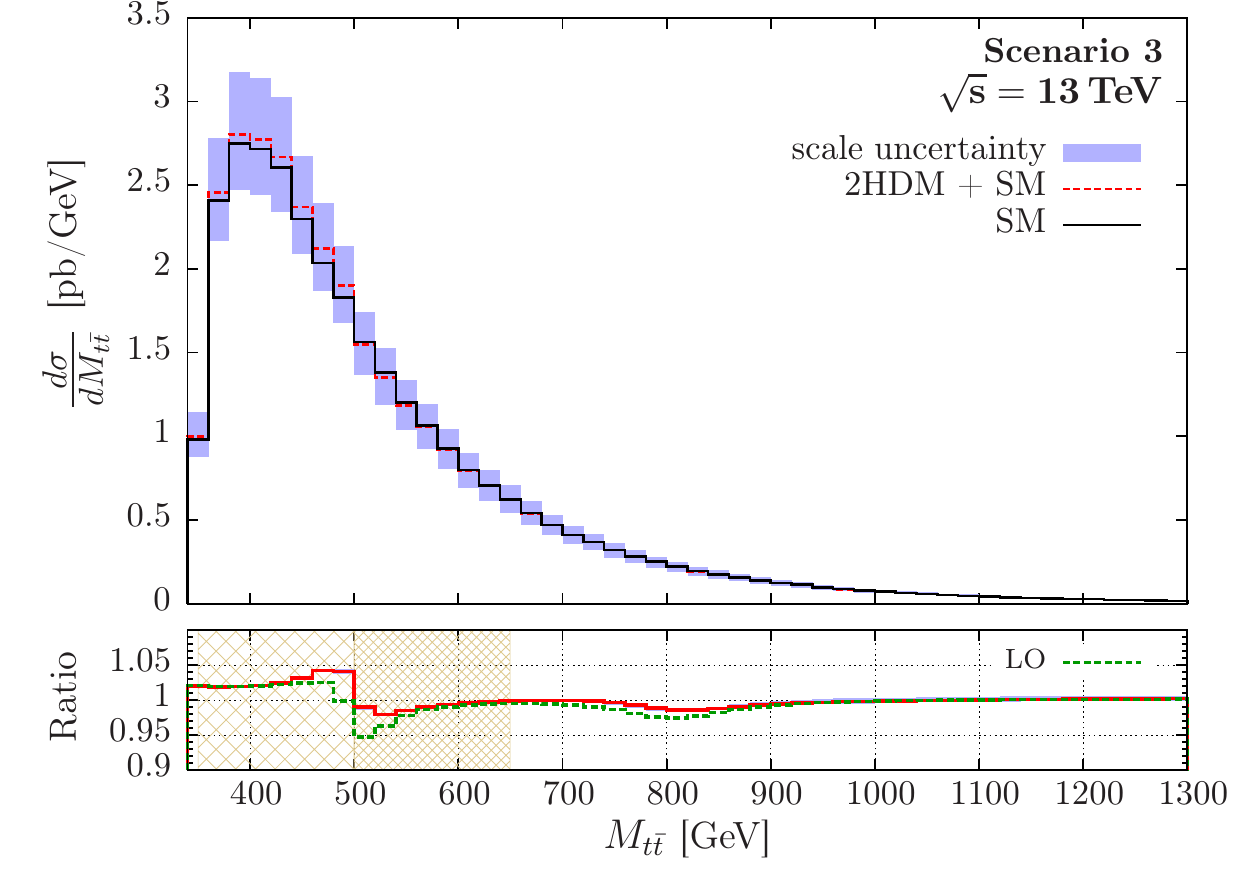}
\caption{Same as \Fig{mttsc1}, but for scenario 3.}
\label{mttsc3}
\end{figure}

In the following we present results for proton-proton collisions at a
hadronic center-of-mass energy of $\sqrt{s}=13$\,TeV. The $\ttbar$ cross section differential in the
top-quark pair invariant mass is displayed  in Figs.~\ref{mttsc1}--\ref{mttsc3}
for scenarios 1--3.  The ratios in the lower panes of these figures show that the corrections
with respect to the leading-order Higgs contribution are sizeable but the
overall effect of resonant $\phi_2, \phi_3$ production is rather small: In the \mtt distribution it
ranges from below 5\% in the \CP-violating case with significantly different  $\phi_2, \phi_3$ masses
(scenario~3) to about 6\% in the nearly mass-degenerate \CP-conserving
case (scenario~1). The effects on the inclusive $\ttbar$ cross section are even smaller, as can be seen in
\Tab{tab:totxs}. As a general feature we observe that in all three scenarios the QCD
corrections lead to a positive shift in the vicinity of the resonances
with respect to the leading-order results. The potential cancellation
due to the peak-dip structure, when the resonance region is integrated over,
is thus reduced at next-to-leading order.\\
Because the validity of the predictions is
restricted in the pole approximation  to the resonant region, we apply \mtt cuts around the
resonance which are indicated as hatched regions in the lower plots in
Figs.~\ref{mttsc1}--\ref{mttsc3}. In order to enhance the signal we divide the \mtt
cuts into two regions. One region is associated with the peak
\begin{equation}
m^*-150\text{ GeV}\quad\le\quad\mtt\quad\le\quad m^*
\label{eq:peakwindow}
\end{equation}
and the other one with the dip
\begin{equation}
m^*\quad\le\quad\mtt\quad\le\quad m^*+150\text{ GeV},
\label{eq:dipwindow}
\end{equation}
where $m^*$ is defined as the smallest value of \mtt where the NLO
ratio (red curve in \Figs{mttsc1}--\ref{mttsc3}) crosses 1.  These
cuts enhance the signal-to-background ratio because they prevent 
the partial cancellation due to the peak-dip structure. As already
mentioned in the introduction  experimental
analyses may use a sliding \mtt window.  
{\renewcommand{\arraystretch}{1.25}
\renewcommand{\tabcolsep}{0.2cm}
\begin{table}[h!]
\caption{Inclusive $\ttbar$ production cross sections for $pp$ collisions
at the hadronic center-of-mass energy of
$\sqrt{s}=13$ TeV in different type-II 2HDM
scenarios. $\sigma_{\text{QCDW}}$ denotes the cross section 
 for $pp\rightarrow\ttbar$ including NLO QCD and weak
corrections, but without contributions from $\phi_2,\phi_3.$  Note that due to our choice of the renormalization and factorization
scale $\mu=\mu_0=(m_2+m_3)/4$ the  cross section $\sigma_{\text{QCDW}}$ depends
on the scenario. $\sigma_{\text{2HDM}}$ denotes the cross section 
 for $pp\rightarrow\phi_{2,3}\rightarrow\ttbar$ including the interference with
the QCD background at NLO QCD.
The superscripts (subscripts) correspond to $\mu=\mu_0/2$ ($\mu=2\mu_0$). The scale
variation changes the ratio by less than $\pm0.001$ and is not displayed in the table.}
\centering
\begin{tabular}{l|lll}
& Scenario 1 & Scenario 2 & Scenario 3\\
\hline
$\sigma_{\text{QCDW}}$ [pb] & $643.22^{+81.23}_{-77.71}$& $624.25^{+80.98}_{-76.19}$& $619.56^{+81.05}_{-75.72}$\\
$\sigma_{\text{2HDM}}$ [pb] & $13.59^{+1.85}_{-1.64}$& $7.4^{+0.77}_{-0.78}$& $7.21^{+0.81}_{-0.77}$\\
\hline
$\sigma_{\text{2HDM}}/\sigma_{\text{QCDW}}$ & 0.021& 0.012& 0.012\end{tabular}
\label{tab:totxs}
\end{table}}

\begin{figure}
\centering
\includegraphics{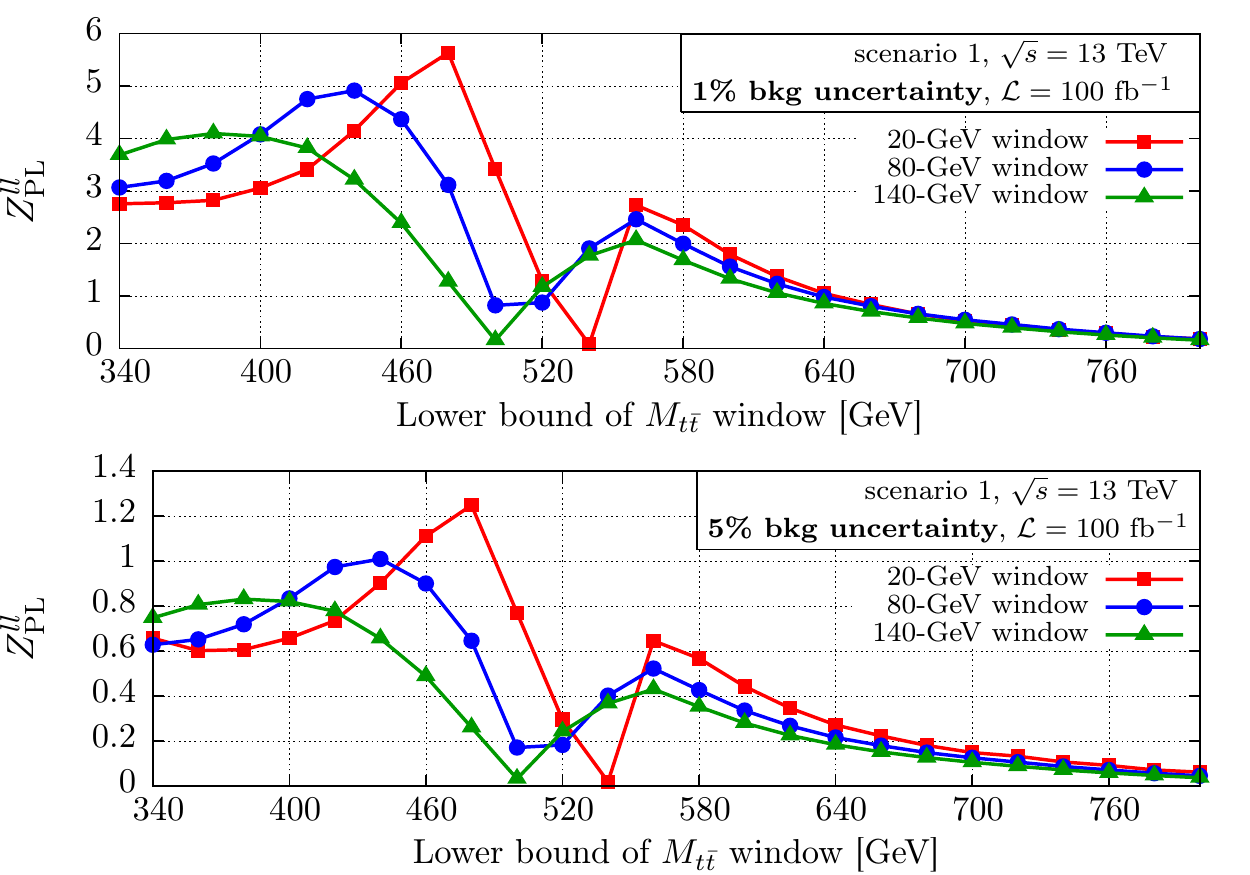}
\caption{Significance \ZPLll for different cuts on \mtt. The solid red (dashed blue, dotted green)
curve shows the behavior for \mtt windows of 20 GeV (80 GeV, 140 GeV) width. On the abscissa
the lower boundaries of the \mtt windows are given. The upper (lower) pane shows the
results for \ZPLll where a 1\% (5\%) uncertainty on the background is assumed.}
\label{ZPLscan}
\end{figure}

We investigate the sensitivity of such an analysis (\Fig{ZPLscan}) by calculating
the number of heavy Higgs signal and background events for
the dileptonic $\ttbar$ final states $(l=e,\mu)$
\begin{equation}
N_{s,b}^{ll} = \mathcal{L}\,\sigma_{s,b}\,\text{Br}(\ttbar\rightarrow 2l)\,\epsilon_{ll}
\end{equation}
and for the lepton plus jets final states $(l=e,\mu)$
\begin{equation}
N_{s,b}^{lj} = \mathcal{L}\,\sigma_{s,b}\,\text{Br}(\ttbar\rightarrow l+\text{jets})\,\epsilon_{lj}
\end{equation}
assuming the  branching ratios
\begin{align}
\text{Br}(\ttbar\rightarrow 2l)  =  \frac{4}{81} \, , & \qquad 
\text{Br}(\ttbar\rightarrow l+\text{jets})  = \frac{24}{81}\label{Brlj},
\end{align}
an integrated  luminosity of $\mathcal{L}=100$ fb$^{-1}$, and selection efficiencies
\cite{Chatrchyan:2013faa,Khachatryan:2015oqa}
  $\epsilon_{ll}=0.22$ and $\epsilon_{lj}=0.12$. We assume that the observed number of events $N_{\text{obs}}$
 is given by $N_{\text{obs}}=N_s+N_b$,
 where $N_b$ is the number of background events without the heavy Higgs bosons and
  $N_{\text{obs}}$ is the total number of events including those from resonant
 $\phi_2, \phi_3$ production  including the Higgs-QCD interference. These events are
calculated for \mtt windows of different width (20, 80 and 140 GeV). We consider for the lower
boundary of the \mtt windows  the range 340--800 GeV in steps of 20 GeV. We determine 
from the number of signal
and background events the significance \ZPL \cite{Li:1983fv,Cousins:2008zz}
based on the profile likelihood method for each \mtt window. In the calculation of
\ZPL we include an uncertainty on the background of 1\% (upper pane in \Fig{ZPLscan}) and\footnote{
According to \Tab{tab:totxs} the theory uncertainty of the background calculated at NLO
is about 13\%. The uncertainties of the NLO background cross sections within the \mtt
windows used in \Fig{ZPLscan} are roughly the same.  These uncertainties would reduce the
significance \ZPL to a value below 1. However, we expect that the uncertainty of 13\%
can be reduced by using higher order QCD predictions of the
background, in particular the next-to-next-to-leading order QCD corrections to $\ttbar$ production \cite{Czakon:2013goa,Czakon:2015owf}
and by using a sideband analysis to determine the background in the
signal region.} 5\% (lower pane in \Fig{ZPLscan}), respectively.
 Only the significance of the dileptonic channel \ZPLll is shown in \Fig{ZPLscan}. The results for the significance
of the lepton plus jets channel \ZPLlj are very similar. Figure \ref{ZPLscan} shows 
 that the significance increases if the \mtt window becomes narrower. In wider \mtt windows
peak values for the signal-to-background ratio are averaged out whereas for smaller windows peaks in the
signal-to-background ratio show up as peaks in the significance. Because the experimental resolution
is limited we show in \Fig{ZPLscan} also results for realistic \mtt window widths of 80 GeV and
140 GeV. Comparing the upper and lower plot in \Fig{ZPLscan} where we have assumed a 1\% and 5\%
background uncertainty, respectively, we observe that increasing the assumed uncertainty by a factor of
5 leads to a reduction of the significance by a factor of 1/5. This is due to
\begin{equation}
\lim_{N_b\rightarrow\infty}\ZPL=\frac{\sqrt{2(\eta-\ln(1+\eta))}}{\varepsilon},
\label{ZPLasymp}
\end{equation}
where $\eta$ is the signal-to-background ratio and $\varepsilon$ is the relative
background uncertainty. This emphasizes the importance of reducing background uncertainties.

In \Tab{tab:significance} we show numerical values of the significances \ZPLll and \ZPLlj for
all three scenarios using \mtt windows of 80 GeV width. The values for $Z_{\text{PL,1\%}}^{ll}$
and $Z_{\text{PL,5\%}}^{ll}$ for scenario 1 can be directly compared to \Fig{ZPLscan}. 
 The lepton plus jets decay channel has considerably higher
event numbers than the dileptonic channel because of the higher branching ratio. However, the
significance in both channels is almost the same. This is because we fixed the
relative background uncertainty and the signal-to-background ratio is independent of the
channel. A further increase of the luminosity would thus only marginally increase the
significance, given a constant relative uncertainty of the background [see also \Eq{ZPLasymp}].
We have also calculated the significances based on the $p$-value of the Poisson distribution
from the number of observed events $N_{\text{obs}}$ and background events $N_b$ ignoring the
uncertainty. The resulting values for these significances become unreasonably high and cannot
be trusted.

{\renewcommand{\arraystretch}{1.25}
\renewcommand{\tabcolsep}{0.2cm}
\begin{table}
\caption{Number of events ($N$) and significance ($Z$) for the dileptonic
(superscript $ll$) and lepton plus jets (superscript $lj$) channels $(l=e,\mu)$  at $\sqrt{s}=13$ TeV
assuming an integrated luminosity of $\mathcal{L}=100$ fb$^{-1}$ at the LHC.}
\centering
\begin{tabular}{l|l|l|l|l|l|l}
&\multicolumn{2}{|c}{scenario 1} & \multicolumn{2}{|c}{scenario 2} & \multicolumn{2}{|c}{scenario 3}\\
\hline
$M_{t\bar{t}}$ cut (GeV) & 460--540 & 540--620 & 480--560 & 560--640 & 420--500 & 500--580 \\
\hline
$N_b^{ll}$ &153517 & 90050 & 130718 & 76566 & 190486 & 113261\\
$N_{\text{obs}}^{ll}$ &160536 & 88250 & 135112 & 76357 & 196907 & 111712\\
\hline
$Z_{\text{PL,1\%}}^{ll}$ &4.4 & 1.9 & 3.2 & 0.3 & 3.3 & 1.3\\
$Z_{\text{PL,5\%}}^{ll}$ &0.9 & 0.4 & 0.7 & 0.1 & 0.7 & 0.3\\
\hline
\hline
$N_b^{lj}$ &502422 & 294711 & 427805 & 250582 & 623410 & 370675\\
$N_{\text{obs}}^{lj}$ &525393 & 288819 & 442186 & 249897 & 644423 & 365603\\
\hline
$Z_{\text{PL,1\%}}^{lj}$ &4.5 & 2 & 3.3 & 0.3 & 3.3 & 1.4\\
$Z_{\text{PL,5\%}}^{lj}$ &0.9 & 0.4 & 0.7 & 0.1 & 0.7 & 0.3\\
\end{tabular}
\label{tab:significance}
\end{table}}

\begin{table}
\caption{\mtt windows used in the computation of the  $y_t$ and $\cos\theta_{CS}$ distributions.}
\centering
\begin{tabular}{l|l|l}
& Lower \mtt window & Upper \mtt window\\
\hline
Scenario 1 & $390\text{ GeV}\le\mtt\le540\text{ GeV}$ & $540\text{ GeV}\le\mtt\le690\text{ GeV}$\\
Scenario 2 & $410\text{ GeV}\le\mtt\le560\text{ GeV}$ & $560\text{ GeV}\le\mtt\le710\text{ GeV}$\\
Scenario 3 & $350\text{ GeV}\le\mtt\le500\text{ GeV}$ & $500\text{ GeV}\le\mtt\le650\text{ GeV}$
\end{tabular}
\label{tab:cuts}
\end{table}

\begin{figure}
\centering
\includegraphics[]{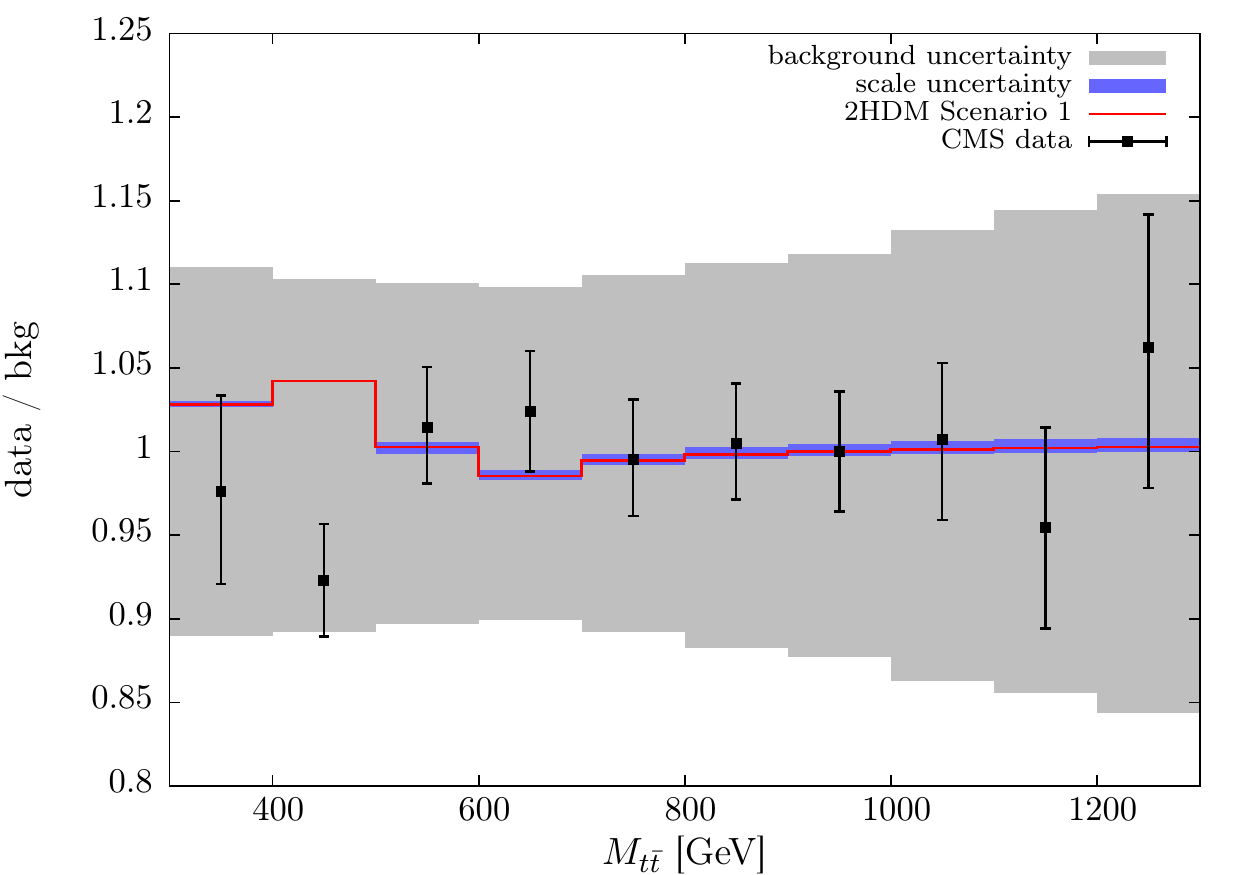}
\caption{\mtt distribution for $pp\to \ttbar X$  at the LHC (8 TeV).  The black data points were
extracted from  Fig.~4(b) of \Ref{Chatrchyan:2013lca} using the tool {\tt EasyNData}
\cite{Uwer:2007rs}. For error bars  not visible in the original plot
the radius of the black circle representing the respective data point is used as the error bar
length. The  shaded grey area was also extracted from  Fig.~4(b) of \Ref{Chatrchyan:2013lca}.
It displays the theoretical uncertainty of the background Monte Carlo prediction.
The solid red curve in this plot shows our result for scenario 1 including NLO QCD corrections.
The shaded blue area represents the scale uncertainty due to the variation of
$\mu_R=\mu_F=\mu_0$ by a factor of 2 or 1/2, respectively. }
\label{mttbar_CMS}
\end{figure}

In \Fig{mttbar_CMS} we show for the \mtt distribution a comparison of
8 TeV data from the CMS experiment \cite{Chatrchyan:2013lca} with
theoretical predictions obtained within scenario 1. One can see that
the current experimental and theoretical accuracy is insufficient to
establish or exclude scenario~1. Furthermore, the limits on the cross
section for a spin-zero resonance stated in \Ref{Chatrchyan:2013lca} were 
 derived without taking interference effects between signal and background into
account. There is a  more recent ATLAS analysis on resonances in the \mtt
distribution, \Ref{Aad:2015fna}, but it is also not 
sensitive enough to establish or exclude our scenarios.


\begin{figure}
\centering
\includegraphics{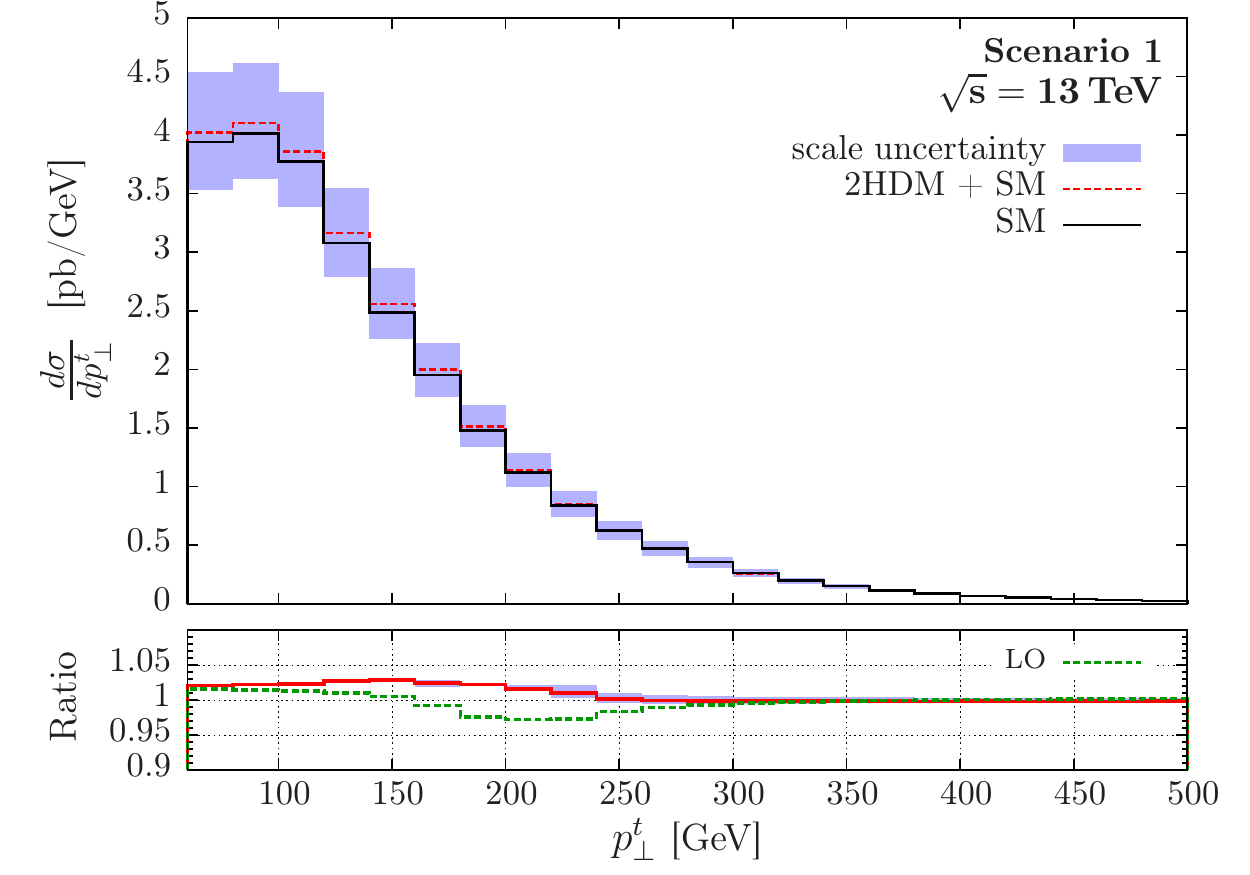}
\caption{Transverse momentum distribution of the top quark $p_{\perp}^t$ in scenario 1 of the 2HDM at NLO.
The upper pane shows the SM contribution (solid black) which includes NLO QCD
and weak corrections. The dashed red line displays the sum of the SM contribution and
2HDM contribution at NLO QCD. The shaded blue area represents the scale uncertainty
when varying $\mu_R=\mu_F=\mu_0$ by a factor of 2 or 1/2, respectively, in the combined
SM + 2HDM result. The plot in the lower pane shows the sum of SM and 2HDM contributions normalized
to the SM contribution at NLO (solid red) and LO (dashed green). The scale uncertainty (shaded blue)
of the ratio is very small and is invisible for most of the bins in this plot.}
\label{pTsc1}
\end{figure}

\begin{figure}
\centering
\includegraphics{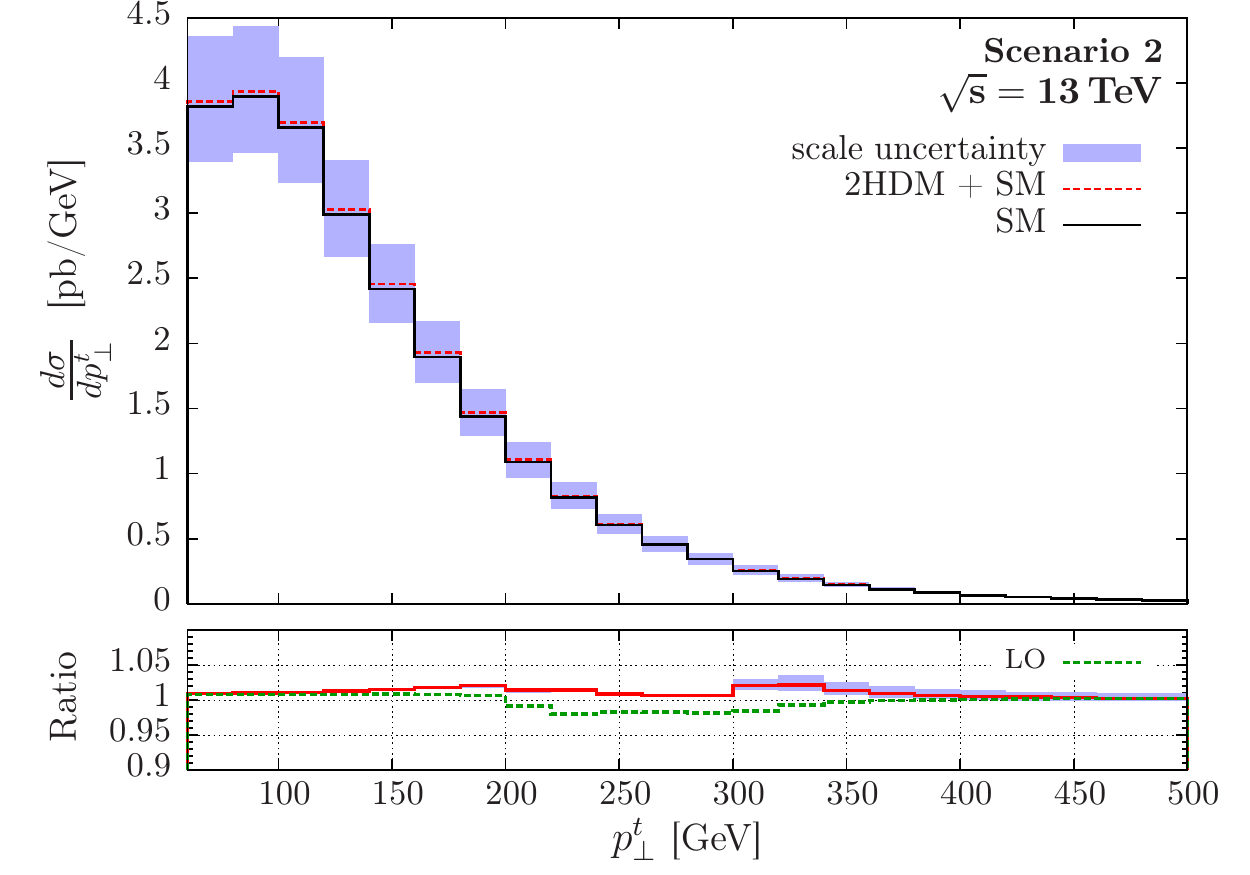}
\caption{Same as \Fig{pTsc1}, but for scenario 2.}
\label{pTsc2}
\end{figure}

\begin{figure}
\centering
\includegraphics{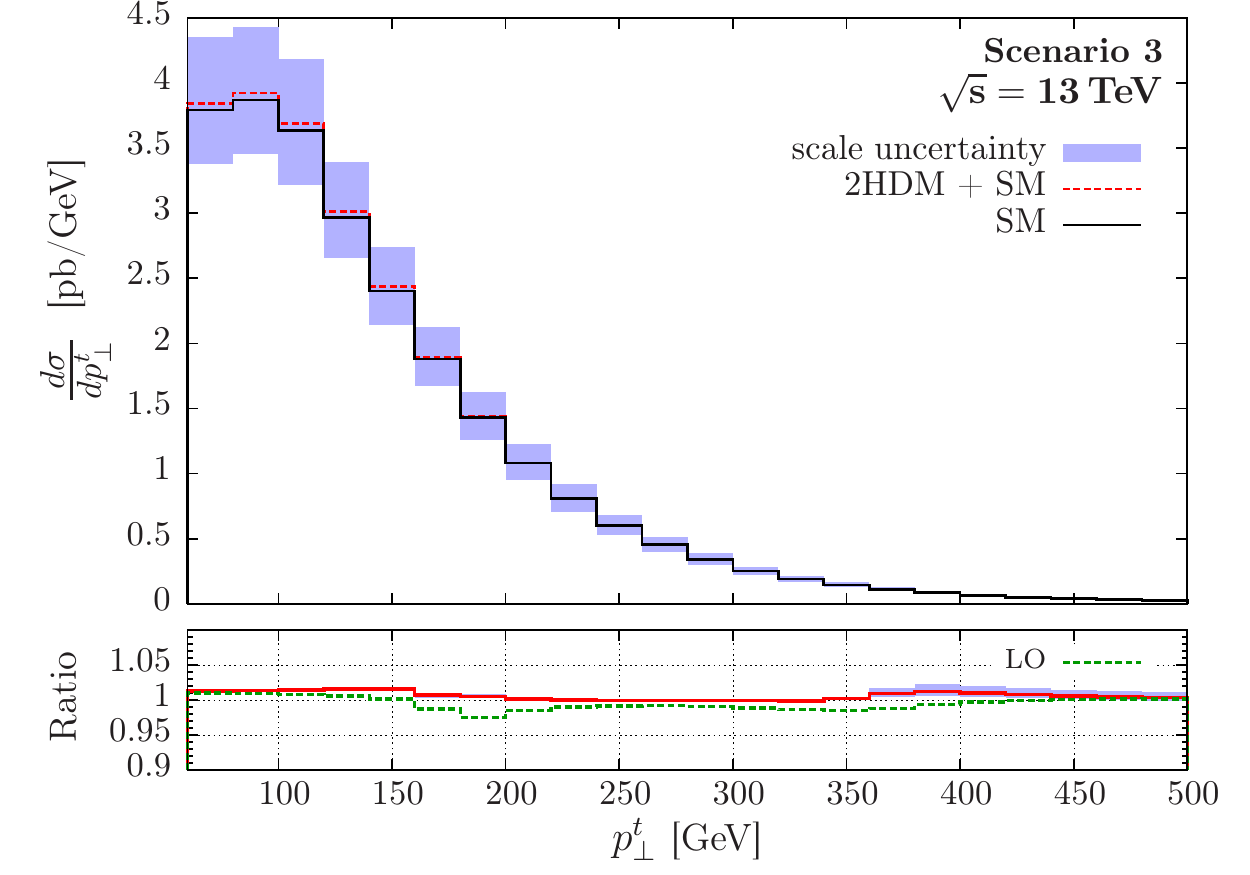}
\caption{Same as \Fig{pTsc1}, but for scenario 3.}
\label{pTsc3}
\end{figure}

\newpage
The distributions of the top-quark transverse momentum $p_{\perp}^t$ are displayed in Figs.~\ref{pTsc1}--\ref{pTsc3}
for scenarios 1--3, respectively. These distributions are computed without \mtt cuts. A cut on
the \mtt value in the $p_{\perp}^t$ distribution restricts the Born contribution
and the virtual corrections to values $p_{\perp}^t\le\sqrt{(\mtt^{\text{cut}})^2/4-m_t}$.
As a consequence the results are only leading order in the strong coupling constant
for $p_{\perp}^t>\sqrt{(\mtt^{\text{cut}})^2/4-m_t}$.
As in the case of the \mtt distribution the QCD corrections to
the 2HDM scenarios are mainly positive such that the ratios of the $p_{\perp}^t$ distributions in
all three scenarios stay above one. However, the deviations from one are small. As a remnant of the resonance
structure an enhancement around $p_\perp^t\approx m_i/2$ can be observed. 

\begin{figure}
\centering
\includegraphics{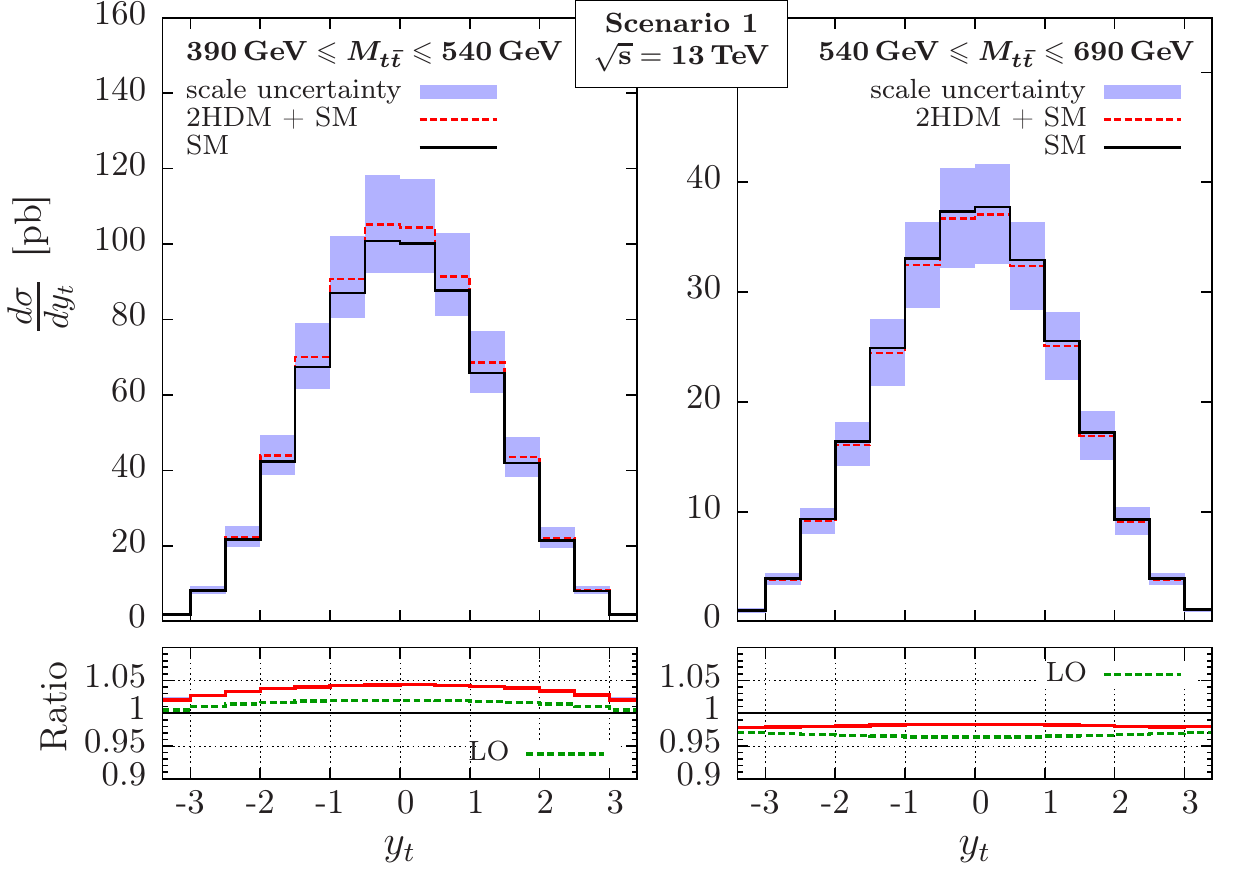}
\caption{Top-quark rapidity distribution $y_t$ for scenario 1 of the 2HDM at NLO.
The  plots in the upper panes show the SM contribution (solid black) which includes NLO QCD
and weak corrections. The dashed red line shows the sum of the SM contribution and
2HDM contribution at NLO QCD. The shaded blue area represents the scale uncertainty
when varying $\mu_R=\mu_F=\mu_0$ by a factor of 2 or 1/2, respectively, in the combined
SM + 2HDM result. The plots in the lower panes show the sum of SM and 2HDM contributions normalized
to the SM contribution at NLO (solid red) and LO (dashed green). The scale uncertainties (shaded blue)
of the ratios are very small and are invisible in these plots. The plots on the
left-  and right-hand side correspond to different  \mtt windows.}
\label{ytsc1}
\end{figure}

\begin{figure}
\centering
\includegraphics{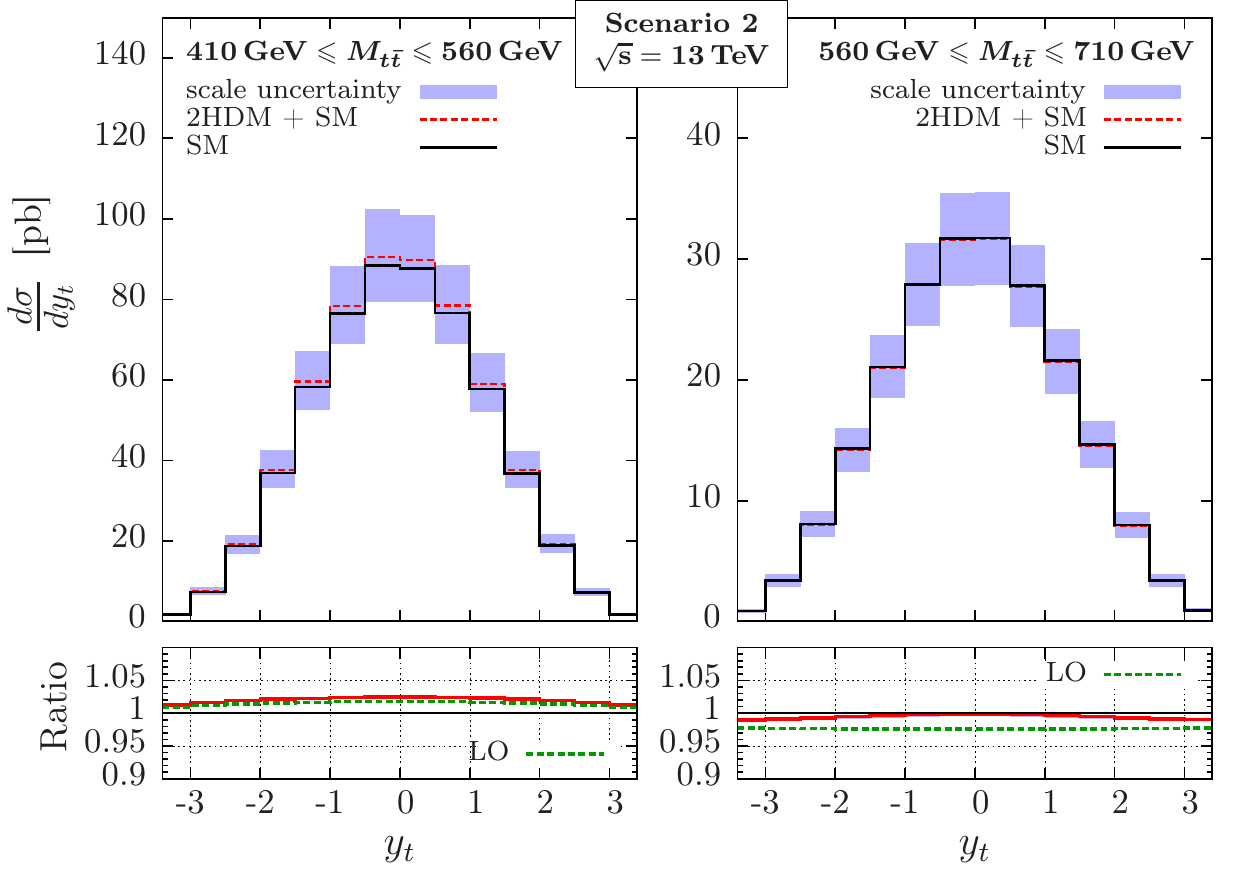}
\caption{Same as \Fig{ytsc1}, but for scenario 2 and different \mtt windows.}
\label{ytsc2}
\end{figure}

\begin{figure}
\centering
\includegraphics{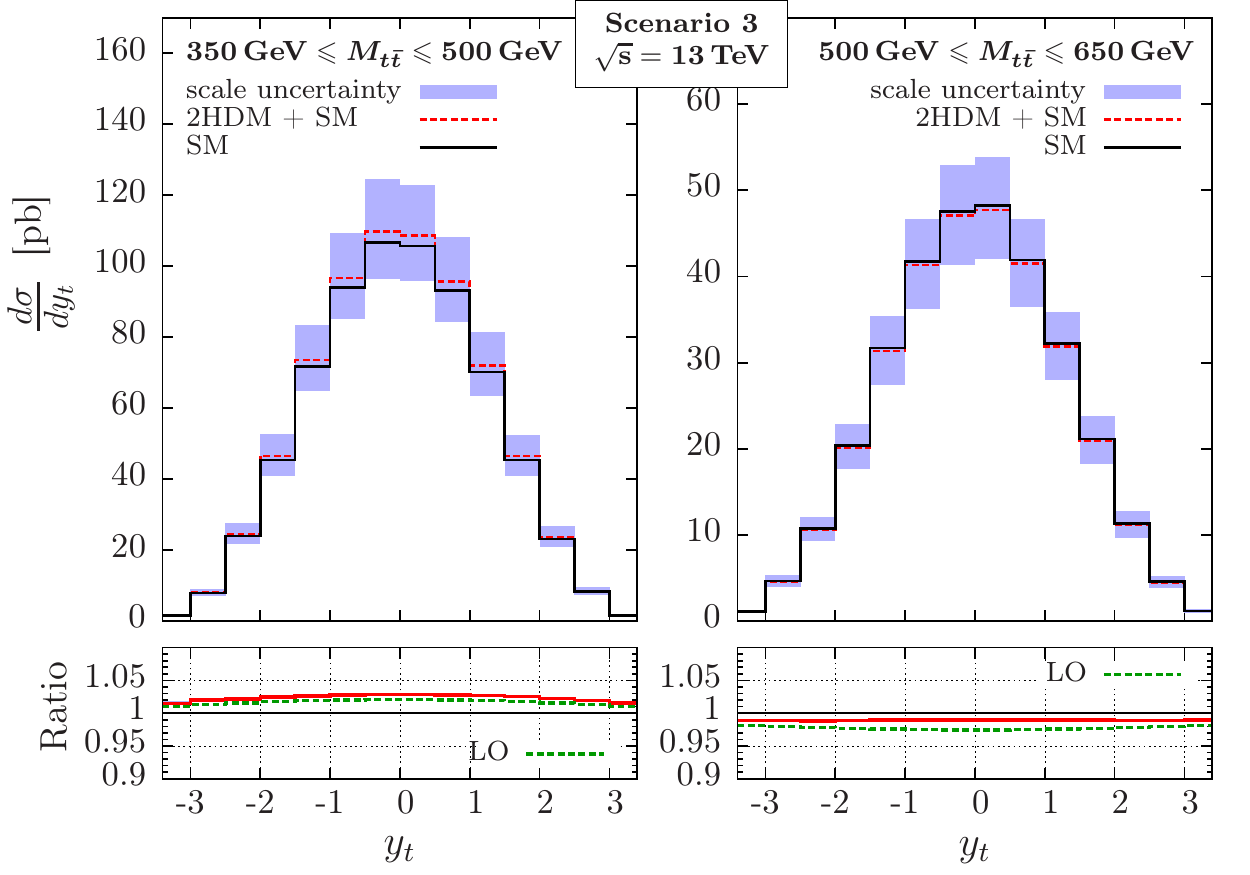}
\caption{Same as \Fig{ytsc1}, but for scenario 3 and different \mtt windows.}
\label{ytsc3}
\end{figure}

In the following we present, for the LHC (13 TeV), a number of 
 distributions for $\ttbar$ events obeying the  \mtt cuts given in \Tab{tab:cuts}.
The top-quark rapidity distributions in the laboratory frame, $y_t$,  are displayed in Figs.~\ref{ytsc1}--\ref{ytsc3}
for scenarios 1--3, respectively. As expected the $y_t$ distributions---and the rapidity distributions of the
 $\bar t$ quarks, which are not shown---are (within small statistical fluctuations) symmetric around 
 $y_t=0$. The impact of the heavy Higgs bosons is rather small, it is maximally $\sim 5 \%$ in the central region 
  for scenario 1 in the low  \mtt window. Top-quark pair production by $q\bar q$ annihilation and 
  (anti)quark gluon fusion at NLO QCD 
   induces a small asymmetry in the distribution of the moduli difference $\Delta|y|=|y_t|-|y_{\bar t}|.$
  The inclusive $\ttbar$ charge asymmetry $A_C=[N(\Delta|y|>0)-N(\Delta|y|<0)]/N_{\rm tot}$ at 13 TeV (integrated over $\mtt$)
   is $A_C=0.75(5)\times 10^{-2}$ at NLO QCD including electroweak corrections. 
   The heavy Higgs boson contributions, which are symmetric with respect to the interchange of $t$ and $\bar t$, do not change this result.

\begin{figure}
\centering
\includegraphics{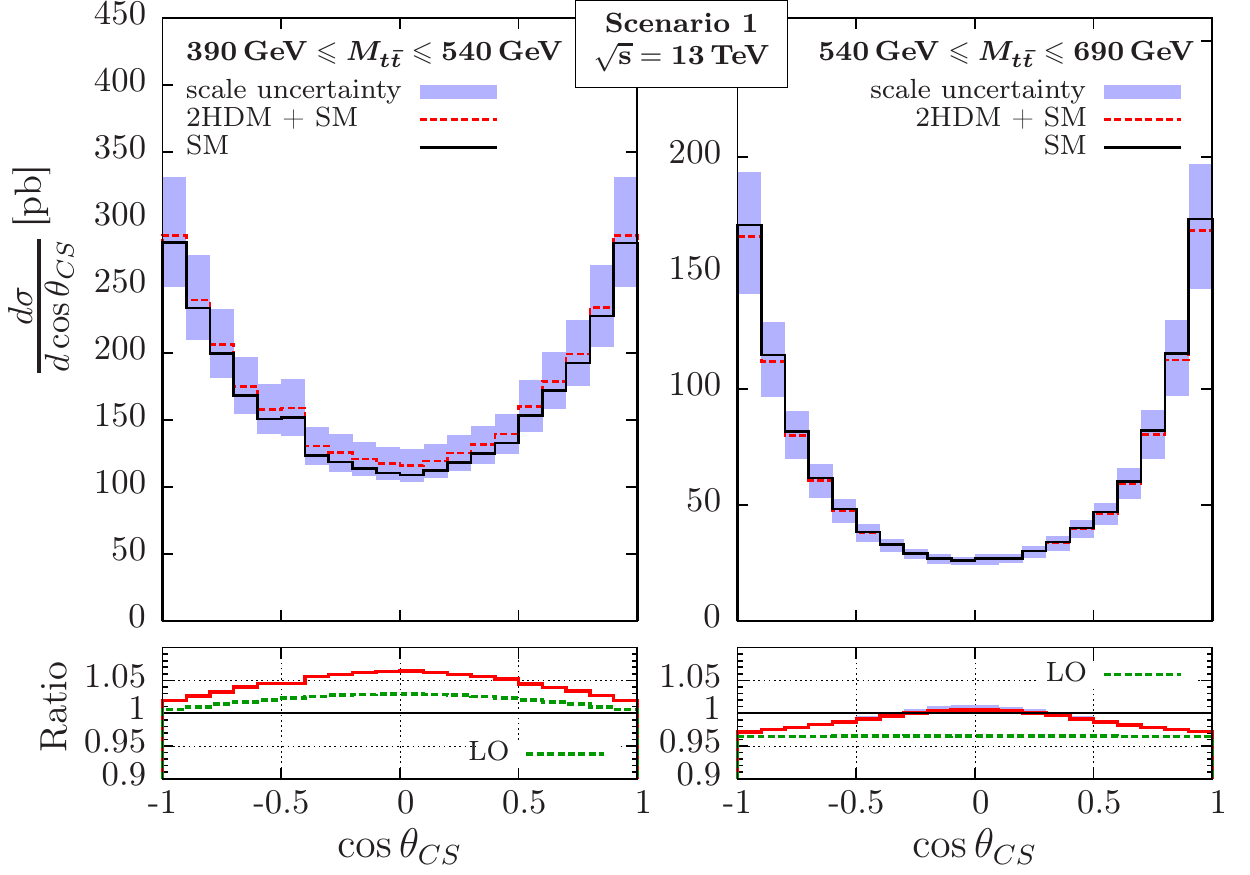}
\caption{Distribution of the cosine of the Collins-Soper angle, $\cos\theta_{CS}$,
for the top quark in scenario 1 of the 2HDM at NLO.
The plots in the upper panes show the SM contribution (solid black) which includes NLO QCD
and weak corrections. The dashed red line displays the sum of the SM contribution and
2HDM contribution at NLO QCD. The shaded blue area represents the scale uncertainty
when varying $\mu_R=\mu_F=\mu_0$ by a factor of 2 or 1/2, respectively, in the combined
SM + 2HDM result. The plots in the lower panes show the sum of SM and 2HDM contributions normalized
to the SM contribution at NLO (solid red) and LO (dashed green).The scale uncertainties (shaded blue)
of the ratios are very small and are invisible for most of the bins in these plots. The plots on the
   left-  and right-hand side   correspond to different \mtt windows.}
\label{costhetasc1}
\end{figure}

\begin{figure}
\centering
\includegraphics{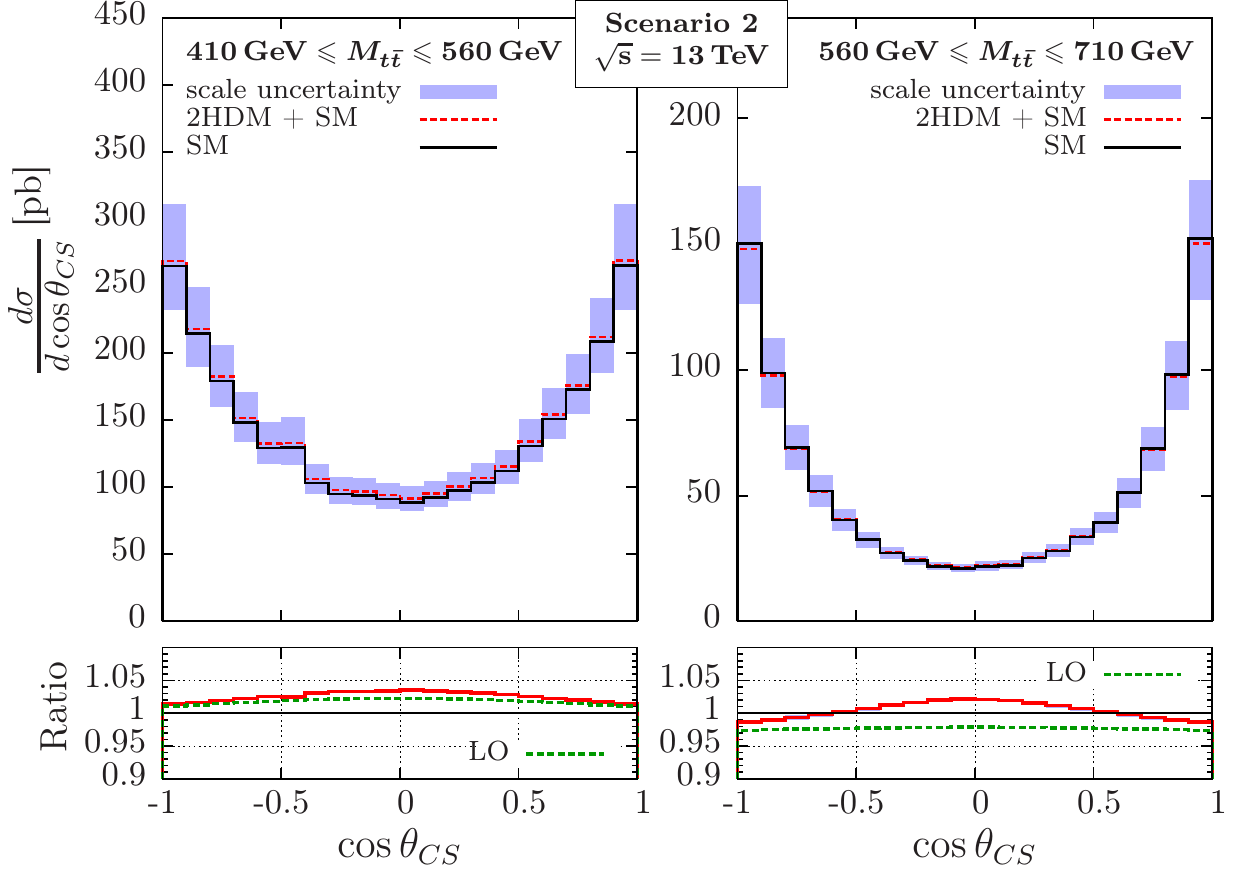}
\caption{Same as \Fig{costhetasc1}, but for scenario 2 and different \mtt windows.}
\label{costhetasc2}
\end{figure}

\begin{figure}
\centering
\includegraphics{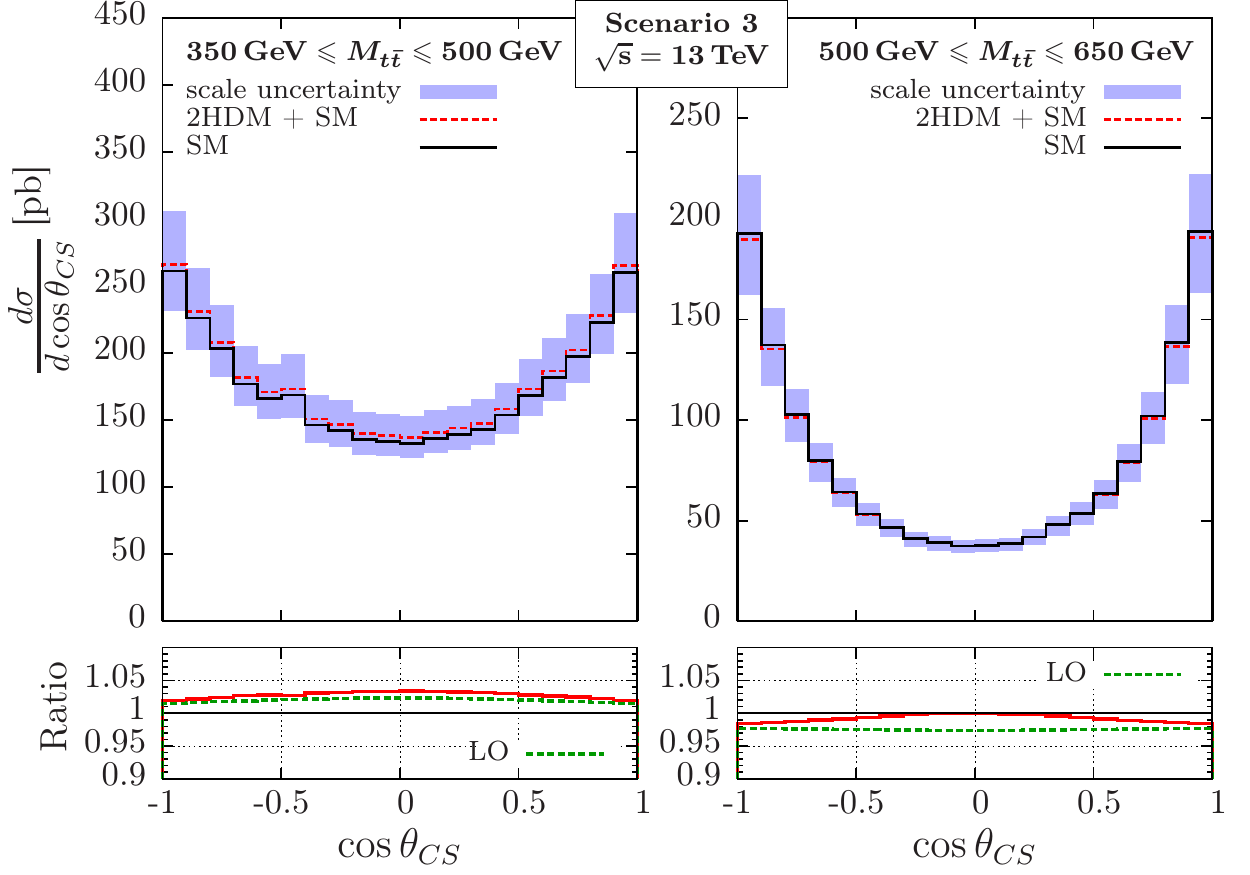}
\caption{Same as \Fig{costhetasc1}, but for scenario 3 and different \mtt windows.}
\label{costhetasc3}
\end{figure}

The distributions of the cosine of the Collins-Soper angle \cite{Collins:1977iv} $\cos\theta_{CS}$ 
 are shown in \Figs{costhetasc1}--\ref{costhetasc3} for scenarios 1--3, respectively. 
 We recall that $\theta_{CS}$ is similar to the angle between the top quark and the  direction of one of the beams 
  in  the $\ttbar$ zero-momentum frame (ZMF)---in fact, at leading order they are identical---, 
  but $\theta_{CS}$ is less affected by initial state radiation. Let us denote the momenta of the two proton beams in the $\ttbar$ ZMF
   by $p_1$ and $p_2$. If the transverse momentum of the $\ttbar$ pair is nonzero the three-momenta ${\bf p}_1$ and ${\bf p}_2$ are not collinear,
    and $\theta_{CS}$ is defined to be the angle between the top-quark direction in the $\ttbar$ ZMF and the bisecting
line between ${\bf p}_1$ and $-{\bf p}_2$.
For the production of a spin-zero resonance and its decay to $\ttbar$ the $\cos\theta_{CS}$ distribution is flat at LO.
 For the background at LO QCD, one gets for $q{\bar q}\to \ttbar$ a distribution proportional to ${\hat s}(1+\cos^2\theta_{CS})+4m_t^2(1-\cos^2\theta_{CS})$.
  For $gg\to \ttbar$, where all partial waves are present, the $\cos\theta_{CS}$ distribution is more complicated, but is also minimal in the central region and maximal in the 
   forward and backward region. The background distributions computed at NLO QCD including weak corrections, which are displayed 
   in Figs.~\ref{costhetasc1}--\ref{costhetasc3}, corroborate this behavior. Adding the contributions of the heavy Higgs bosons and the interference 
    with the background changes the shape of the $\cos\theta_{CS}$ distributions only moderately. 
  The effect is largest in the low $\mtt$ window of scenario 1, where the distribution changes by $\sim 6\%$ in the central region.



Finally, we have also calculated the above distributions for 14 TeV.  We do 
 not display them here because these results are
similar to those for 13 TeV and do not bear significant new insights.

\newpage
\section{Summary and conclusions}
\label{sec:summary}
This paper is based on the hypothesis that 
 in addition to the 125 GeV Higgs boson additional heavy neutral Higgs bosons with unsuppressed Yukawa couplings to top quarks  exist, which 
 can be resonantly produced at the LHC in the $\ttbar$ channel.
The analysis of $\ttbar$ data at the LHC, which so far has been rather rudimentary with regard to this possibility, 
requires precise theoretical predictions. We have therefore investigated  effects of heavy, neutral Higgs boson resonances in 
$t\bar t$ production at the LHC at next-to-leading order in the strong coupling. 
A  new ingredient of our analysis is the 
incorporation of the interferences of the heavy Higgs production amplitudes with the respective nonresonant 
QCD $t \bar t$ amplitudes at NLO QCD. The NLO corrections to heavy Higgs boson production and the Higgs-QCD interference were computed 
in the large $m_t$ limit with an effective K-factor rescaling. The nonresonant $\ttbar$ background was incorporated at NLO QCD 
 including weak-interaction corrections.

Our NLO QCD computation of heavy Higgs production and decay into $\ttbar$ and the interference with the nonresonant $\ttbar$ 
amplitude is in principle rather model independent. However, essential ingredients in our analysis are the total 
decay widths of the heavy Higgs bosons, which must be computed in a concrete model in order to maintain the unitarity of the $S$-matrix. 
 To be specific, we used the type-II two-Higgs-doublet model 
 where three neutral Higgs bosons appear in the physical particle spectrum. We investigated three phenomenologically viable
 parameter scenarios in more detail, 
where one of these particles is identified with the 125 GeV Higgs boson.
 We considered parameter sets of this model where this boson has SM-like couplings to the quarks, leptons, and weak gauge bosons. 
The masses of the other two neutral Higgs bosons were assumed to lie above the $t\bar t$ threshold, with top-quark Yukawa couplings that are 
slightly larger than the SM top-quark Yukawa coupling. 
 
For these three parameter sets we investigated heavy Higgs boson production and decay to $\ttbar$ at the LHC operating at 13 TeV.
  We computed  the inclusive $\ttbar$ cross section  and 
   the $\ttbar$ invariant mass distribution with and without the two heavy Higgs resonances. Assuming an integrated luminosity of 100 fb$^{-1}$,
    we estimated,  for appropriately chosen $\mtt$ intervals and our three parameter scenarios,
    the significances for detecting a heavy Higgs signal  in the $\ttbar$ dileptonic and 
     lepton plus jets decay channels. 
     In addition we computed the top-quark transverse momentum distribution and, for selected $\mtt$ windows, the top-quark rapidity  distribution and that
    of the cosine of the Collins-Soper angle. We showed also 
  that the so-far negative results by the CMS and ATLAS experiment  on searches for heavy spin-zero resonances 
   in the $\ttbar$ channel  at the LHC (8 TeV) do not exclude our parameter scenarios.
    
  Because the masses of the heavy Higgs resonances are not known, we propose to scan the $\ttbar$ invariant mass distribution with a sliding $\mtt$ window
   of width $\Delta\mtt \sim 80$ GeV. Mass bins of this size are experimentally feasible, at least for $\mtt\lesssim 1$ TeV.
   Our analysis of Sec.~\ref{sec:results} shows that, for our parameter scenarios, the significances  for detecting a heavy Higgs resonance 
   exceed $3\sigma$ only if the uncertainties involved in measuring $\mtt$ windows and modeling the nonresonant $\ttbar$ background can be pushed down
    to a level of $\sim 1\%$. Such a precision is unrealistic at present, but since both theoretical and experimental uncertainties in measuring the 
     (binned) $\ttbar$    cross section have been significantly reduced in recent years,  one may be optimistic and expect further improvements also in the future. 
      Full multivariate experimental analyses, which increase the sensitivity, would also use information from other distributions, including those considered
        in Sec.~\ref{sec:results}. A further set of observables, which we plan to investigate in future work, includes $\ttbar$ spin correlation and $t$, $\bar t$
         polarization observables. Such observables show, in selected $\mtt$ windows, also some sensitivity to heavy Higgs resonances and potentially allow one
          to discriminate between a scalar, a pseudoscalar, and a \CP mixture. 

\subsubsection*{Acknowledgments}

 W.B. thanks A. Brandenburg  for a
  discussion.  This work was supported in part by Deutsche Forschungsgemeinschaft (DFG) Grant No. SFB-TR09.
   P. G.  was supported by DFG through Graduiertenkolleg 
  Grant No. GRK 1504 and C. M. through DFG Gra\-duiertenkolleg Grant No. GRK 1675.  The work  of Z.G. S.   
  was supported by  National Natural Science Foundation of China and by Natural Science Foundation of
Shandong Province.

\appendix
\section{Description of Higgs resonances}
\label{app:higgs_resonances} 

The unstable particles which are involved in the reactions considered
in this paper are  the weak gauge bosons, and the three
neutral Higgs bosons of the 2HDM extension of the SM. The charged
Higgs boson does not play any role here---see the 2HDM scenarios
considered in Sec.~\ref{sec:3-2hdmsc}.  The top quark, too, is of course unstable, but
 in this paper we consider reactions at the level of $t$ and $\bar t$ final states.
 The following discussion applies also if top-quark decays are incorporated
  into the analysis.
 While we take higher order QCD
corrections to the parton reactions $ab \to t{\bar t}+X$ into account
 we emphasize that we
incorporate the weak gauge boson and Yukawa interactions only
to leading order.  This is important to keep in mind in the following
discussion.

As argued above, the top quark can be treated in the narrow width
approximation. The $Z$ boson appears in $t{\bar t}+X$ production as
$s$-channel resonance only far off shell, $Z^* \to t{\bar t}+ X$.
Hence its width can be safely neglected. The same statement applies to
the light Higgs boson $\phi_1$ with\footnote{The total width of $\phi_1$ is SM-like, i.e.
  very small in the 2HDM scenarios considered below, $\Gamma_1 \sim 4$
  MeV.} $m_1 = 125$ GeV. The $W^\pm$ boson is resonantly produced in $t$ and ${\bar
  t}$ decay, respectively. As we work to lowest order in the weak
gauge couplings this can be incorporated  by using the width
$\Gamma_W$ in the $W$-boson propagator.
     
In the resonant production of the heavy Higgs bosons, $ab \to
\phi_{2,3} \to t{\bar t} +X$, the finite width effects must be accounted for.
 The adequate method which respects gauge invariance also
with respect to higher order electroweak corrections is the so-called
complex mass scheme \cite{Denner:1999gp,Denner:2005fg,Nowakowski:1993iu}. Let us
illustrate this method for the case at hand, using only one $\phi$. We
consider the amplitude for the process $ab \to \phi \to t{\bar t}+ X$,
including only factorizable QCD corrections.  The
nonfactorizable QCD corrections are analyzed in
Sec.~\ref{subsec:NLOQCD}  and Appendix~\ref{app:SGA}. In the complex mass scheme the amplitude
is of the form
 \begin{equation}\label{eq:resH1}
  {\cal M}_{ij\to f} = S_{ij}\left(k^2,\cdots\right) \,i P(k^2) \,S_f\left(k^2,\cdots\right) + {\cal N} \, ,  
  \qquad P(k^2) = \frac{1}{k^2 - \mu^2_\phi} \, ,
 \end{equation}
 where the complex mass parameter $\mu^2_{\phi}$ is the pole of the
 full $\phi$ propagator, i.e., the point in the complex $s$ plane
 where the renormalized Higgs boson self energy $\Pi(s)$ vanishes:
\begin{equation}\label{eq:resH2}
  \Pi(\mu^2_\phi) = 0  \, .
\end{equation}
The symbol ${\cal N}$ in Eq.~\eqref{eq:resH1} denotes the nonresonant
contribution to the scattering matrix element.  The complex mass
parameter must be used wherever the $\phi$ mass squared occurs in the
scattering amplitude.  However, as we compute \eqref{eq:resH1} only to
lowest order in the weak and Yukawa interactions, $\mu^2_\phi$
does not appear in the loops contributing to $S_{ij}$ and $S_f$, but
only in $P(s)$. We parametrize
\begin{equation}\label{eq:resH3}
  \mu^2_\phi  =  m^2 - i m\Gamma \, .
\end{equation}
The meaning of $m$ and $\Gamma$ is as follows. The relation between
the renormalized and bare Higgs boson self energy is, to one-loop
order in the electroweak and Yukawa couplings,
\begin{equation}\label{eq:resH4}
  \Pi(k^2) = \Pi_0(k^2) - \delta\mu^2_\phi +(k^2-\mu^2_\phi)\delta{\cal Z}_\phi \, ,
\end{equation}
where ${\cal Z}_\phi =1 + \delta{\cal Z}_\phi$ is the complex Higgs
wave-function renormalization constant and $\delta\mu^2_\phi$ is the
complex mass counterterm. The relation between the {\it real} bare
Higgs mass squared, $m_0^2$, and the complex renormalized mass squared
$\mu^2_\phi$ is
\begin{equation}\label{eq:resH5}   
  m_0^2 = \mu^2_\phi + \delta\mu^2_\phi \, . 
\end{equation}
The condition \eqref{eq:resH2} is the complex-mass-scheme
generalization of the usual on-shell renormalization condition. Using
\eqref{eq:resH4} it implies that
\begin{equation} 
 \delta\mu^2_\phi = \Pi_0(\mu^2_\phi)  \, . 
\end{equation}
Now we use the parametrization \eqref{eq:resH3} and the relation
\eqref{eq:resH5}, i.e.,  $\delta\mu^2_\phi =m_0^2 - \mu^2_\phi$.
Taking the real part of this relation shows that $m$ can be
interpreted as the on-shell mass of $\phi$. Taking the imaginary part
gives the equation
\begin{equation}\label{eq:resH6}   
  m\Gamma = {\rm Im}{ \Pi_0(m^2 - i m\Gamma) } \, ,
\end{equation}
which in general can be solved iteratively for $\Gamma$.  Expanding
this relation around $m$, one gets to one-loop order in the
electroweak and Yukawa couplings \cite{Denner:2005fg}:
\begin{equation}\label{eq:resH7} 
  m\Gamma = {\rm Im}{\Pi_0(m^2)} - m \Gamma\, {\rm Re}{ \Pi'_0(m^2)} ,
\end{equation}
where the prime denotes the derivative with respect to $k^2$. Again,
this equation may be solved iteratively for $\Gamma$. Because we work
to lowest order in the weak and Yukawa couplings we can neglect
the second term in \eqref{eq:resH7}. Then $\Gamma$ is the total
$\phi$-boson width in the usual on-shell scheme.

In summary we use \eqref{eq:resH1} where $\mu^2_\phi$ is determined by
the on-shell mass and total width of the respective Higgs boson
$\phi_2, \phi_3$.

\section{Cancellation of real and virtual nonfactorizable corrections in the
soft gluon approximation}
\label{app:SGA}

As an example for the cancellation of real and virtual nonfactorizable corrections in
the SGA we take the amplitude depicted in
\Fig{fig:SoftGluonCancelation}. The dotted red  cut in
\Fig{fig:SoftGluonCancelation} corresponds to the interference of two
Feynman diagrams contributing to the real corrections and the 
 dashed green  cut indicates the interference of the one-loop diagram
shown in \Fig{fig:nonresonant} with the $s$-channel QCD Born diagram.
We start with the latter interference.  In the SGA the loop
integral for the amplitude shown in \Fig{fig:nonresonant} reduces to the
scalar integral
\begin{equation}
\int\frac{d^4\ell}{(2\pi)^4}\frac{\Nv}{[\ell^2+i\varepsilon]
[-2\ell\cdot p_1+i\varepsilon][-2\ell\cdot (p_1+p_2)+s-m_{\phi}^2+i\varepsilon]
[-2\ell\cdot k_1+i\varepsilon]} \, , 
\label{loopintegral}
\end{equation}
where $\ell$ is the loop momentum which is required to be soft within
the SGA. The numerator $\Nv$ in the SGA is obtained from the exact 
 numerator by setting $\ell=0$.  The momenta of the incoming gluons 
 are denoted $p_1$ and $p_2$, and  $k_1$ is the momentum
of the outgoing top quark.  In the last three square brackets of 
 the denominator of \eqref{loopintegral} the $\ell^2$ terms are neglected, which is also
  part of the SGA.
   The integration over the $\ell^0$
component is performed in the complex $\ell^0$ plane using the residue
theorem. Closing the contour in the lower half of the plane the
residue of the pole of the first propagator in \Eq{loopintegral}
is picked up. The location of the poles of the other propagators is in the upper half
of the    $\ell^0$ plane. Ignoring the numerator the following result is obtained:
\begin{eqnarray}
&&-i\int\frac{d^3\ell}{(2\pi)^3}\frac{1}{2|\vl|}
\frac{1}{\big[-2|{\vl}|p_1^0+2\vl\cdot\mathbf{p}_1+i\varepsilon\big]}\nonumber\\
&&\times\frac{1}{\big[-2|\vl|(p_1^0+p_2^0)+2\vl(\mathbf{p}_1+\mathbf{p}_2)
+s-m_{\phi}^2+i\varepsilon\big]\big[-2|\vl|k_1^0+2\vl\cdot\mathbf{k}_1
+i\varepsilon\big]}\nonumber\\
&&=i\int\frac{d^3\ell}{(2\pi)^3}\frac{1}{2\ell^0}
\frac{1}{\big[-2\ell\cdot p_1+i\varepsilon\big]\big[-2\ell\cdot(p_1+p_2)
+s-m_{\phi}^2+i\varepsilon\big]\big[2\ell\cdot k_1-i\varepsilon\big]}
\label{psi_loop}
\end{eqnarray}
with $|\vl|=\ell^0$.  The integration measure
${d^3\ell}/[{(2\pi)^3}{2\ell^0}]$ is identical to the 
Lorentz invariant phase-space measure of the gluon. 
What remains to be done is to
show that \Eq{psi_loop} including the numerator indeed  
corresponds to the phase-space integral
over the soft gluon in the corresponding real correction contribution. Thus we
 analyze the interference term indicated in \Fig{fig:SoftGluonCancelation}
by the dotted red  cut.  After applying the SGA this interference
leads to the following propagator structure,
\begin{equation}
\frac{1}{\big[-2q\cdot p_1+i\varepsilon\big]\big[-2q\cdot(p_1+p_2)+s-m_{\phi}^2
+i\varepsilon\big]\big[2q\cdot k_1-i\varepsilon\big]} \, ,
\end{equation}
where $q$ is the four-momentum of the  (soft) gluon.
Taking the phase-space integration over the soft gluon momentum $q$
into account,
\begin{equation}
\int\frac{d^3q}{(2\pi)^3}\frac{1}{2q^0}\frac{1}{\big[-2q\cdot p_1
+i\varepsilon\big]\big[-2q\cdot(p_1+p_2)+s-m_{\phi}^2+i\varepsilon\big]
\big[2q\cdot k_1-i\varepsilon\big]} \, ,
\label{psi_real}
\end{equation}
the two expressions  \Eq{psi_loop} and \Eq{psi_real} are identical up to 
a factor of $i$.\\
So far only the denominators were investigated and the numerators of 
the amplitudes were ignored.
It remains to show that they are identical up to a sign. The real radiation
amplitude leads to
\begin{eqnarray}
\Nr & = & \Tr[(\cancel{k}_1+m_t)V_{\phi tt}(\cancel{k}_2-m_t)
V_{gtt,\alpha_2}^{c_2}(\cancel{k}_1+m_t)V_{gtt}^{\rho',c_3}]\nonumber\\
& \times & V_{ggg}^{\mu'\nu'\alpha_2,abc_2}(p_1,p_2,-p_1-p_2)
V_{ggg}^{\mu\alpha_1\rho,ac_1c_3}(p_1,-p_1,0)
V_{gg\phi,\alpha_1}^{\nu,bc_1}(p_2,p_1)\nonumber\\
& \times & \frac{\mathcal{P}_{\mu\mu'}(p_1,p_2)\mathcal{P}_{\nu\nu'}(p_2,p_1)
\mathcal{P}_{\rho\rho'}(q,n)}{(p_1+p_2)^2} \, ,
\end{eqnarray}
where
\begin{eqnarray}
V_{\phi tt} & = & -i\frac{m_t}{\vev}(a_t-ib_t\gamma_5) \, , \\
V_{gtt}^{\mu,a} & = & -ig_st^a\gamma^{\mu} \, , \\
V_{ggg}^{\mu\nu\rho,abc}(p_1,p_2,p_3) & = & -g_sf^{abc}(
g^{\mu\nu}(p_1 - p_2)^{\rho}+g^{\nu\rho}(p_2 - p_3)^{\mu}
+ g^{\rho\mu}(p_3 - p_1)^{\nu}) \, ,\\
V_{gg\phi}^{\mu\nu,ab}(p_1,p_2) & = & -4i\delta^{ab}[
f_S^{(0)}(g^{\mu\nu}p_1\cdot p_2 - p_1^{\nu}p_2^{\mu})
-2f_P^{(0)}\epsilon^{\mu\nu\rho\sigma}p_{1\rho}p_{2\sigma}] \, , 
\end{eqnarray}
and $\mathcal{P}_{\alpha\alpha'}$ is the gluon polarization sum
\begin{equation}
\mathcal{P}_{\alpha\alpha'}(p,n)=-g_{\alpha\alpha'}
+\frac{p_{\alpha}n_{\alpha'}+p_{\alpha'}n_{\alpha}}{p\cdot n},\quad
n^2=0 \, ,
\label{polsum}
\end{equation}
where $n$ is an arbitrary lightlike vector. Instead of the polarization sum in \Eq{polsum} 
we use
\begin{equation}
\mathcal{P}_{\rho\rho'}(q,n)=-g_{\rho\rho'} \, ,
\end{equation}
because in the present case there are no diagrams where the external soft gluon could
be replaced by a ghost. With this relation we obtain:
\begin{eqnarray}
\Nr & = & -\Tr[(\cancel{k}_1+m_t)V_{\phi tt}(\cancel{k}_2-m_t)
V_{gtt,\alpha_2}^{c_2}(\cancel{k}_1+m_t)V_{gtt}^{\alpha_3,c_3}]\nonumber\\
& \times & V_{ggg}^{\mu'\nu'\alpha_2,abc_2}(p_1,p_2,-p_1-p_2)
V_{ggg,\alpha_3}^{\mu\alpha_1ac_1c_3}(p_1,-p_1,0)
V_{gg\phi,\alpha_1}^{\nu,bc_1}(p_2,p_1)\nonumber\\
& \times & \frac{\mathcal{P}_{\mu\mu'}(p_1,p_2)
\mathcal{P}_{\nu\nu'}(p_2,p_1)}{(p_1+p_2)^2} \, . 
\end{eqnarray}
The  numerator of $\Nv$ of  the virtual correction contribution  \eqref{loopintegral}
 is computed in straightforward fashion in the SGA. One obtains
\begin{equation}
\Nv = i\Nr.
\end{equation}
Therefore 
\begin{eqnarray}
&&i\int\frac{d^3\ell}{(2\pi)^3}\frac{1}{2\ell^0}
\frac{\Nv}{\big[-2\ell\cdot p_1+i\varepsilon\big]\big[-2\ell\cdot(p_1+p_2)
+s-m_{\phi}^2+i\varepsilon\big]\big[2\ell\cdot k_1-i\varepsilon\big]}\nonumber\\
&&=-\int\frac{d^3\ell}{(2\pi)^3}\frac{1}{2\ell^0}
\frac{\Nr}{\big[-2\ell\cdot p_1+i\varepsilon\big]\big[-2\ell\cdot(p_1+p_2)
+s-m_{\phi}^2+i\varepsilon\big]\big[2\ell\cdot k_1-i\varepsilon\big]}
\label{eq:appvicorr}
\end{eqnarray}
which exactly cancels the contribution from the real correction,
\begin{equation}
\int\frac{d^3q}{(2\pi)^3}\frac{1}{2q^0}\frac{\Nr}{\big[-2q\cdot p_1
+i\varepsilon\big]\big[-2q\cdot(p_1+p_2)+s-m_{\phi}^2+i\varepsilon\big]
\big[2q\cdot k_1-i\varepsilon\big]}.
\label{eq:apprecorr}
\end{equation}
Note that
\begin{equation}
\delta^{(4)}(p_1+p_2-k_1-k_2-q)\quad\stackrel{\text{SGA}}{\longrightarrow}
\quad\delta^{(4)}(p_1+p_2-k_1-k_2)
\end{equation}
in the phase-space integration of \eqref{eq:apprecorr}. Thus the four-momentum
conservation matches the one in the phase-space integration of \eqref{eq:appvicorr}.

\bibliography{heavy_Higgs_ttbar_v2}

\providecommand{\href}[2]{#2}\begingroup\raggedright\begin{thebibliography}{10}

\bibitem{Aad:2012tfa}
{ ATLAS} Collaboration, G.~Aad {et~al.}, {\em {Observation of a new particle in
  the search for the Standard Model Higgs boson with the ATLAS detector at the
  LHC}}, \href{http://dx.doi.org/10.1016/j.physletb.2012.08.020}{Phys. Lett.
  {\bfseries B716} (2012) 1--29},
\href{http://arxiv.org/abs/1207.7214}{{\ttfamily arXiv:1207.7214 [hep-ex]}}.

\bibitem{Chatrchyan:2012xdj}
{ CMS} Collaboration, S.~Chatrchyan {et~al.}, {\em {Observation of a new boson
  at a mass of 125 GeV with the CMS experiment at the LHC}},
  \href{http://dx.doi.org/10.1016/j.physletb.2012.08.021}{Phys. Lett.
  {\bfseries B716} (2012) 30--61},
\href{http://arxiv.org/abs/1207.7235}{{\ttfamily arXiv:1207.7235 [hep-ex]}}.

\bibitem{Khachatryan:2014jba}
{ CMS} Collaboration, V.~Khachatryan {et~al.}, {\em {Precise determination of
  the mass of the Higgs boson and tests of compatibility of its couplings with
  the standard model predictions using proton collisions at 7 and 8 $\,\text
  {TeV}$}}, \href{http://dx.doi.org/10.1140/epjc/s10052-015-3351-7}{Eur. Phys.
  J. {\bfseries C75} no.~5, (2015) 212},
\href{http://arxiv.org/abs/1412.8662}{{\ttfamily arXiv:1412.8662 [hep-ex]}}.

\bibitem{Aad:2015gba}
{ ATLAS} Collaboration, G.~Aad {et~al.}, {\em {Measurements of the Higgs boson
  production and decay rates and coupling strengths using pp collision data at
  $\sqrt{s}=7$ and 8 TeV in the ATLAS experiment}},
  \href{http://dx.doi.org/10.1140/epjc/s10052-015-3769-y}{Eur. Phys. J.
  {\bfseries C76} no.~1, (2016) 6},
\href{http://arxiv.org/abs/1507.04548}{{\ttfamily arXiv:1507.04548 [hep-ex]}}.

\bibitem{Djouadi:2005gj}
A.~Djouadi, {\em {The Anatomy of electro-weak symmetry breaking. II. The Higgs
  bosons in the minimal supersymmetric model}},
  \href{http://dx.doi.org/10.1016/j.physrep.2007.10.005}{Phys. Rept. {\bfseries
  459} (2008) 1--241},
\href{http://arxiv.org/abs/hep-ph/0503173}{{\ttfamily arXiv:hep-ph/0503173
  [hep-ph]}}.

\bibitem{Branco:2011iw}
G.~C. Branco, P.~M. Ferreira, L.~Lavoura, M.~N. Rebelo, M.~Sher, and J.~P.
  Silva, {\em {Theory and phenomenology of two-Higgs-doublet models}},
  \href{http://dx.doi.org/10.1016/j.physrep.2012.02.002}{Phys. Rept. {\bfseries
  516} (2012) 1--102},
\href{http://arxiv.org/abs/1106.0034}{{\ttfamily arXiv:1106.0034 [hep-ph]}}.

\bibitem{ATLAS:2013nma}
{ ATLAS} Collaboration,
{\em {Measurements of the properties of the Higgs-like boson in the four lepton
  decay channel with the ATLAS detector using 25 fb$^{-1}$ of proton-proton
  collision data, ATLAS-CONF-2013-013}},.

\bibitem{Khachatryan:2015lba}
{ CMS} Collaboration, V.~Khachatryan {et~al.}, {\em {Search for a pseudoscalar
  boson decaying into a Z boson and the 125 GeV Higgs boson in $\ell^+\ell^-
  b\overline{b}$ final states}},
  \href{http://dx.doi.org/10.1016/j.physletb.2015.07.010}{Phys. Lett.
  {\bfseries B748} (2015) 221--243},
\href{http://arxiv.org/abs/1504.04710}{{\ttfamily arXiv:1504.04710 [hep-ex]}}.

\bibitem{Aad:2014yja}
{ ATLAS} Collaboration, G.~Aad {et~al.}, {\em {Search For Higgs Boson Pair
  Production in the $\gamma\gamma b\bar{b}$ Final State using $pp$ Collision
  Data at $\sqrt{s}=8$ TeV from the ATLAS Detector}},
  \href{http://dx.doi.org/10.1103/PhysRevLett.114.081802}{Phys. Rev. Lett.
  {\bfseries 114} no.~8, (2015) 081802},
\href{http://arxiv.org/abs/1406.5053}{{\ttfamily arXiv:1406.5053 [hep-ex]}}.

\bibitem{Aad:2015wra}
{ ATLAS} Collaboration, G.~Aad {et~al.}, {\em {Search for a CP-odd Higgs boson
  decaying to Zh in pp collisions at $\sqrt{s} = 8$ TeV with the ATLAS
  detector}}, \href{http://dx.doi.org/10.1016/j.physletb.2015.03.054}{Phys.
  Lett. {\bfseries B744} (2015) 163--183},
\href{http://arxiv.org/abs/1502.04478}{{\ttfamily arXiv:1502.04478 [hep-ex]}}.

\bibitem{Khachatryan:2015cwa}
{ CMS} Collaboration, V.~Khachatryan {et~al.}, {\em {Search for a Higgs Boson
  in the Mass Range from 145 to 1000 GeV Decaying to a Pair of W or Z Bosons}},
  \href{http://dx.doi.org/10.1007/JHEP10(2015)144}{JHEP {\bfseries 10} (2015)
  144},
\href{http://arxiv.org/abs/1504.00936}{{\ttfamily arXiv:1504.00936 [hep-ex]}}.

\bibitem{Nagai:2013xwa}
{ ATLAS, CMS} Collaboration, Y.~Nagai, {\em {Higgs Search in $b\bar{b}$
  Signatures at ATLAS and CMS}}, PoS {\bfseries Beauty2013} (2013) 001,
\href{http://arxiv.org/abs/1306.1784}{{\ttfamily arXiv:1306.1784 [hep-ex]}}.

\bibitem{Khachatryan:2015yea}
{ CMS} Collaboration, V.~Khachatryan {et~al.}, {\em {Search for resonant pair
  production of Higgs bosons decaying to two bottom quark–antiquark pairs in
  proton–proton collisions at 8 TeV}},
  \href{http://dx.doi.org/10.1016/j.physletb.2015.08.047}{Phys. Lett.
  {\bfseries B749} (2015) 560--582},
\href{http://arxiv.org/abs/1503.04114}{{\ttfamily arXiv:1503.04114 [hep-ex]}}.

\bibitem{Khachatryan:2014wca}
{ CMS} Collaboration, V.~Khachatryan {et~al.}, {\em {Search for neutral MSSM
  Higgs bosons decaying to a pair of tau leptons in pp collisions}},
  \href{http://dx.doi.org/10.1007/JHEP10(2014)160}{JHEP {\bfseries 10} (2014)
  160},
\href{http://arxiv.org/abs/1408.3316}{{\ttfamily arXiv:1408.3316 [hep-ex]}}.

\bibitem{Aad:2014vgg}
{ ATLAS} Collaboration, G.~Aad {et~al.}, {\em {Search for neutral Higgs bosons
  of the minimal supersymmetric standard model in pp collisions at $\sqrt{s}$ =
  8 TeV with the ATLAS detector}},
  \href{http://dx.doi.org/10.1007/JHEP11(2014)056}{JHEP {\bfseries 11} (2014)
  056},
\href{http://arxiv.org/abs/1409.6064}{{\ttfamily arXiv:1409.6064 [hep-ex]}}.

\bibitem{Gaemers:1984sj}
K.~J.~F. Gaemers and F.~Hoogeveen, {\em {Higgs Production and Decay Into Heavy
  Flavors With the Gluon Fusion Mechanism}},
\href{http://dx.doi.org/10.1016/0370-2693(84)91711-8}{Phys. Lett. {\bfseries
  B146} (1984) 347}.

\bibitem{Dicus:1994bm}
D.~Dicus, A.~Stange, and S.~Willenbrock, {\em {Higgs decay to top quarks at
  hadron colliders}},
  \href{http://dx.doi.org/10.1016/0370-2693(94)91017-0}{Phys. Lett. {\bfseries
  B333} (1994) 126--131},
\href{http://arxiv.org/abs/hep-ph/9404359}{{\ttfamily arXiv:hep-ph/9404359
  [hep-ph]}}.

\bibitem{Bernreuther:1993hq}
W.~Bernreuther and A.~Brandenburg, {\em {Tracing CP violation in the production
  of top quark pairs by multiple TeV proton proton collisions}},
  \href{http://dx.doi.org/10.1103/PhysRevD.49.4481}{Phys. Rev. {\bfseries D49}
  (1994) 4481--4492},
\href{http://arxiv.org/abs/hep-ph/9312210}{{\ttfamily arXiv:hep-ph/9312210
  [hep-ph]}}.

\bibitem{Bernreuther:1997gs}
W.~Bernreuther, M.~Flesch, and P.~Haberl, {\em {Signatures of Higgs bosons in
  the top quark decay channel at hadron colliders}},
  \href{http://dx.doi.org/10.1103/PhysRevD.58.114031}{Phys. Rev. {\bfseries
  D58} (1998) 114031},
\href{http://arxiv.org/abs/hep-ph/9709284}{{\ttfamily arXiv:hep-ph/9709284
  [hep-ph]}}.

\bibitem{Bernreuther:1998qv}
W.~Bernreuther, A.~Brandenburg, and M.~Flesch, {\em {Effects of Higgs sector CP
  violation in top quark pair production at the LHC}},
\href{http://arxiv.org/abs/hep-ph/9812387}{{\ttfamily arXiv:hep-ph/9812387
  [hep-ph]}}.

\bibitem{Barger:2006hm}
V.~Barger, T.~Han, and D.~G.~E. Walker, {\em {Top Quark Pairs at High Invariant
  Mass: A Model-Independent Discriminator of New Physics at the LHC}},
  \href{http://dx.doi.org/10.1103/PhysRevLett.100.031801}{Phys. Rev. Lett.
  {\bfseries 100} (2008) 031801},
\href{http://arxiv.org/abs/hep-ph/0612016}{{\ttfamily arXiv:hep-ph/0612016
  [hep-ph]}}.

\bibitem{Frederix:2007gi}
R.~Frederix and F.~Maltoni, {\em {Top pair invariant mass distribution: A
  Window on new physics}},
  \href{http://dx.doi.org/10.1088/1126-6708/2009/01/047}{JHEP {\bfseries 01}
  (2009) 047},
\href{http://arxiv.org/abs/0712.2355}{{\ttfamily arXiv:0712.2355 [hep-ph]}}.

\bibitem{Barger:2011pu}
V.~Barger, W.-Y. Keung, and B.~Yencho, {\em {Azimuthal Correlations in Top Pair
  Decays and The Effects of New Heavy Scalars}},
  \href{http://dx.doi.org/10.1103/PhysRevD.85.034016}{Phys. Rev. {\bfseries
  D85} (2012) 034016},
\href{http://arxiv.org/abs/1112.5173}{{\ttfamily arXiv:1112.5173 [hep-ph]}}.

\bibitem{Craig:2015jba}
N.~Craig, F.~D'Eramo, P.~Draper, S.~Thomas, and H.~Zhang, {\em {The Hunt for
  the Rest of the Higgs Bosons}},
  \href{http://dx.doi.org/10.1007/JHEP06(2015)137}{JHEP {\bfseries 06} (2015)
  137},
\href{http://arxiv.org/abs/1504.04630}{{\ttfamily arXiv:1504.04630 [hep-ph]}}.

\bibitem{Jung:2015gta}
S.~Jung, J.~Song, and Y.~W. Yoon, {\em {Dip or nothingness of a Higgs resonance
  from the interference with a complex phase}},
  \href{http://dx.doi.org/10.1103/PhysRevD.92.055009}{Phys. Rev. {\bfseries
  D92} no.~5, (2015) 055009},
\href{http://arxiv.org/abs/1505.00291}{{\ttfamily arXiv:1505.00291 [hep-ph]}}.

\bibitem{Bhattacherjee:2015sga}
B.~Bhattacherjee, A.~Chakraborty, and A.~Choudhury, {\em {Status of the MSSM
  Higgs sector using global analysis and direct search bounds, and future
  prospects at the High Luminosity LHC}},
  \href{http://dx.doi.org/10.1103/PhysRevD.92.093007}{Phys. Rev. {\bfseries
  D92} no.~9, (2015) 093007},
\href{http://arxiv.org/abs/1504.04308}{{\ttfamily arXiv:1504.04308 [hep-ph]}}.

\bibitem{Chatrchyan:2013lca}
{ CMS} Collaboration, S.~Chatrchyan {et~al.}, {\em {Searches for new physics
  using the $t\bar{t}$ invariant mass distribution in pp collisions at
  $\sqrt{s}$=8  TeV}},
  \href{http://dx.doi.org/10.1103/PhysRevLett.111.211804,
  10.1103/PhysRevLett.112.119903}{Phys. Rev. Lett. {\bfseries 111} no.~21,
  (2013) 211804}, \href{http://arxiv.org/abs/1309.2030}{{\ttfamily
  arXiv:1309.2030 [hep-ex]}}.
[Erratum: Phys. Rev. Lett.112,119903(2014)].

\bibitem{Aad:2015fna}
{ ATLAS} Collaboration, G.~Aad {et~al.}, {\em {A search for $ t\overline{t} $
  resonances using lepton-plus-jets events in proton-proton collisions at $
  \sqrt{s}=8 $ TeV with the ATLAS detector}},
  \href{http://dx.doi.org/10.1007/JHEP08(2015)148}{JHEP {\bfseries 08} (2015)
  148},
\href{http://arxiv.org/abs/1505.07018}{{\ttfamily arXiv:1505.07018 [hep-ex]}}.

\bibitem{Kramer:1996iq}
M.~Kramer, E.~Laenen, and M.~Spira, {\em {Soft gluon radiation in Higgs boson
  production at the LHC}},
  \href{http://dx.doi.org/10.1016/S0550-3213(97)00679-2}{Nucl. Phys. {\bfseries
  B511} (1998) 523--549},
\href{http://arxiv.org/abs/hep-ph/9611272}{{\ttfamily arXiv:hep-ph/9611272
  [hep-ph]}}.

\bibitem{Collins:1977iv}
J.~C. Collins and D.~E. Soper, {\em {Angular Distribution of Dileptons in
  High-Energy Hadron Collisions}},
\href{http://dx.doi.org/10.1103/PhysRevD.16.2219}{Phys. Rev. {\bfseries D16}
  (1977) 2219}.

\bibitem{ElKaffas:2006nt}
A.~W. El~Kaffas, W.~Khater, O.~M. Ogreid, and P.~Osland, {\em {Consistency of
  the two Higgs doublet model and CP violation in top production at the LHC}},
  \href{http://dx.doi.org/10.1016/j.nuclphysb.2007.03.041}{Nucl. Phys.
  {\bfseries B775} (2007) 45--77},
\href{http://arxiv.org/abs/hep-ph/0605142}{{\ttfamily arXiv:hep-ph/0605142
  [hep-ph]}}.

\bibitem{Grzadkowski:2014ada}
B.~Grzadkowski, O.~M. Ogreid, and P.~Osland, {\em {Measuring CP violation in
  Two-Higgs-Doublet models in light of the LHC Higgs data}},
  \href{http://dx.doi.org/10.1007/JHEP11(2014)084}{JHEP {\bfseries 11} (2014)
  084},
\href{http://arxiv.org/abs/1409.7265}{{\ttfamily arXiv:1409.7265 [hep-ph]}}.

\bibitem{Bernreuther:1992dz}
W.~Bernreuther, T.~Schroder, and T.~N. Pham, {\em {CP violating dipole
  form-factors in $e^+ e^- \to t\bar t$}},
\href{http://dx.doi.org/10.1016/0370-2693(92)90410-6}{Phys. Lett. {\bfseries
  B279} (1992) 389--396}.

\bibitem{Inoue:2014nva}
S.~Inoue, M.~J. Ramsey-Musolf, and Y.~Zhang, {\em {CP-violating phenomenology
  of flavor conserving two Higgs doublet models}},
  \href{http://dx.doi.org/10.1103/PhysRevD.89.115023}{Phys. Rev. {\bfseries
  D89} no.~11, (2014) 115023},
\href{http://arxiv.org/abs/1403.4257}{{\ttfamily arXiv:1403.4257 [hep-ph]}}.

\bibitem{Chen:2015gaa}
C.-Y. Chen, S.~Dawson, and Y.~Zhang, {\em {Complementarity of LHC and EDMs for
  Exploring Higgs CP Violation}},
  \href{http://dx.doi.org/10.1007/JHEP06(2015)056}{JHEP {\bfseries 06} (2015)
  056},
\href{http://arxiv.org/abs/1503.01114}{{\ttfamily arXiv:1503.01114 [hep-ph]}}.

\bibitem{Mell15}
C.~Mellein, {\em {Doctoral Thesis, RWTH Aachen University (2015),
  unpublished}},.

\bibitem{Mahmoudi:2009zx}
F.~Mahmoudi and O.~Stal, {\em {Flavor constraints on the two-Higgs-doublet
  model with general Yukawa couplings}},
  \href{http://dx.doi.org/10.1103/PhysRevD.81.035016}{Phys. Rev. {\bfseries
  D81} (2010) 035016},
\href{http://arxiv.org/abs/0907.1791}{{\ttfamily arXiv:0907.1791 [hep-ph]}}.

\bibitem{Hermann:2012fc}
T.~Hermann, M.~Misiak, and M.~Steinhauser, {\em {$\bar{B}\to X_s \gamma$ in the
  Two Higgs Doublet Model up to Next-to-Next-to-Leading Order in QCD}},
  \href{http://dx.doi.org/10.1007/JHEP11(2012)036}{JHEP {\bfseries 11} (2012)
  036},
\href{http://arxiv.org/abs/1208.2788}{{\ttfamily arXiv:1208.2788 [hep-ph]}}.

\bibitem{Eberhardt:2013uba}
O.~Eberhardt, U.~Nierste, and M.~Wiebusch, {\em {Status of the
  two-Higgs-doublet model of type II}},
  \href{http://dx.doi.org/10.1007/JHEP07(2013)118}{JHEP {\bfseries 07} (2013)
  118},
\href{http://arxiv.org/abs/1305.1649}{{\ttfamily arXiv:1305.1649 [hep-ph]}}.

\bibitem{Djouadi:1997yw}
A.~Djouadi, J.~Kalinowski, and M.~Spira, {\em {HDECAY: A Program for Higgs
  boson decays in the standard model and its supersymmetric extension}},
  \href{http://dx.doi.org/10.1016/S0010-4655(97)00123-9}{Comput. Phys. Commun.
  {\bfseries 108} (1998) 56--74},
\href{http://arxiv.org/abs/hep-ph/9704448}{{\ttfamily arXiv:hep-ph/9704448
  [hep-ph]}}.

\bibitem{Eriksson:2009ws}
D.~Eriksson, J.~Rathsman, and O.~Stal, {\em {2HDMC: Two-Higgs-Doublet Model
  Calculator Physics and Manual}},
  \href{http://dx.doi.org/10.1016/j.cpc.2009.09.011}{Comput. Phys. Commun.
  {\bfseries 181} (2010) 189--205},
\href{http://arxiv.org/abs/0902.0851}{{\ttfamily arXiv:0902.0851 [hep-ph]}}.

\bibitem{Braaten:1980yq}
E.~Braaten and J.~P. Leveille, {\em {Higgs Boson Decay and the Running Mass}},
\href{http://dx.doi.org/10.1103/PhysRevD.22.715}{Phys. Rev. {\bfseries D22}
  (1980) 715}.

\bibitem{Drees:1990dq}
M.~Drees and K.-i. Hikasa, {\em {NOTE ON QCD CORRECTIONS TO HADRONIC HIGGS
  DECAY}}, \href{http://dx.doi.org/10.1016/0370-2693(90)91130-4}{Phys. Lett.
  {\bfseries B240} (1990) 455}.
[Erratum: Phys. Lett.B262,497(1991)].

\bibitem{Spira:1995rr}
M.~Spira, A.~Djouadi, D.~Graudenz, and P.~M. Zerwas, {\em {Higgs boson
  production at the LHC}},
  \href{http://dx.doi.org/10.1016/0550-3213(95)00379-7}{Nucl. Phys. {\bfseries
  B453} (1995) 17--82},
\href{http://arxiv.org/abs/hep-ph/9504378}{{\ttfamily arXiv:hep-ph/9504378
  [hep-ph]}}.

\bibitem{Dabelstein:1991ky}
A.~Dabelstein and W.~Hollik, {\em {Electroweak corrections to the fermionic
  decay width of the standard Higgs boson}},
\href{http://dx.doi.org/10.1007/BF01625912}{Z. Phys. {\bfseries C53} (1992)
  507--516}.

\bibitem{Fleischer:1980ub}
J.~Fleischer and F.~Jegerlehner, {\em {Radiative Corrections to Higgs Decays in
  the Extended Weinberg-Salam Model}},
\href{http://dx.doi.org/10.1103/PhysRevD.23.2001}{Phys. Rev. {\bfseries D23}
  (1981) 2001--2026}.

\bibitem{Grimus:2007if}
W.~Grimus, L.~Lavoura, O.~M. Ogreid, and P.~Osland, {\em {A Precision
  constraint on multi-Higgs-doublet models}},
  \href{http://dx.doi.org/10.1088/0954-3899/35/7/075001}{J. Phys. {\bfseries
  G35} (2008) 075001},
\href{http://arxiv.org/abs/0711.4022}{{\ttfamily arXiv:0711.4022 [hep-ph]}}.

\bibitem{Grimus:2008nb}
W.~Grimus, L.~Lavoura, O.~M. Ogreid, and P.~Osland, {\em {The Oblique
  parameters in multi-Higgs-doublet models}},
  \href{http://dx.doi.org/10.1016/j.nuclphysb.2008.04.019}{Nucl. Phys.
  {\bfseries B801} (2008) 81--96},
\href{http://arxiv.org/abs/0802.4353}{{\ttfamily arXiv:0802.4353 [hep-ph]}}.

\bibitem{Haber:2010bw}
H.~E. Haber and D.~O'Neil, {\em {Basis-independent methods for the
  two-Higgs-doublet model III: The CP-conserving limit, custodial symmetry, and
  the oblique parameters S, T, U}},
  \href{http://dx.doi.org/10.1103/PhysRevD.83.055017}{Phys. Rev. {\bfseries
  D83} (2011) 055017},
\href{http://arxiv.org/abs/1011.6188}{{\ttfamily arXiv:1011.6188 [hep-ph]}}.

\bibitem{Baker:2006ts}
C.~A. Baker {et~al.}, {\em {An Improved experimental limit on the electric
  dipole moment of the neutron}},
  \href{http://dx.doi.org/10.1103/PhysRevLett.97.131801}{Phys. Rev. Lett.
  {\bfseries 97} (2006) 131801},
\href{http://arxiv.org/abs/hep-ex/0602020}{{\ttfamily arXiv:hep-ex/0602020
  [hep-ex]}}.

\bibitem{Baron:2013eja}
{ ACME} Collaboration, J.~Baron {et~al.}, {\em {Order of Magnitude Smaller
  Limit on the Electric Dipole Moment of the Electron}},
  \href{http://dx.doi.org/10.1126/science.1248213}{Science {\bfseries 343}
  (2014) 269--272},
\href{http://arxiv.org/abs/1310.7534}{{\ttfamily arXiv:1310.7534
  [physics.atom-ph]}}.

\bibitem{Kuhn:2013zoa}
J.~H. Kuhn, A.~Scharf, and P.~Uwer, {\em {Weak Interactions in Top-Quark Pair
  Production at Hadron Colliders: An Update}},
  \href{http://dx.doi.org/10.1103/PhysRevD.91.014020}{Phys. Rev. {\bfseries
  D91} no.~1, (2015) 014020},
\href{http://arxiv.org/abs/1305.5773}{{\ttfamily arXiv:1305.5773 [hep-ph]}}.

\bibitem{Nason:1987xz}
P.~Nason, S.~Dawson, and R.~K. Ellis, {\em {The Total Cross-Section for the
  Production of Heavy Quarks in Hadronic Collisions}},
\href{http://dx.doi.org/10.1016/0550-3213(88)90422-1}{Nucl. Phys. {\bfseries
  B303} (1988) 607}.

\bibitem{Beenakker:1988bq}
W.~Beenakker, H.~Kuijf, W.~L. van Neerven, and J.~Smith, {\em {QCD Corrections
  to Heavy Quark Production in p anti-p Collisions}},
\href{http://dx.doi.org/10.1103/PhysRevD.40.54}{Phys. Rev. {\bfseries D40}
  (1989) 54--82}.

\bibitem{Denner:1999gp}
A.~Denner, S.~Dittmaier, M.~Roth, and D.~Wackeroth, {\em {Predictions for all
  processes $e^+e^-\to 4~\mbox{fermions} +\gamma$}},
  \href{http://dx.doi.org/10.1016/S0550-3213(99)00437-X}{Nucl. Phys. {\bfseries
  B560} (1999) 33--65},
\href{http://arxiv.org/abs/hep-ph/9904472}{{\ttfamily arXiv:hep-ph/9904472
  [hep-ph]}}.

\bibitem{Denner:2005fg}
A.~Denner, S.~Dittmaier, M.~Roth, and L.~H. Wieders, {\em {Electroweak
  corrections to charged-current $e^+e^-\to 4~\mbox{fermion}$ processes:
  Technical details and further results}},
  \href{http://dx.doi.org/10.1016/j.nuclphysb.2011.09.001,
  10.1016/j.nuclphysb.2005.06.033}{Nucl. Phys. {\bfseries B724} (2005)
  247--294}, \href{http://arxiv.org/abs/hep-ph/0505042}{{\ttfamily
  arXiv:hep-ph/0505042 [hep-ph]}}.
[Erratum: Nucl. Phys.B854,504(2012)].

\bibitem{Nowakowski:1993iu}
M.~Nowakowski and A.~Pilaftsis, {\em {On gauge invariance of Breit-Wigner
  propagators}}, \href{http://dx.doi.org/10.1007/BF01650437}{Z. Phys.
  {\bfseries C60} (1993) 121--126},
\href{http://arxiv.org/abs/hep-ph/9305321}{{\ttfamily arXiv:hep-ph/9305321
  [hep-ph]}}.

\bibitem{Harlander:2005rq}
R.~Harlander and P.~Kant, {\em {Higgs production and decay: Analytic results at
  next-to-leading order QCD}},
  \href{http://dx.doi.org/10.1088/1126-6708/2005/12/015}{JHEP {\bfseries 12}
  (2005) 015},
\href{http://arxiv.org/abs/hep-ph/0509189}{{\ttfamily arXiv:hep-ph/0509189
  [hep-ph]}}.

\bibitem{Djouadi:1991tka}
A.~Djouadi, M.~Spira, and P.~M. Zerwas, {\em {Production of Higgs bosons in
  proton colliders: QCD corrections}},
\href{http://dx.doi.org/10.1016/0370-2693(91)90375-Z}{Phys. Lett. {\bfseries
  B264} (1991) 440--446}.

\bibitem{Harlander:2001is}
R.~V. Harlander and W.~B. Kilgore, {\em {Soft and virtual corrections to proton
  proton to H + x at NNLO}},
  \href{http://dx.doi.org/10.1103/PhysRevD.64.013015}{Phys. Rev. {\bfseries
  D64} (2001) 013015},
\href{http://arxiv.org/abs/hep-ph/0102241}{{\ttfamily arXiv:hep-ph/0102241
  [hep-ph]}}.

\bibitem{Chetyrkin:1998mw}
K.~G. Chetyrkin, B.~A. Kniehl, M.~Steinhauser, and W.~A. Bardeen, {\em
  {Effective QCD interactions of CP odd Higgs bosons at three loops}},
  \href{http://dx.doi.org/10.1016/S0550-3213(98)00594-X}{Nucl. Phys. {\bfseries
  B535} (1998) 3--18},
\href{http://arxiv.org/abs/hep-ph/9807241}{{\ttfamily arXiv:hep-ph/9807241
  [hep-ph]}}.

\bibitem{Harlander:2002vv}
R.~V. Harlander and W.~B. Kilgore, {\em {Production of a pseudoscalar Higgs
  boson at hadron colliders at next-to-next-to leading order}},
  \href{http://dx.doi.org/10.1088/1126-6708/2002/10/017}{JHEP {\bfseries 10}
  (2002) 017},
\href{http://arxiv.org/abs/hep-ph/0208096}{{\ttfamily arXiv:hep-ph/0208096
  [hep-ph]}}.

\bibitem{Fadin:1993dz}
V.~S. Fadin, V.~A. Khoze, and A.~D. Martin, {\em {Interference radiative
  phenomena in the production of heavy unstable particles}},
\href{http://dx.doi.org/10.1103/PhysRevD.49.2247}{Phys. Rev. {\bfseries D49}
  (1994) 2247--2256}.

\bibitem{Melnikov:1995fx}
K.~Melnikov and O.~I. Yakovlev, {\em {Final state interaction in the production
  of heavy unstable particles}},
  \href{http://dx.doi.org/10.1016/0550-3213(96)00151-4}{Nucl. Phys. {\bfseries
  B471} (1996) 90--120},
\href{http://arxiv.org/abs/hep-ph/9501358}{{\ttfamily arXiv:hep-ph/9501358
  [hep-ph]}}.

\bibitem{Beenakker:1997ir}
W.~Beenakker, A.~P. Chapovsky, and F.~A. Berends, {\em {Nonfactorizable
  corrections to W pair production: Methods and analytic results}},
  \href{http://dx.doi.org/10.1016/S0550-3213(97)00628-7}{Nucl. Phys. {\bfseries
  B508} (1997) 17--63},
\href{http://arxiv.org/abs/hep-ph/9707326}{{\ttfamily arXiv:hep-ph/9707326
  [hep-ph]}}.

\bibitem{Dittmaier:2014qza}
S.~Dittmaier, A.~Huss, and C.~Schwinn, {\em {Mixed QCD-electroweak
  $\mathcal{O}(\alpha_s\alpha)$ corrections to Drell-Yan processes in the
  resonance region: pole approximation and non-factorizable corrections}},
  \href{http://dx.doi.org/10.1016/j.nuclphysb.2014.05.027}{Nucl. Phys.
  {\bfseries B885} (2014) 318--372},
\href{http://arxiv.org/abs/1403.3216}{{\ttfamily arXiv:1403.3216 [hep-ph]}}.

\bibitem{Nason:1989zy}
P.~Nason, S.~Dawson, and R.~K. Ellis, {\em {The One Particle Inclusive
  Differential Cross-Section for Heavy Quark Production in Hadronic
  Collisions}}, \href{http://dx.doi.org/10.1016/0550-3213(89)90286-1}{Nucl.
  Phys. {\bfseries B327} (1989) 49--92}.
[Erratum: Nucl. Phys.B335,260(1990)].

\bibitem{Beenakker:1990maa}
W.~Beenakker, W.~L. van Neerven, R.~Meng, G.~A. Schuler, and J.~Smith, {\em
  {QCD corrections to heavy quark production in hadron hadron collisions}},
\href{http://dx.doi.org/10.1016/S0550-3213(05)80032-X}{Nucl. Phys. {\bfseries
  B351} (1991) 507--560}.

\bibitem{Bernreuther:2001rq}
W.~Bernreuther, A.~Brandenburg, Z.~G. Si, and P.~Uwer, {\em {Top quark spin
  correlations at hadron colliders: Predictions at next-to-leading order QCD}},
  \href{http://dx.doi.org/10.1103/PhysRevLett.87.242002}{Phys. Rev. Lett.
  {\bfseries 87} (2001) 242002},
\href{http://arxiv.org/abs/hep-ph/0107086}{{\ttfamily arXiv:hep-ph/0107086
  [hep-ph]}}.

\bibitem{Bernreuther:2004jv}
W.~Bernreuther, A.~Brandenburg, Z.~G. Si, and P.~Uwer, {\em {Top quark pair
  production and decay at hadron colliders}},
  \href{http://dx.doi.org/10.1016/j.nuclphysb.2004.04.019}{Nucl. Phys.
  {\bfseries B690} (2004) 81--137},
\href{http://arxiv.org/abs/hep-ph/0403035}{{\ttfamily arXiv:hep-ph/0403035
  [hep-ph]}}.

\bibitem{Passarino:1978jh}
G.~Passarino and M.~J.~G. Veltman, {\em {One Loop Corrections for $e^+ e^-$
  Annihilation Into $\mu^+ \mu^-$ in the Weinberg Model}},
\href{http://dx.doi.org/10.1016/0550-3213(79)90234-7}{Nucl. Phys. {\bfseries
  B160} (1979) 151}.

\bibitem{Bernreuther:2005gw}
W.~Bernreuther, R.~Bonciani, T.~Gehrmann, R.~Heinesch, P.~Mastrolia, and
  E.~Remiddi, {\em {Decays of scalar and pseudoscalar Higgs bosons into
  fermions: Two-loop QCD corrections to the Higgs-quark-antiquark amplitude}},
  \href{http://dx.doi.org/10.1103/PhysRevD.72.096002}{Phys. Rev. {\bfseries
  D72} (2005) 096002},
\href{http://arxiv.org/abs/hep-ph/0508254}{{\ttfamily arXiv:hep-ph/0508254
  [hep-ph]}}.

\bibitem{Catani:1996vz}
S.~Catani and M.~H. Seymour, {\em {A General algorithm for calculating jet
  cross-sections in NLO QCD}},
  \href{http://dx.doi.org/10.1016/S0550-3213(96)00589-5}{Nucl. Phys. {\bfseries
  B485} (1997) 291--419}, \href{http://arxiv.org/abs/hep-ph/9605323}{{\ttfamily
  arXiv:hep-ph/9605323 [hep-ph]}}.
[Erratum: Nucl. Phys.B510,503(1998)].

\bibitem{Catani:2002hc}
S.~Catani, S.~Dittmaier, M.~H. Seymour, and Z.~Trocsanyi, {\em {The Dipole
  formalism for next-to-leading order QCD calculations with massive partons}},
  \href{http://dx.doi.org/10.1016/S0550-3213(02)00098-6}{Nucl. Phys. {\bfseries
  B627} (2002) 189--265},
\href{http://arxiv.org/abs/hep-ph/0201036}{{\ttfamily arXiv:hep-ph/0201036
  [hep-ph]}}.

\bibitem{Bernreuther:2010ny}
W.~Bernreuther and Z.-G. Si, {\em {Distributions and correlations for top quark
  pair production and decay at the Tevatron and LHC.}},
  \href{http://dx.doi.org/10.1016/j.nuclphysb.2010.05.001}{Nucl. Phys.
  {\bfseries B837} (2010) 90--121},
\href{http://arxiv.org/abs/1003.3926}{{\ttfamily arXiv:1003.3926 [hep-ph]}}.

\bibitem{Lai:2010vv}
H.-L. Lai, M.~Guzzi, J.~Huston, Z.~Li, P.~M. Nadolsky, J.~Pumplin, and C.~P.
  Yuan, {\em {New parton distributions for collider physics}},
  \href{http://dx.doi.org/10.1103/PhysRevD.82.074024}{Phys. Rev. {\bfseries
  D82} (2010) 074024},
\href{http://arxiv.org/abs/1007.2241}{{\ttfamily arXiv:1007.2241 [hep-ph]}}.

\bibitem{Dittmaier:2011ti}
{ LHC Higgs Cross Section Working Group} Collaboration, S.~Dittmaier {et~al.},
  {\em {Handbook of LHC Higgs Cross Sections: 1. Inclusive Observables}},
\href{http://arxiv.org/abs/1101.0593}{{\ttfamily arXiv:1101.0593 [hep-ph]}}.

\bibitem{Chatrchyan:2013faa}
{ CMS} Collaboration, S.~Chatrchyan {et~al.}, {\em {Measurement of the $t
  \bar{t}$ production cross section in the dilepton channel in pp collisions at
  $\sqrt{s}$ = 8 TeV}}, \href{http://dx.doi.org/10.1007/JHEP02(2014)024,
  10.1007/JHEP02(2014)102}{JHEP {\bfseries 02} (2014) 024},
  \href{http://arxiv.org/abs/1312.7582}{{\ttfamily arXiv:1312.7582 [hep-ex]}}.
[Erratum: JHEP02,102(2014)].

\bibitem{Khachatryan:2015oqa}
{ CMS} Collaboration, V.~Khachatryan {et~al.}, {\em {Measurement of the
  differential cross section for top quark pair production in pp collisions at
  $\sqrt{s} = 8\,\text {TeV} $}},
  \href{http://dx.doi.org/10.1140/epjc/s10052-015-3709-x}{Eur. Phys. J.
  {\bfseries C75} no.~11, (2015) 542},
\href{http://arxiv.org/abs/1505.04480}{{\ttfamily arXiv:1505.04480 [hep-ex]}}.

\bibitem{Li:1983fv}
T.~P. Li and Y.~Q. Ma, {\em {Analysis methods for results in gamma-ray
  astronomy}},
\href{http://dx.doi.org/10.1086/161295}{Astrophys. J. {\bfseries 272} (1983)
  317--324}.

\bibitem{Cousins:2008zz}
R.~D. Cousins, J.~T. Linnemann, and J.~Tucker, {\em {Evaluation of three
  methods for calculating statistical significance when incorporating a
  systematic uncertainty into a test of the background-only hypothesis for a
  Poisson process}},
\href{http://dx.doi.org/10.1016/j.nima.2008.07.086}{Nucl. Instrum. Meth.
  {\bfseries A595} (2008) 480--501}.

\bibitem{Czakon:2013goa}
M.~Czakon, P.~Fiedler, and A.~Mitov, {\em {Total Top-Quark Pair-Production
  Cross Section at Hadron Colliders Through $O(\alpha_s^4)$}},
  \href{http://dx.doi.org/10.1103/PhysRevLett.110.252004}{Phys. Rev. Lett.
  {\bfseries 110} (2013) 252004},
\href{http://arxiv.org/abs/1303.6254}{{\ttfamily arXiv:1303.6254 [hep-ph]}}.

\bibitem{Czakon:2015owf}
M.~Czakon, D.~Heymes, and A.~Mitov, {\em {High-precision differential
  predictions for top-quark pairs at the LHC}},
\href{http://arxiv.org/abs/1511.00549}{{\ttfamily arXiv:1511.00549 [hep-ph]}}.

\bibitem{Uwer:2007rs}
P.~Uwer, {\em {EasyNData: A Simple tool to extract numerical values from
  published plots}},
\href{http://arxiv.org/abs/0710.2896}{{\ttfamily arXiv:0710.2896
  [physics.comp-ph]}}.

\end{thebibliography}\endgroup
\bibliographystyle{utphysmod}

\end{document}